\documentclass[useAMS,usenatbib]{mn2e}
\usepackage[pdftex]{graphicx}
\usepackage{subfigure,aas_macros}

\newcommand\msun{{M_{\odot}}}

\def\stacksymbols #1#2#3#4{\def\theguybelow{#2}
        \def\verticalposition{\lower#3pt}
        \def\spacingwithinsymbol{\baselineskip0pt\lineskip#4pt}
        \mathrel{\mathpalette\intermediary#1}}
\def\intermediary #1#2{\verticalposition\vbox{\spacingwithinsymbol
        \everycr={}\tabskip0pt
        \halign{$\mathsurround0pt#1\hfil##\hfil$\crcr#2\crcr
                \theguybelow\crcr}}}

\def\lta{\stacksymbols{<}{\sim}{2.5}{.2}}
\def\gta{\stacksymbols{>}{\sim}{2.5}{.2}}

\title[Solving the cooling flow problem via mechanical AGN feedback]
{Mechanical AGN Feedback: Controlling the Thermodynamical Evolution of Elliptical Galaxies}
\author[Gaspari, Brighenti \& Temi]{M. Gaspari$^{1}$\thanks{E-mail:
massimo.gaspari4@unibo.it},
F. Brighenti$^1$ and P. Temi$^2$\\
$^{1}$Astronomy Department, University of Bologna, Via Ranzani 1, 40127 Bologna, Italy\\
$^{2}$Astrophysics Branch, NASA/Ames Research Center, MS 245-6, Moffett Field, CA 94035\\
}

\begin{document}

%\date{Accepted. Received; in original form}

\pagerange{\pageref{firstpage}--\pageref{lastpage}} \pubyear{2012}

\maketitle

\label{firstpage}

%modificato gran parte

\begin{abstract}
A fundamental gap in the current understanding of galaxies concerns the thermodynamical evolution of the ordinary, baryonic matter. On one hand, radiative emission drastically decreases the thermal energy content of the interstellar plasma (ISM), inducing a slow cooling flow toward the centre. On the other hand, the active galactic nucleus (AGN) struggles to prevent the runaway cooling catastrophe, injecting huge amount of energy in the ISM. The present study intends to deeply investigate the role of mechanical AGN feedback in (isolated or massive) elliptical galaxies, extending and completing the mass range of tested cosmic environments. Our previously successful feedback models, in galaxy clusters and groups, demonstrated that AGN outflows, self-regulated by cold gas accretion, are able to properly quench the cooling flow, without destroying the cool core. Via 3D hydrodynamic simulations (FLASH 3.3), including also stellar evolution, we show that massive mechanical AGN outflows can indeed solve the cooling flow problem for the entire life of the galaxy, at the same time reproducing typical observational features and constraints, such as buoyant underdense bubbles, elliptical shock cocoons, sonic ripples, dredge-up of metals, subsonic turbulence, and extended filamentary or nuclear cold gas. In order to avoid overheating and totally emptying the isolated galaxy, the frequent mechanical AGN feedback should be less powerful and efficient ($\epsilon\sim10^{-4}$), compared to the heating required for more massive and bound ellipticals surrounded by the intragroup medium ($\epsilon\sim10^{-3}$).

\end{abstract}


\section{Introduction}

The present paper is the third of a series aimed to investigate how massive
black holes (BHs) control the thermal and dynamical evolution of the gaseous
component of galaxies, groups and clusters. The focus of this investigation
is the effect of AGN feedback on the interstellar medium of
isolated and massive elliptical galaxies.

The key role of active
galactic nuclei outbursts in
sterilising the parent galaxy -- quenching star formation and gas cooling -- 
has become blatant in the last decade. 
The need for sustained gas heating in a range of astronomical systems
comes from both observations and theory.
High spatial resolution X-ray
and radio images of clusters, groups and elliptical galaxies reveal a
clear connection between the nuclear activity and the large scale disturbances
in the hot gas, such as X-ray cavities often filled with radio
emission (e.g. \citealt{Churazov:2000,Diehl:2008a,Dunn:2010,Giacintucci:2011}), and shocks (\citealt{Baldi:2009}).
These features are thought to be the manifestation of the heating process
necessary to prevent the excessive gas cooling
predicted by the classic cooling flow theory (\citealt{Fabian:1994,Mathews:2003}). 
X-ray and UV spectroscopy indicate indeed cooling rates at least one order of magnitude lower
than simple expectations, both in galaxies than in clusters/groups -- this is the
so-called {\it cooling flow problem}
(e.g. \citealt{Peterson:2001,Peterson:2003,Oegerle:2001,Xu:2002,Tamura:2003,Bregman:2001,Bregman:2005,Bregman:2006}). Excellent reviews on this
subject are provided by \citet{Peterson:2006}, \citet{McNamara:2007},
and \citet{Bohringer:2010}.

Another piece of evidence for the reduced cooling rates comes from the lack
of significant star formation in most massive elliptical galaxies
(e.g. \citealt{Ferreras:2000,Trager:2000, Graves:2009,Jeong:2009}),
%Scott et al. 2009 -- v_esc
which are typically located in the quiescent part of the infrared colour-colour
diagram (\citealt{Temi:2009}). 
%an important indicator of the star formation rate.

The problem of ISM heating and the related absence of cold gas in elliptical
galaxies is long-lasting. 
This was investigated forty years ago by \citet{Mathews:1971} seminal paper 
to explain the dearth of cold ISM in typical ellipticals. %, which should accumulate due to stellar evolution. 
%if the gas shed by the stellar population accumulates. 
They proposed that type Ia supernovae (SNIa) explosions
heat and eject the ISM, which is continuously supplied by stellar winds
from evolving stars. Subsequent investigations (\citealt{MacDonald:1981, 
Mathews:1986,Loewenstein:1987,Ciotti:1991,David:1991}) have analysed 
in depth the secular evolution of the ISM in isolated ellipticals. 
%under the action of SNIa heating. 
%These studies have shown that 
SNIa heating is indeed able to prevent gas cooling
and drive global winds in small or intermediate ellipticals but not
in massive galaxies, which usually reside at the centre of a massive
dark halo.
These systems necessitate a different energy source.
 
Furthermore, galaxy formation theory requires a mechanism
to substantially reduce star formation in massive galaxies in order to agree
with the observed sharp cut-off of the luminosity function at the
high-mass end 
%(\citealt{Benson:2003,Balogh:2006,Croton:2006,Faber:2007,Cattaneo:2008,Cattaneo:2009}).
(\citealt{Benson:2003,Balogh:2006,Croton:2006,Cattaneo:2009}), although ellipticals at the centre
of cool-core systems show sometimes significant star formation rates (\citealt{ODea:2008}).
It is now clear that supernova heating alone is insufficient to prevent
the gas from cooling and forming stars (e.g. \citealt{Tornatore:2003,Piontek:2011}). 
Thus, the black hole at the centre of massive galaxies becomes 
the natural suspect for providing the required feedback heating
(\citealt{Binney:1995,Churazov:2002}).
%%[[CITARE SCHAWINSKI ET AL. 2007 FOR EMPIRICAL EVIDENCE OF AGN
%%SF-QUENCHING?]]

From an energetic point of view, AGNs are able to satisfy galaxies
or clusters heating demand. 
For a typical black hole mass of $M_{\rm BH} \approx 10^9$
$M_\odot$ (e.g. \citealt{Ferrarese:2000,Gebhardt:2000}\footnote{The most massive
BH at the centre of brightest clusters can be even an order of
magnitude heavier, like the BH in NCG 4889 
%the brightest cluster galaxy of the Coma cluster, 
with $M_{\rm BH} \sim 2\times 10^{10}$ $M_\odot$ (\citealt{McConnell:2011}).}), 
an amount of energy $E_{\rm BH} \approx 0.1\, M_{\rm BH} c^2 \approx 2\times 10^{62}$ erg could in
principle be injected in the surrounding gas. This is a
sizeable fraction of the whole energy radiated away by the
intracluster medium (ICM) in a massive cluster lifetime, and is significantly
larger than the binding energy of the interstellar medium 
in a typical elliptical
galaxy or group. For example, the gas binding energy for the massive, 
X-ray bright group NGC 5044 is $\approx 9\times 10^{61}$ erg (\citealt{Mathews:2005}).

Many fundamental questions about the physics of the AGN feedback, however,
remain unanswered (see also \citealt{Ostriker:2010}). 
A key issue is to identify the physical process
through which the AGN interact with the surrounding gas.
Several mechanisms have been proposed and numerically investigated, 
including radiative heating by quasars %QSOs
(\citealt{Ciotti:1997}), cavities generated through the injection
of thermal energy or cosmic rays
(\citealt{DallaVecchia:2004,Bruggen:2005,Mathews:2008,Mathews:2009,Guo:2010}), 
or bipolar mechanical outflows/jets (\citealt{Omma:2004,Soker:2005,Zanni:2005,Brighenti:2006, 
Vernaleo:2006,Bruggen:2007,Gaspari:2009,Gaspari:2011a,Gaspari:2011b,Gaspari:2012}). 

In low-redshift systems %ellipticals and clusters 
it seems likely that massive black holes mostly
accrete mass and return energy in a radiatively inefficient way 
(\citealt{Fabian:1995,Loewenstein:2001,DiMatteo:2003,Taylor:2006}). 
Thus, while luminous quasars might have been
an important source of heating at high redshift, with the peak of the QSO distribution at $z\sim2$, 
(\citealt{Nesvadba:2008,Moe:2009,DunnJP:2010}), 
observations point to advection dominated accretion flow 
(ADAF)-like systems as a primary
source of local AGN feedback. These accretors naturally generate 
powerful outflows (\citealt{Narayan:2008} and references therein).

Recent X-ray and radio observations (see the reviews by 
\citealt{McNamara:2007} and \citealt{Gitti:2012})
provide a crucial guidance
to narrow down the range of the possible heating scenarios.
The common presence of radio-filled X-ray cavities, ellipsoidal weak shocks
and metal (iron) enhancements along the radio jet path
(e.g. \citealt{Gitti:2010,Kirkpatrick:2011,Randall:2011})
strongly suggests that local AGNs introduce
energy directionally and in mechanical form, likely as bipolar massive outflows or jets.

For this reason in \citeauthor{Gaspari:2011a} (\citeyear{Gaspari:2011a},b -- G11a,b) 
we tested a variety of outflow models in clusters
and groups in order to verify whether this mechanical feedback is able to
quench cooling without overheating the core, an often
unwelcome byproduct of central energy injection (Brighenti \& Mathews 2002, 2003, 2006). 
We showed that massive, subrelativistic outflows are indeed
a viable mechanism to solve the cooling flow problem in clusters and groups. 
In lighter systems the AGN feedback must be however more gentle and continuous, 
or, in other words, in massive clusters a possible
quasi continuous low-power activity must be intermixed with
sporadic powerful events. 

In the present work we study how the AGN feedback self-regulates
in a galactic environment and test the models against several observational constraints.
Massive elliptical galaxies are in fact important
laboratories to study the AGN feedback process in the local universe.
Their ISM shines in X-ray and many giant ellipticals are relatively nearby, within
a distance of $\sim 15-30$ Mpc -- a valuable virtue compared to X-ray
bright, but distant, massive clusters. The relative proximity allows thus
the investigation of a region closer to the AGN, where the feedback
engine might manifest itself in a clearer way. 

In Section 2, we present our theoretical modelling and the details
of the numerical methods.
%As discussed in Section 2, 
We adopt a simple recipe to activate
the outflows, based on gas cooling near the central black hole.
We discuss in Sec. 2.1.1 why it is not feasible to investigate
with appropriate accuracy the accretion and the outflow generation processes.
We turn instead our attention to the effect of the outflows 
on large scales. We analyse the observable consequences of
the assumed feedback scenario on the environment of an isolated elliptical (Section 3)
and a galaxy with circumgalactic gas (Section 4). %, i.e. mechanical anisotropic feedback.   
In Section 5, we discuss the key features of the results
presented in this work, and summarise the effects of mechanical AGN feedback on every scale,
from galaxies to massive clusters.
%of all masses (from galaxies to clusters).
%wrapping up our entire study
%on mechanical AGN jet feedback from the large scales of massive clusters down to isolated galaxies.

\section[]{Theoretical and Numerical Setup}

\subsection{AGN feedback}
\subsubsection[]{What is the accretion rate?}  %% and black hole growth}

As remarked in the Introduction, massive black holes in 
nearby X-ray bright ellipticals likely accrete gas through a
radiatively inefficient mechanism. ADAF models
naturally predict outflows and jets (\citealt{Narayan:2008}),
which might provide the mechanical feedback we study
in the present article.

Our feedback scheme assumes that the black hole reaction 
is proportional to the accretion rate (Sec. 2.1.2). This simple conjecture 
leads, however, to a substantial difficulty in building a self-consistent
model, i.e. properly estimating the BH accretion rate. This is such a complex
problem that we should not be ashamed to accept our ignorance.
We give below some reasons why the accretion rate 
(and the BH growth) can not be investigated in depth
within the frame of our models, and must therefore be estimated 
with some subgrid scheme, which usually depends on the uncertain
gas properties in the nuclear region. 
Because the simulated ISM in the central few hundreds pc is quite different from that of real
ellipticals, our calculated accretion rate should not be trusted with high confidence.
We suspect that this is true for most numerical works on AGN feedback, either
in galaxies or clusters.
Specifically, the motivations which lead to a rather uncertain
accretion rate are the following.
\\

%\begin{enumerate}
(i) In order to self-consistently model the three-dimensional 
accretion flow, using magneto-hydrodynamics, the necessary condition is to
numerically resolve a region with size a few Schwarzschild radii, $R_{\rm S} \approx 3\times 10^{14}
(M/ 10^9{\it M}_\odot)$ cm $\approx \, 10^{-5} R_{\rm B}$ (where
$R_{\rm B}$ is the usual Bondi radius). This is currently not feasible for
(3D) simulations aimed to investigate galactic-scale flows.
Because the direct calculation is impractical,
we are left to rely on a sub-grid prescription to estimate the accretion rate.
%of the central black hole. 
However, this is a formidable task; %, in our opinion,
the exact quantitative of material that is considered  `accreted' must be taken
with serious caution and only as a rough approximation.
%so uncertain to make the results hard to swallow.}

(ii) Many studies -- including ours (G11a,b) -- use the Bondi
(1952) theory to estimate the instantaneous mass accretion rate onto the BH 
(e.g. \citealt{Springel:2005,Cattaneo:2007,Sijacki:2007,Puchwein:2008,
Booth:2009,Dubois:2010}).
In order to make results compatible with observations,
the Bondi accretion rate is often artificially and arbitrarily boosted
by a factor of $\sim 100$ (see \citealt{Booth:2009} for a 
related discussion).
While this is a convenient parametrisation, it seems unsafe to rely
on it for a realistic representation of the accretion process.
Beside complications such as turbulence, rotation and magnetic fields
(see \citealt{Krumholz:2005,Krumholz:2006,Igumenshchev:2006,Moscibrodzka:2008,
Moscibrodzka:2009,Narayan:2011}),
which can easily change the classic accretion rate by a factor of a few,
we believe that cooling (and heating) 
%%and the complex multiphase character of the central ISM
make the Bondi assumption particularly hazardous. 
The central ISM can be prone to thermal instabilities which can drastically
change the nature of the accretion process (\citealt{Pizzolato:2005,Gaspari:2012,McCourt:2012,Sharma:2012}).
In fact, in the presence of
a classical cooling flow the mass inflow rate at some small radius is determined
by the gas cooling process rather than the BH gravity. Thus, we expect
$\dot M_{\rm BH} \approx 0.1-1$ M$_\odot$ yr$^{-1}$ for an isolated
elliptical galaxy.
Accretion scenarios suggested by \citet{King:2007} and \citet{Pope:2007,Pope:2009}
%\citet{King:2009} 
also differ significantly from the Bondi prediction.

The common assumption in numerical
models is to take as Bondi boundary conditions
some average of density and temperature within
a radius larger than $R_{\rm B}$, usually a simulated region filled with smooth
hot gas.
However, the real ISM at the fiducial Bondi radius for the hot gas
($\approx 50$ pc) 
is often constituted by a multiphase, dusty $\sim 10^4$ K gas
(\citealt{Heckman:1989,Shields:1991,Buson:1993,Goudfrooij:1994,Macchetto:1996,
Ferrari:1999,Martel:2000,Tran:2001,Colbert:2001,Verdoes:2002}),
%Patil et al. 2007,
which can be both dynamically relaxed or in chaotic motion
(\citealt{Caon:2000,Sarzi:2006,Sarzi:2010}). This warm gas is likely 
intermixed with hot ISM, and the interaction between the two phases
is poorly understood. With such a complex gas conditions it is
difficult to estimate the accretion rate.
With 1D or 2D simulations it is relatively easy to resolve the Bondi
radius in galactic flows
(e.g. \citealt{Ciotti:1997,Brighenti:1999a,Quataert:2000,Novak:2011}), yet the necessarily
simplified gas physics adopted and the limits imposed by the
geometry prevent these models to catch the complexity of the
ISM near the black hole.

Because of these (not easily quantifiable) uncertainties we
prefer to not assume any sophisticated subgrid modelling
for the accretion process.

(iii) Finally, the assumptions made for stellar source terms
(smooth injection of mass and energy from evolved stars and SNIa),
described in Section 2.2.1 (see also \citealt{Mathews:1971,
Loewenstein:1987,Mathews:2003}), clearly
break down within the galactic core. A mature SNIa remnant would
have a size of $\approx 50$ pc if the external medium has a
number density of $\sim 0.1$ cm$^{-1}$ and a temperature
$\sim 10^7$ K, comparable with the galactic core radius.
Stellar winds from orbiting stars likely generate long tails
of warm gas, several tens of pc in size (\citealt{Mathews:1990,Parriott:2008,
Bregman:2009}), giving rise to inhomogeneities.
The stellar material can indeed help explaining the ubiquitous
presence of emission line gas at the centre of early-type
galaxies (\citealt{Mathews:2003}).
These intrinsic limitations, along with the lack of merging as
possible source of gas, hamper a proper modelling of the ISM
in the central $\approx 100-200$ pc.
%\end{enumerate}
\\

For all the previous reasons, we believe that it is futile to investigate the details
of BH accretion history in the present work. Therefore, we realistically limit the goal of our study
to the effects of (large scale) outflows on the ISM
evolution. 
%The investigation is aimed to test the mechanical outflow mechanism as the dominant
%AGN feedback mode.

\subsubsection[]{Self-regulated mechanical outflows}
Encouraged by the results in G11a and G11b for clusters and groups,
we consider here subrelativistic collimated outflows as the main ingredient of the
AGN mechanical feedback in elliptical galaxies.
In considering massive slow outflows, 
we are implicitly assuming that the relativistic
jet entrains some ISM mass (the active mass $M_{\rm act}$, defined
below) or perhaps that a wind originate from the accretion disk
(e.g. \citealt{Crenshaw:2003,Morganti:2007,Nesvadba:2008,Cappi:2009,Cappi:2011,Tombesi:2010a,Tombesi:2010b,Tombesi:2012}; see also
the references in G11a,b). 
Moreover, radio jets are highly relativistic on pc scale, 
but rapidly decrease to subrelativistic velocities within 
few kpc from the black hole (\citealt{Giovannini:2004}).

In G11b we showed that, on galaxy group scales, the self-regulation
imposed by cold gas accretion tends to generate a quasi-continuous, gentle
feedback, qualitatively similar to what expected for a boosted (by a
factor of $\sim 100$) Bondi accretion.
However, on top of this low level AGN activity, several powerful
outbursts can occur, which produce visible cavities and shocks, a
desirable feature of those models.
Therefore, in the present work on elliptical galaxies, we preferred to
adopt a cold feedback mechanism (see also \citealt{Pizzolato:2005}), 
where the assumed accretion
rate is linked to the cooled gas in a given central region. 
As pointed out in Section 2.1.1, there are several reasons 
to believe that a purely Bondi accretion scenario (that is, the hot
feedback mode) is inappropriate.

%There are also physical reason
%to dislike Bondi accretion. First, the ISM has never zero angular
%momentum, as required by the theory (\citealt{bon52}) and thus reducing
%the net accreted mass. Second, we have carried out a very high resolution
%simulation ($\sim1$ pc) and discovered that the usual `Bondi accretion rate'
%calculated on kpc scale
%is always at least two orders of magnitude
%below the cooling rate (dropout mass), in
%the presence of a radiative cooling flow (see a forthcoming paper).
%Third, its intrinsic continuous nature poses a great riddle in 
%explaining any kind of AGN quiescence or duty cycle.

We slightly improved the feedback self-regulation
mechanism compared to G11a,b. 
The radius of the region used to estimate the accreted cold gas -- usually several kpc --
was rather arbitrary, and we assumed
that the jet is instantaneously
triggered only by the gas cooled near the centre of the system.
This lengthscale
was reasonable because almost all the cooling occurred inside the chosen
radius.

In order to eliminate this parameter, we adopt here the physical assumption that
the cooled gas (dropped out at radius $r_{\rm i}$; Sec. 2.3.2) goes into free fall and accretes onto
the black hole in a free-fall time, given by
\begin{equation}\label{tff}
t_{\rm ff}=\int^{0}_{r_{\rm i}} \frac{dr}{\sqrt{2\,|\phi(r)-\phi(r_{\rm i})|}}\,,
\end{equation}
where  $\phi$ is the gravitational potential (associated with the stellar and dark matter mass distribution).
%is given by the stellar
%(de Vaucouleurs) and dark matter (NFW) mass distribution.} 
We also impose
that the infalling gas must have a low angular momentum, 
i.e. less than $v_{\rm c}(\Delta r)\,\Delta r$ 
(circular velocity times the minimum radial grid size). 
Most of the cooled gas tends to have nevertheless a negligible 
angular momentum (see also \citealt{Pizzolato:2010}).
This new method does not require the fixed radius used in G11a,b, 
since the accretion is no more instantaneous while it is set by the free-fall time delay; 
the latter is dependent on where the cold clouds locally condense out of the hot phase.

The cold mass reaching the centre at a given time, $\Delta M_{\rm c, tot}$,
triggers an outflow with kinetic energy given by
\begin{equation}\label{cold}
\Delta E_{\rm jet} = \epsilon\,\Delta M_{\rm c, tot}\, c^2,
\end{equation}
where $\epsilon$ is the parametrized mechanical efficiency of the feedback.
It is likely that the efficiency depends on the ratio between
the actual accretion rate and the Eddington rate,
$\dot{M}_{\rm Edd}\approx22\,(M_{\rm BH}/10^9\msun)$ $\msun\,$yr$^{-1}$
%defined as 
%$\dot M_{\rm Edd}\,=\,4\pi G m_pM_{\rm BH}/\eta c \sigma_{\rm T}$
(\citealt{Churazov:2005,Merloni:2008}).
However, given the uncertainties in estimating the accretion rate, 
we prefer to assume a constant efficiency.

This kinetic energy (and the associated momentum)
is given to the gas located in a small region at the centre of the grid
(the `active' jet region), whose size is usually\footnote{A
slightly smaller or larger active region does not drastically alter the global evolution (see G11a,b). 
The same is true if we inject the energy from the internal boundaries of
the grid.} 
$\Delta R \times \Delta z \,=\, 2\times1$ cells (about 300 pc wide, 150 pc high),
containing some hot gas mass $M_{\rm act}$. 
The outflow velocity was retrieved as usual (G11a,b), imposing
\begin{equation}\label{cold2}
\frac{1}{2} M_{\rm act} v_{\rm jet}^2 = \Delta E_{\rm jet}.
\end{equation}

%\subsection[]{Galaxy models: ``isolated'' and with circumgalactic gas}
\subsection{Galaxy models}
\subsubsection{Isolated elliptical}

In this Section we describe the modelling of the elliptical galaxy, %parameters for the galaxy model
providing the initial conditions for the simulations.
We model the massive elliptical galaxy with a de Vaucouleurs'
stellar density profile (approximated as in \citealt{Mellier:1987}),
with effective radius $r_{\rm eff}=8.5$ kpc and total stellar mass
$M_{\ast} = 7\times 10^{11}$ $M_\odot$. Assuming a stellar mass-to-light ratio
of 8 (in the B band), appropriate for an old stellar population,
the resulting luminosity is $L_B = 8.75 \times 10^{10}$ $L_{\odot}$.
Thus, our model represents a typical elliptical galaxy similar to NGC 4649, NGC 4472
and many others in the local universe, albeit without the intragroup medium.
The dark matter is modelled as a NFW halo (\citealt{Navarro:1996})
with virial mass $M_{\rm vir} = 4\times 10^{13}$ $M_\odot$ and a 
concentration $c=8.8$ (\citealt{Bullock:2001,Humphrey:2006b}).

Several ingredients characterise the stellar mass and energy sources
(stellar winds
and SNIa). The specific stellar wind mass return rate is assumed to be
$\alpha_{\ast} = 4.7 \times 10^{-20} (t/t_{\rm n})^{-1.3}$ s$^{-1}$, where
$t_{\rm n} = 13$ Gyr is the galaxy age (results are not sensitive
to the adopted $t_{\rm n}$). This formula is a good approximation for
a single stellar population older than $~ 100$ Myr. %(further details can be found in
The SNIa mass return rate depends on the  
SNIa rate, $r_{\rm Ia} = 0.06\, (t/t_n)^{-1.1}$
SNu (\citealt{Cappellaro:1999,Greggio:2005,Mannucci:2005}).  %,Greggio:2008}). 
With the chosen SNIa rate, the specific SNIa mass return rate is
$\alpha_{\rm Ia} = 3.17 \times 10^{-20}\, r_{\rm Ia}(t) (M_{\rm ej}/M_\odot)/(M_\ast/L_B)$,
where $M_{\rm ej}=1.4$ $M_\odot$ is the ejecta mass of a SNIa and
$M_\ast/L_B = 8$ is the stellar mass-to-light ratio in the B band.
The adopted $r_{\rm Ia}$ results in gas abundances consistent with observations 
(\citealt{Humphrey:2006a}). 

The energy injection associated to stellar winds and SNIa is treated as
outlined in \citet{Mathews:2003}. The stellar velocity dispersion,
which determines the temperature of the injected wind gas, is calculated by
solving the Jeans equation for a spherically symmetric, isotropic stellar
system.
Fortunately, the stellar wind heating has not a large impact on the
flow and it is not necessary to calculate $\sigma_{\ast}(r)$ using more sophisticated
dynamical models. 

Concerning the gaseous component, we decided to simulate two kinds
of systems. In the first one, indicated as the `isolated galaxy',
the ISM is produced by internal processes alone (stellar mass loss
and SNIa ejecta). Although the hot ISM of massive, X-ray bright ellipticals 
can not be realistically explained without a circumgalactic gas component
(\citealt{Brighenti:1998,Brighenti:1999a}), we decided to use this approach for its
simplicity and to allow a direct comparison with previous calculations
of hot gas flows in ellipticals (e.g. \citealt{Loewenstein:1987,Ciotti:1991, 
Ciotti:1997}). Within $\sim r_{\rm eff}$ most
of the interstellar medium is indeed likely to come from stellar mass loss
(\citealt{Brighenti:1999a}).
Thus, the terminology `isolated galaxy' does not refer to
the presence of other (large) galaxies nearby, but %it is `isolated'
that the ISM is not contaminated by the group or 
cluster gas. 

The simulations of the isolated galaxy start at cosmic time
$t=1$ Gyr % (lasting about 12 Gyr),
with the galaxy essentially devoid of gas, conforming to the usual
assumption that the gas has been
cleared by a SNII-driven wind.
The ISM is then gradually supplied by the stars.
After few $10^8$ yr, the system loses memory of the initial conditions
and, if not heated, approaches a quasi-steady state, with secular changes
controlled by the slow variation of the stellar mass return rate
(\citealt{Loewenstein:1987,Ciotti:1991}). The simulations are evolved
to final time, $13$ Gyr.

\subsubsection{Elliptical with circumgalactic gas}

In the second class of models, more realistic and appropriated for a detailed
comparison with well-observed galaxies, we take into account the presence of
the circumgalactic gas (CGG). Many of the famous X-ray ellipticals belong to
this breed. A distinctive characteristic of classic cooling flow models
with CGG is the presence, often observed in real systems, of a relatively
broad cool core, in perfect analogy with galaxy clusters (\citealt{Brighenti:1998,Brighenti:1999a,
Humphrey:2006b,Diehl:2008b}). 
The separation between galaxies with CGG and galaxy groups is largely
semantic, so these models are tightly linked to those calculated in G11b.
However, here we focus mainly on the region close to the galaxy 
($r\lta r_{\rm eff}$). At variance with G11b, we used an improved feedback
scheme (Sec. 2.1.2) and adopted a significantly higher numerical resolution.

The galaxy parameters are chosen to agree with NGC 5044,
a X-ray bright galaxy in the centre of the homonymous group, with
$r_{\rm eff}=10$ kpc, $M_{\ast} = 3.4\times 10^{11}$ $M_\odot$
and $M_\ast/L_B = 7.5$ (see \citealt{Buote:2004}). 
In these simulations we use the observed $T(r)$ and $n(r)$ profiles
(\citealt{David:1994,Buote:2003}; dotted lines in Figure \ref{fig:cgg}) to retrieve the total gravitational potential under the assumption of hydrostatic equilibrium.

We run the CGG models for 7 Gyr. As emphasised in G11a, it is crucial 
to check the long-term behaviour of a heated model in order to assess its merit.
Unfortunately, short-term simulations can in fact return misleading results
on a particular feedback model as a solution of the cooling flow problem.

\subsection[]{Numerical techniques}

\subsubsection[]{Code setup}
As in G11a,b we have used a substantially modified version
of the 3D hydrodynamic code FLASH 3.3 (\citealt{Fryxell:2000}).
The main advantages include 
the adaptive mesh refinement (AMR) block structure,
suited for a very efficient scalability 
trough the Message-Passing Interface (MPI).
The simulations were carried out on the parallel 
high-performance clusters
SP6 (CINECA) and Pleiades (NASA).
%or on a private small cluster. 
We employed the PPM split scheme to treat the transport terms in the conservation equations
(in principle third-order accurate).
A full description of the
hydrodynamic equations can be found in G11a,b, with the additional
source terms representing AGN feedback, radiative cooling, and stellar winds plus SNIa.

The computational rectangular 3D box in almost every model extends up to
150 kpc. We have simulated the $z > 0$ half-space with reflection boundary
condition at $z=0$, setting elsewhere the usual outflow
condition with inflow prohibited.
Despite the AMR capability of FLASH, we have 
decided to use a number of concentric
fixed grids in cartesian coordinates. This ensures a proper resolution of 
the waves and cavities generated in the core by the AGN outflows.
We employed a set of 7 grid levels (basic blocks of $16\times16\times8$ points),
with the zone linear size doubling among adjacent levels. 
The most distant regions from the centre are covered by at least 
level 4.
The finest, inner grid (level 7)
has a resolution of $\Delta x = \Delta y = \Delta z =146$ pc for the
isolated galaxy models, and covers a spherical 
region of $\sim8$ kpc in radius.
For the simulations with circumgalactic gas we were forced to lower the
resolution to save computational time, $\Delta x = 293$ pc (the inner grid covers $r\sim16$ kpc); the box is instead four times bigger (8 levels).
In general, grids of every level extend radially for about 55 cells.
This is arguably the best way to cover very large spatial scales
(hundreds pc up to hundreds kpc) and at 
the same time integrating the system for several Gyr.

\subsubsection{Radiative cooling}
The radiative cooling is a necessary ingredient for studying the cooling flow problem.
We adopt the cooling function by \citet{Sutherland:1993}, dependent on gas temperature
and metallicity,
$\Lambda(T,Z)$, for a fully ionised plasma.
The iron abundance is consistently calculated 
by solving a passive advection equation for the iron density,
with the appropriate sources representing the injection of metals by
stellar winds and SNIa (\citealt{Mathews:2003}). We assume
that the abundance of all heavy elements in solar units is the same
as the iron one. Although this is not strictly true, it has a negligible
effect on the cooling process.

We numerically handle the gas cooling at very low temperatures
by using a dropout (mass sink) term: $-q(T)\rho/t_{\rm cool}$.
This is a convenient scheme to remove cold gas from the numerical grid
without altering the calculated cooling rate (see G11a for
more details).

The numerical implementation has been improved.
As commonly done, we are using the splitting method for
adding the source/sink terms, i.e. first solving the
standard hydro-equations and then,
adopting the updated flow variables,
integrating the ordinary differential equation (ODE)
%(e.g., ${\rm D}\rho/{\rm D}t$) 
associated with the source term.
The ODE related to cooling
is solved through a second-order Runge-Kutta explicit method.
%accurate and stable enough for our problem. 
%In the whole process, it is
%important to properly `centre' the differentials, 
%In order
%to retain the second-order accuracy.
Expanding the cooling time, $t_{\rm {cool}}=2.5\,n k_{\rm b}T/n_{\rm e}n_{\rm i} \Lambda(T,Z)$, and centring the variables,
%and separating the variables, 
leads to the
the following discretisation for the dropout ODE:
\begin{equation}\label{drop}
\rho^{\rm n+1}=\rho^{\rm n}\left(1+\frac{\mu}{2.5\,k_{\rm b}\mu_{\rm e} \mu_{\rm i} m_{\rm p}}\frac{q^{\rm \,n+1/2}\,
\Lambda^{\rm n +1/2}\rho^{\rm n}}
%\Lambda(T^{\rm \,n+1/2},Z)}
{T^{\rm \,n+1/2}}\,\Delta t \right)^{-1},
\end{equation}
where the indexes refer to the temporal states, 
$\Delta t$ is the timestep, and
the molecular weight per particle, electron and ion
are $\mu\sim0.62$, $\mu_e \sim 1.18$ and $\mu_i\sim 1.30$,
respectively; $T^{n+1/2}$ is conveniently provided by the previous Runge-Kutta midpoint estimate.
The functional form 
of $q\,=\,2\,\exp [-(T/T_{\rm c})^2]$ ensures that the gas is considered
cold and dropped out of the flow as soon as it approaches $T_{\rm c}$.
Setting $T_{\rm c} = 2\times10^4 K$ prevents the formation of very cold clumps
in the flow, whose physical evolution can not be easily followed by the present computations.
It has also the benefit to prevent the drastic
decreasing of the timestep. %associated with very cold gas. %with $T\approx 5\times 10^5$ K.
%This new implementation drops the cold gas in a smoother
%way per timestep, while the mean cooling rate remains approximately the
%same as before. 
Lowering $T_{\rm c}$ permits to follow better the formation
of local thermal instabilities, nevertheless the global evolution of the
cold gas and associated feedback is essentially the same 
on long timescales (see \citealt{Gaspari:2012}). %(see Sec. 3.X).

The dropout term, as any source step, has a time limiter condition associated.
Using the dropout ODE equation, we find that the relative density decrement, $D=1-\rho^{n+1}/\rho^n$,
is linked to the following timescale: 
$t_{\rm drop} =[D/q(1-D)]\,t_{\rm cool}$.
%\begin{equation}\label{limdrop}
%t_{\rm drop} =\left(\frac{D}{1-D}\right)\frac{t_{\rm cool}}{q}.
%\end{equation}
%is important to not change the density value by more than $\sim 50$\%
%per timestep. 
%We usually adopt an even stricter condition,
%limiting the variation to $\sim 20$\%.
It is evident that the dropout limiter is tightly correlated with the cooling limiter, 
%The limiter associated to the radiative cooling is similar:
$\eta\, t_{\rm cool}$ (with $\eta<1$). %(again $\eta$ should be  $\lta 0.5$).
%The comparison with Eq.    
%\ref{limdrop} shows that 
Limiting the variation of internal energy has the consequence to directly reduce the dropped mass,
quantitatively: $D=\eta\, q/(1+\eta\, q)$. 
In order to smoothly couple the hydro-solver to the sink terms, a reasonable
limiter is $\eta=0.4$; %$\sim 0.5$ $t_{\rm cool}$. 
%Fixing our fiducial $\eta=0.4$
since the maximum $q$ value can be 2, the dropout decrement
will never exceed $\approx45$\% per timestep (it is usually much lower).

\section[]{Results: isolated galaxy}

\begin{table}
 \label{params}
%\begin{minipage}{200mm}
\caption{Properties of the most relevant simulations.}
\begin{tabular}{@{}lcccccc}
\hline 
%       &                       &  $\epsilon$         & Notes      \\ 
Model  & Feedback              &  efficiency ($\epsilon$)       & Notes \\ 
\hline
 iso-CF    & no AGN heating        &      -           &   isolated               \\
 cgg-CF    & no AGN heating        &      -           &   CGG               \\
 iso-1em4 & cold & $10^{-4}$       & isolated    \\
 iso-3em4 & cold & $3.3\times10^{-4}$       & isolated    \\
 iso-1em3 & cold & $10^{-3}$       & isolated    \\
 cgg-8em4 & cold & $8\times10^{-4}$       & CGG    \\
\hline
\end{tabular}
%\end{minipage}
\end{table} 

In Section 3 we report the results for models in which the ISM is produced
only by the mass loss of the galactic stellar population. 
We explore the effect of the mechanical feedback varying the key parameter,
the efficiency $\epsilon$ (see Table 1). In order to solve the cooling flow problem, 
AGN outflows must  quench the cooling flow, without
overheating the ISM in a drastic way. The former request is easy to satisfy, if the feedback
is energetic enough. The simultaneous fulfilment of the two conditions above
is instead complicated, and often require an unpleasant fine tuning of the
heating parameters (\citealt{Brighenti:2002,Brighenti:2003}).
A successful feedback needs also to reproduce
other observational features, such as cavities, shocks, turbulence, multiphase gas, and metals dredge-up.

\subsection[]{Pure cooling flow}

\begin{figure} 
\centering
\includegraphics[scale=0.76]{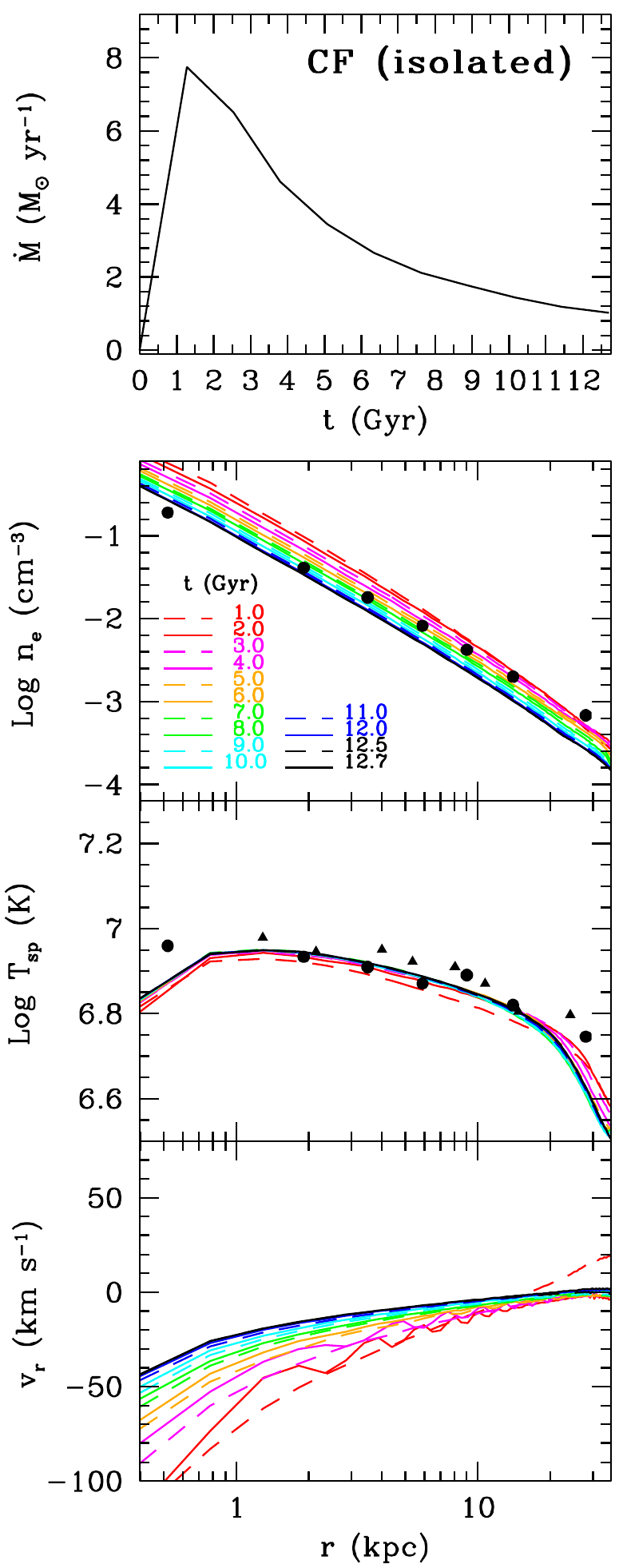}
\caption{Evolution of the cooling flow model (no AGN feedback), for the isolated galaxy (iso-CF). From top
to bottom panel: gas cooling rate ($\dot M$) as a function of time,
radial profiles of electron number density ($n_{\rm e}$),
%at 12 different times as indicated in the figure; 
projected spectroscopic-like temperature ($T_{\rm sp}$), and
radial velocity ($v_{\rm r}$). Times and colours are indicated in the second panel.
Filled points represent observational data for the elliptical galaxy NGC 6482
(circles: \citealt{Khosroshahi:2004}; triangles: \citealt{Diehl:2008b}).
See Section 3.1 for more details.}
\label{fig:iso-CF} 
\end{figure}

In Figure \ref{fig:iso-CF}
we show the relevant global quantities which characterise a pure cooling
flow model, with AGN heating switched off 
and the heating provided mainly by the SNIa.
This is a classic result: the densest gas loses thermal pressure support due to radiative losses,
inducing a slow subsonic inflow ($v_{\rm r}\lta50$ km s$^{-1}$) that further increases the 
plasma emissivity (a detailed discussion can be found in \citealt{Mathews:2003} and references 
therein). We emphasise that when no AGN heating is considered,
the gas cooling rate (top panel), occurring at the very centre of the galaxy,
roughly decreases in pace with the stellar mass loss ($\dot M_{\ast}
\propto t^{-1.3}$), reaching $\dot M_{\rm cool} \approx 1$ $M_\odot$
yr$^{-1}$ at $t=12.7$ Gyr, when the simulation ends. Needless to say,
this is a much larger cooling rate than allowed by the observations
quoted in the Introduction. The rapid initial increase and the
  peak of $\dot M_{\rm cool}$ at
  $t\sim 1$ Gyr reflect the transient phase during which the
  ISM is built up by the stellar mass loss.

The X-ray luminosity also secularly drops, from $L_{\rm X} \sim 5\times10^{42}$
erg s$^{-1}$ at $t=2$ Gyr to $\sim 9 \times 10^{41}$ erg s$^{-1}$ at final time.
We refer here to the bolometric X-ray luminosity, calculated within $r\lta 200$ kpc.  
The reason for the luminosity drop 
is the decline of the ISM density illustrated in the middle panel of Figure \ref{fig:iso-CF}. 
The slope of the density radial profile does not vary significantly
with time and, as expected, is too steep at the centre, a well known
problem of classical cooling flow models (\citealt{Sarazin:1988,Sarazin:1989}).

We also note that classical cooling flows in isolated galaxies lack
the large cool core typical of cooling flow clusters or groups (see the
spectroscopic-like\footnote{To be more adherent to observations, we
avoid the idealised emission-weighted temperature, instead we
use the prescription by \citealt{Vikhlinin:2006} to approximately 
estimate a more realistic projected temperature.
%extract $T$ from the X-ray spectrum that {\it Chandra} would observe
%in the range $T\ga0.5$ keV (see \citealt{Vikhlinin:2006}). 
The two types of profile appear nevertheless similar in the majority of the cases.}
temperature profile in the third panel). Instead, the temperature
monotonically increases toward the centre (except for the very inner region).
This can be understood because
the steep gravitational potential associated with the peaked
de Vaucouleurs' density profile (the dark halo is 
sub-dominant within $r_{\rm eff}$) provides enough gravitational
heating to the inflowing -- thus compressing --
gas to balance the radiative losses; needless to say, this process
terminates when the gas reaches the centre, where it cools catastrophically.

Ideally, we would compare the profiles with a well-observed
elliptical which is at the same time massive {\it and} isolated, in the
sense of Section 2.2.1.
However, most of the known giant ellipticals
reside at the centre of groups 
or clusters and their hot ISM is contaminated by the CGG.
%This strengthen our 
%belief that the galaxy evolution should be studied in conjunction
%with the environment, in order to explain the majority of 
%observed systems (see Sec. 4 or G11b). 
NGC 6482 is one of the few massive and X-ray bright ellipticals with a
negative temperature gradient (\citealt{Khosroshahi:2004}) and we show
the data of this system in Figure \ref{fig:iso-CF} for comparison.  On
the other hand, we note that NGC 6482 presents some differences
compared to our model. The dark halo is less massive
(\citealt{Khosroshahi:2004,Humphrey:2006b}) and it owns a relatively
extended X-ray halo with $L_{\rm X}\sim10^{42}$ erg s$^{-1}$, likely
contaminated by CGG. Nevertheless for $r\lta r_{\rm eff}$, i.e. the
region more relevant to our investigation, the hot gas should be
dominated by the internal component only.

\subsection[]{AGN feedback: $\epsilon = 10^{-4}$}

\begin{figure*} 
\centering
\includegraphics[scale=0.558]{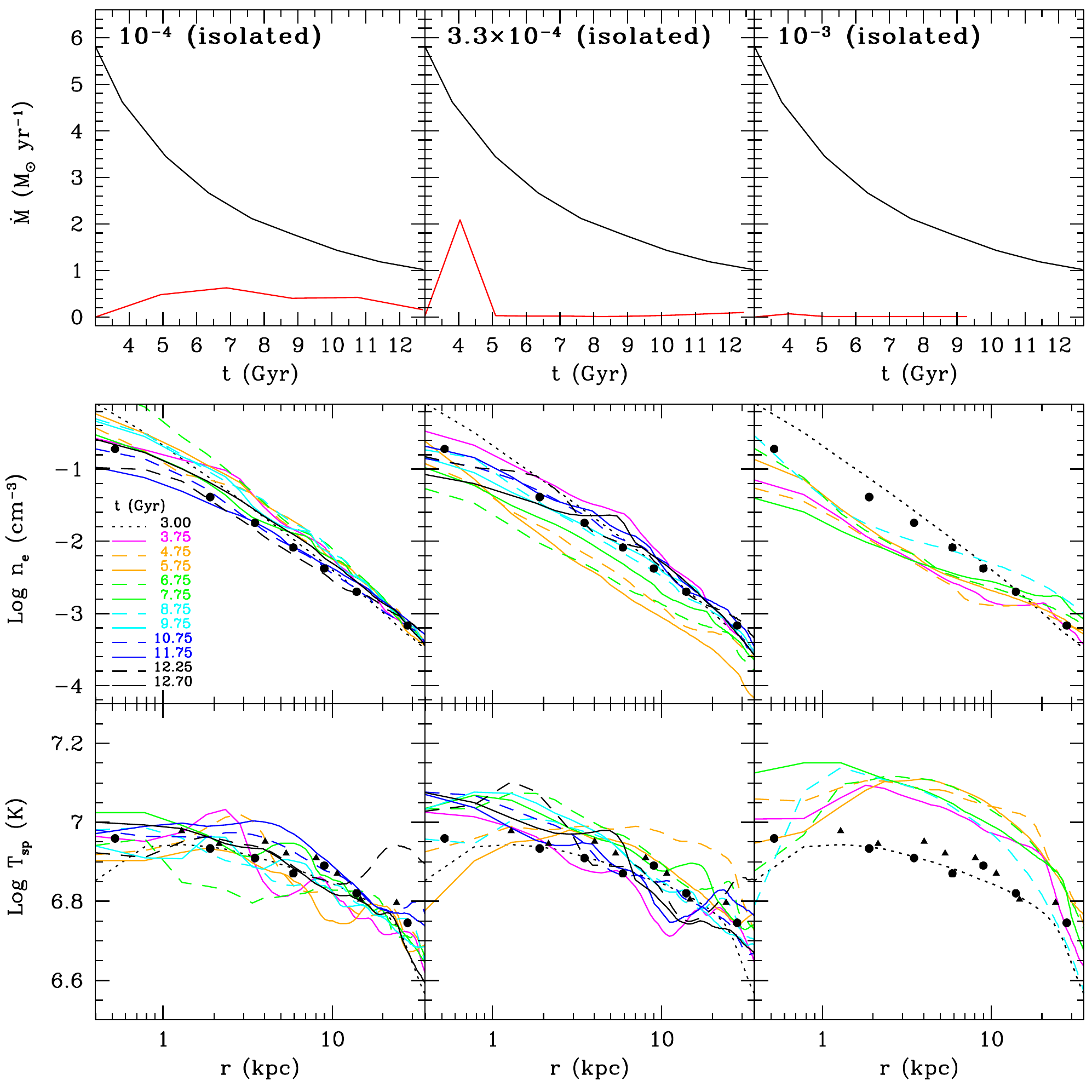}
\includegraphics[scale=0.558]{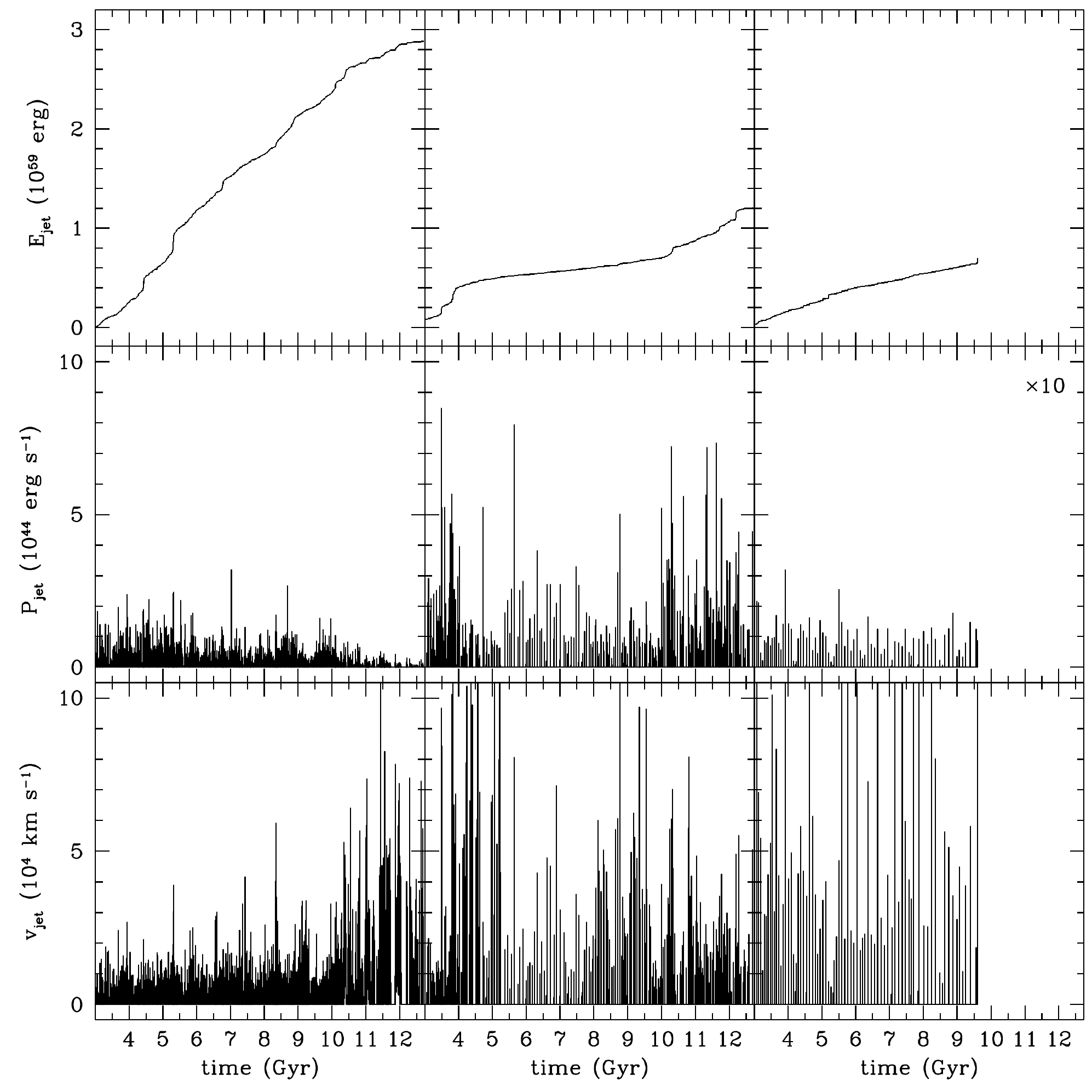}
\caption{Evolution of the isolated galaxy with AGN feedback and
increasing mechanical efficiency (from left to right column): 
$\epsilon=10^{-4}$, $3.3\times 10^{-4}$, and $10^{-3}$. 
First three rows: gas cooling rate versus time (red: AGN feedback; black: pure cooling flow),
%Other panels, from top down:
radial profiles of electron number density and
%at 12 different times as indicated in the figure; 
%azimuthally averaged, emission weighted temperature profiles.
spectroscopic-like temperature (at twelve different times).
Last three rows: injected mechanical energy, 
instantaneous power and velocity of the AGN feedback (for the single outflow), as a function of time.
%The velocity of the outflows is shown in the bottom row.}
%\end{figure*}
%\begin{figure*} \label{fig:hf2} 
}
\label{fig:iso-AGN} 
\end{figure*}

We now describe some results for the simulations with
AGN feedback, progressively increasing the efficiency and 
focusing on their global properties.
We compare them with the cooling flow model discussed above
and perform a critical analysis of the observable quantities.
%, such as the gas cooling rate, temperature and density profiles, metallicity
%distribution and the perturbations induced by the AGN outflows, such as cavities and shocks.
The first three rows of Figure \ref{fig:iso-AGN} show the gas cooling rate, plus the radial profiles of electron number density
and (projected) spectroscopic-like temperature, for
three representative feedback models.

In the first column, top panel, the model 
with $\epsilon=10^{-4}$ exhibits a cooling rate
significantly reduced with respect to the iso-CF simulation (red vs. black line),
with an average value of $\sim 0.4$ $M_\odot$ yr$^{-1}$
and $\dot{M}(12.7\,\rm{Gyr})\sim 0.2$ $M_\odot$ yr$^{-1}$. These numbers
are probably still too high compared to available X-ray and UV observations
(\citealt{Bregman:2001,Bregman:2005,Xu:2002,Tamura:2003})
or star formation estimates in massive ellipticals (\citealt{Temi:2009}).

The density profile (second panel) oscillates in time,
especially in the inner region, but retains reasonable values,
as can be noticed through a comparison with typical observations of ellipticals
(\citealt{Humphrey:2006b}). 

The density cycles correspond to the accumulation
of gas during periods of less intense AGN activity, e.g.
%For instance, 
%in the middle row (first column) it can be seen that 
between 6 and 7 Gyr there is no powerful outburst. This leads
to an increase of the gas density in the central region, a higher
cooling rate and finally a powerful ($\sim 3 \times 10^{44}$ erg s$^{-1}$)
event which quickly expands the hot gas atmosphere.
The density cycles also reflect in the variation of the X-ray luminosity which, in turn,
fluctuates between $10^{41} \lta L_{\rm X} \lta 2 \times 10^{42}$ erg
s$^{-1}$ (with a mean value of $\sim 10^{42}$ erg s$^{-1}$). 
This variation
can in part explain the long-standing problem of the scatter
in the $L_B$ (or $L_K$) - $L_{\rm X}$ diagram (\citealt{Eskridge:1995,Beuing:1999,Ellis:2006}), 
along with the occurrence of SNIa winds
(\citealt{Ciotti:1991,David:2006}) and the variation of the
dark halo masses at a given optical luminosity (\citealt{Mathews:2006}).

The temperature profile is a sensitive probe of AGN heating scenarios,
yet often neglected in the comparison with observations.
As expected, heating processes originating close to the galactic core
often overheat the central ISM, an effect rarely or never observed
in real galaxies (\citealt{Brighenti:2002,Brighenti:2003}). 
The best observed systems (which, unsurprisingly,
are the X-ray brightest) commonly show positive temperature 
gradient, with the centre hosting the coolest X-ray gas of the system
(\citealt{Humphrey:2006a,Diehl:2008b}). 
On the other hand, observed galaxies which are isolated or in a less dense environment
usually show a moderate negative or flat $T$ gradient.
%(\citealt{Kim:2003,Fukazawa:2006,Humphrey:2006b,Sansom:2006,Diehl:2008b,Nagino:2009}).

In the third panel (left column) we show the radial profile
of the spectroscopic-like temperature at twelve different times.
In this feedback model the ISM is never overheated in the central region
and always maintains an acceptable thermal structure 
(\citealt{Kim:2003,Fukazawa:2006,Humphrey:2006b,Sansom:2006,Diehl:2008b,Nagino:2009}), i.e.
{the central negative $T$ gradient is rather shallow and the central
temperature remains at or below $\sim 10^7$ K.
The lack of a substantial cool core must be ascribed to the absence of
the CGG, not to the presence of the AGN feedback.} %(cfr. the CF model above).
Ripples in the temperature profile are caused by weak shock waves generated 
by the AGN outbursts. These perturbations are somewhat more
  evident in the pressure profile, not shown here.
However, these waves would be largely diluted if the
azimuthal average is made in relatively larger radial bins, as in real
observations. We did not try here to exactly generate a fully synthetic observation
of the simulations (see \citealt{Heinz:2011}) and limit our analysis
to spectroscopic-like, projected quantities. 

The last three rows of Figure \ref{fig:iso-AGN} show the detailed evolution of the mechanical feedback (for the single jet). 
The total injected energy ($E_{\rm jet}$) for this model is $3\times10^{59}$ erg, the typical power ($P_{\rm jet}$) is on the order of $3\times10^{42}-10^{43}$ erg s$^{-1}$, with velocities ($v_{\rm jet}$) around $10^4$ km s$^{-1}$.

\subsection[]{AGN feedback: $\epsilon = 10^{-3}$}

The previous model, with efficiency $\epsilon = 10^{-4}$, was found 
in satisfactory agreement with several observational constraints.
However, the cooling rate, although significantly reduced with respect to the
classical iso-CF model, was still too large compared with observational limits.
%for several well observed  galaxies. 
It is thus natural to increase the
heating efficiency, essentially a free parameter of any current
feedback scheme, to test if the cooling rate can be further reduced
without scrambling the variable profiles.

When $\epsilon = 10^{-3}$ (third column in Figure \ref{fig:iso-AGN}), the effect of
the feedback is excessive: the ISM becomes rarefied in the inner
region ($r\lta r_{\rm eff}$) and simultaneously overheated.
Because the $\epsilon=10^{-3}$ model was unsatisfactory, we
stopped this simulation at $t \sim 9.5$ Gyr. It is interesting that
the energy released by the AGN
feedback is lower when the efficiency $\epsilon$ is larger (cf.
the fourth row of Figure \ref{fig:iso-AGN}, $E_{\rm jet}\gta10^{59}$ erg). Evidently the more powerful and faster outflows
(several $10^{44}$ erg s$^{-1}$ and over $5\times10^{4}$ km s$^{-1}$) of iso-1em3 model
are able to stop the cooling flow for a relatively long time. The interval between AGN outbursts is in fact
$\sim6\times 10^7 - 10^8$ yr, while for $\epsilon=10^{-4}$
the AGN activates at least one order of magnitude more frequently, 
with the net result of injecting less energy over the galaxy lifetime. This confirms the same
trend found for galaxy clusters (G11a) and illustrate that the
complexity of the problem makes reasonings based on the energetic 
budget alone rather inaccurate.

The lower gas density implies a generally lower X-ray luminosity,
with mean $\sim 5\times 10^{41}$ erg s$^{-1}$.
With efficiencies $\epsilon > 10^{-3}$ (models not shown), the simulated profiles strongly depart from those
of real ellipticals (like NGC 6482) and the X-ray luminosity drops to undetectable values.
We note here that low-$L_{\rm X}$ early-type galaxies usually show the presence
of some ISM in the central region, which has a relatively low
temperature (see for example NGC 4697, \citealt{Sarazin:2001};
NGC 1291, \citealt{Irwin:2002}; NGC 3379, \citealt{Trinchieri:2008}).
Thus, their low X-ray luminosity may not be associated with strong
AGN feedback generating a galactic wind. %Max: pero' si dovrebbe autoregolare...? 

\subsection[]{AGN feedback: $\epsilon = 3.3\times 10^{-4}$}

Finally, we have calculated a model with intermediate efficiency, 
$\epsilon = 3.3\times 10^{-4}$, which will serve as best model
in the following discussion.
As displayed in the middle column of Figure \ref{fig:iso-AGN}, the cooling rate is now
below the limits placed by current observations ($\lta0.1\ \msun$ yr$^{-1}$), apart a narrow, transient
peak at $t\sim 4$ Gyr. 
The density and temperature 
profiles vary in time slightly more than in model iso-1em4, but remain consistent with observations (cf.
NGC 6482 data points).
The spatial oscillations in the (projected) temperature profiles,
with maximum amplitude $\sim 20$\%, are compatible with those
shown in deep {\it Chandra} observations (see for example
fig. 10 in \citealt{Randall:2011}).

The final injected energy is $10^{59}$ erg (fourth panel).
Typical outburst powers are in the 
range ${\rm several}\;10^{42} - {\rm few}\;10^{44}$ erg s$^{-1}$ (fifth panel),
while the velocity of the outflows (bottom panel) 
vary in the range\footnote{The relativistic factor $\gamma \sim 1.06$ is still low enough to safely
using classic hydrodynamics.}
$10^3 - 10^5$ km s$^{-1}$, in good agreement with nuclear outflow and feedback
observations (see Introduction).
The power values correspond to instantaneous accretion rates
$\dot M_{\rm acc} = P_{\rm jet}/(\epsilon c^2) \sim 0.1 - 10$ $M_\odot$ yr$^{-1}$.
The frequency of the AGN activation is usually in the range
  $2\times 10^{-2}-5\times 10^{-2}$ Myr$^{-1}$,  albeit it varies
  significantly during the flow evolution. 
  We point out that the details of the AGN duty cycle depend on the
  ability to resolve the condensing cold clouds, in space and time and
  thus on the numerical resolution adopted\footnote{The duty cycle is thus linked to the thermal instability time scale, roughly the local cooling time, and also to the free-fall time (see \citealt{Gaspari:2012} for complete multiphase gas simulations with self-regulated AGN jet feedback).}. 
Nevertheless, the cold feedback self-regulation is robust enough that the cooling flow problem is solved, irrespective of the duty cycle details.

We consider iso-3em4 the best model because the cooling rate is in
good agreement with known constraints for many Gyr, and the variable profiles always
resemble those of real galaxies. Thus, this model (and to a lesser extent
iso-1em4) passes the first checkpoint, a necessary
condition for a viable heating mechanism. In order to further test the robustness of the model, in the
next Section we analyse the detailed galaxy appearance in the X-ray band.
We finally note that the value of the $\epsilon$ parameter 
should not be taken as the exact physical value,
given the uncertainties on the simulated accretion rate (Sec. 2.1.1).

\subsubsection[]{X-ray features}

\begin{figure*} 
    \subfigure{\includegraphics[scale=0.51]{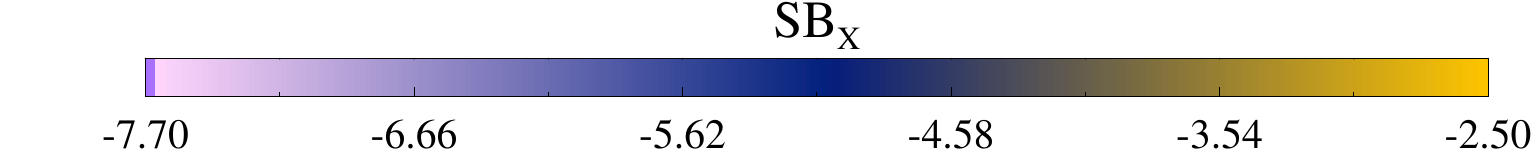}}
    \subfigure{\includegraphics[scale=0.51]{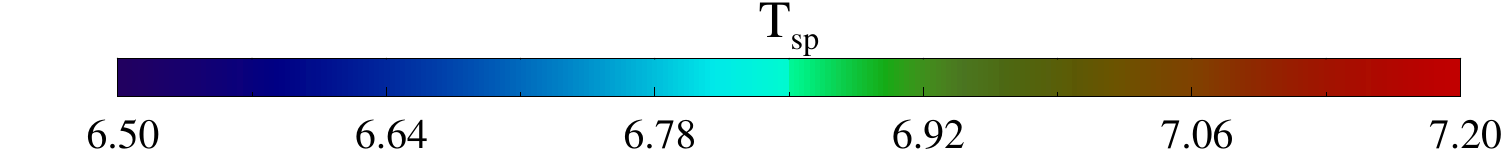}}
    \subfigure{\includegraphics[scale=0.51]{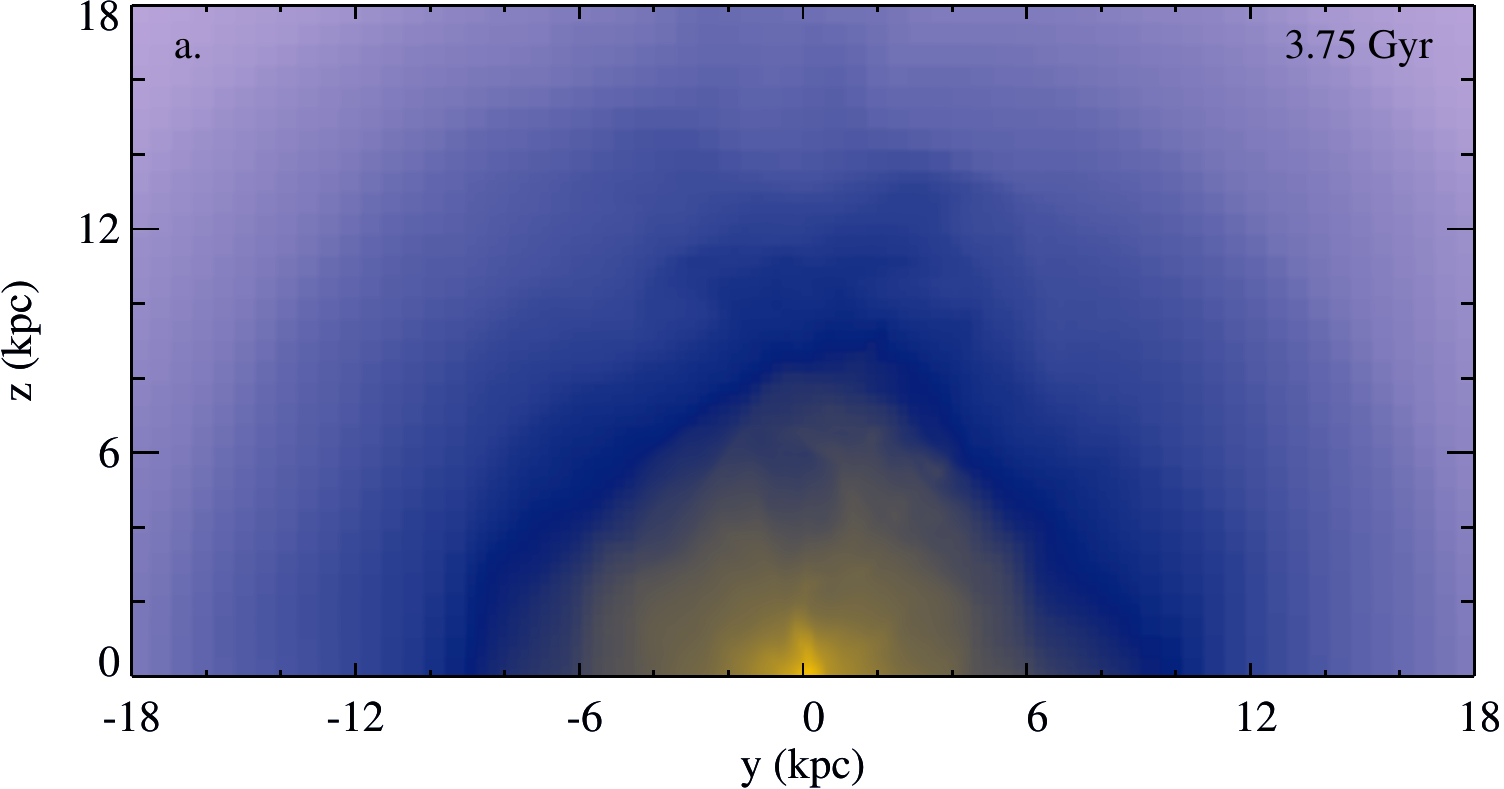}}
    \subfigure{\includegraphics[scale=0.51]{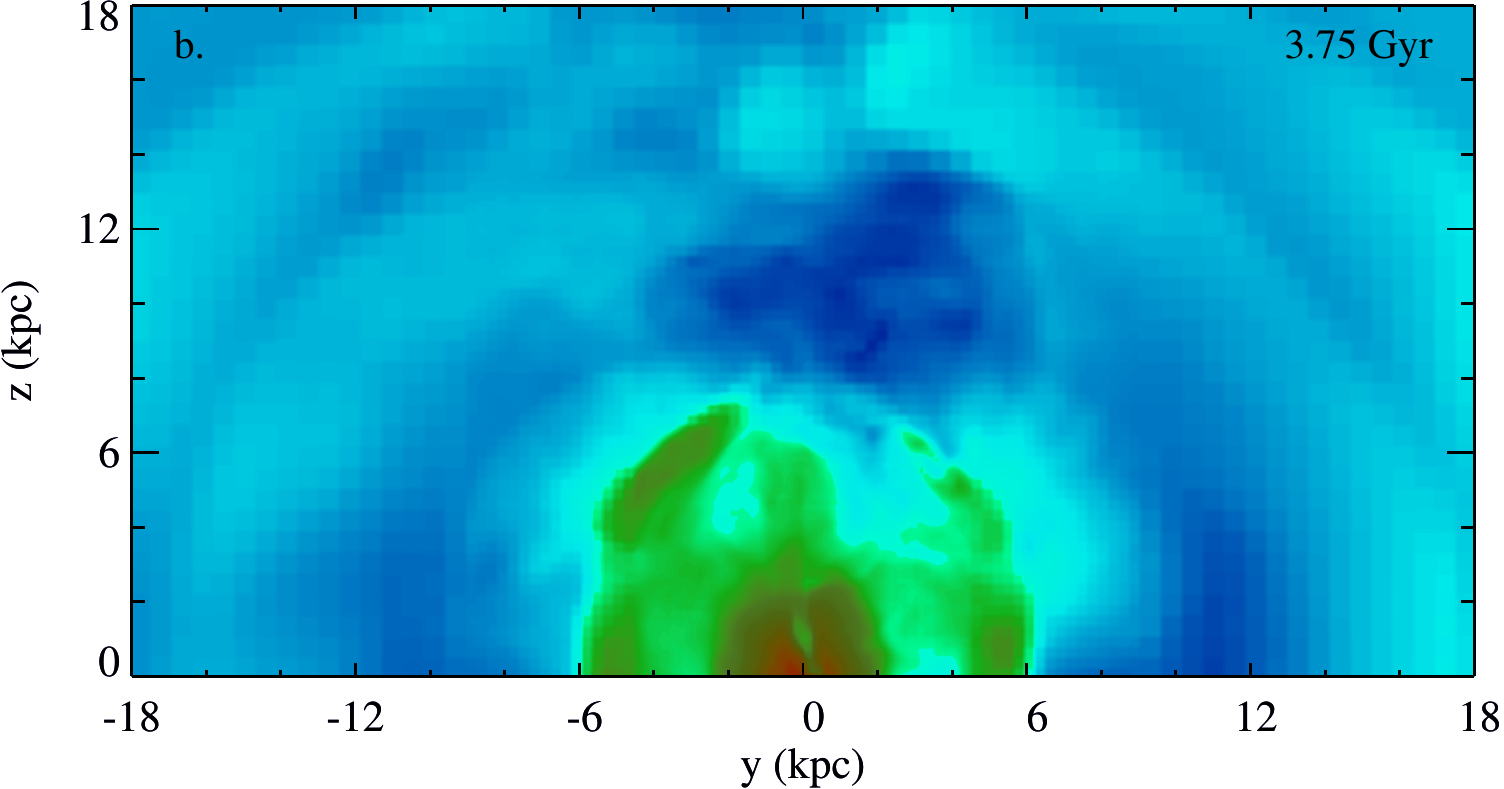}}
    \subfigure{\includegraphics[scale=0.51]{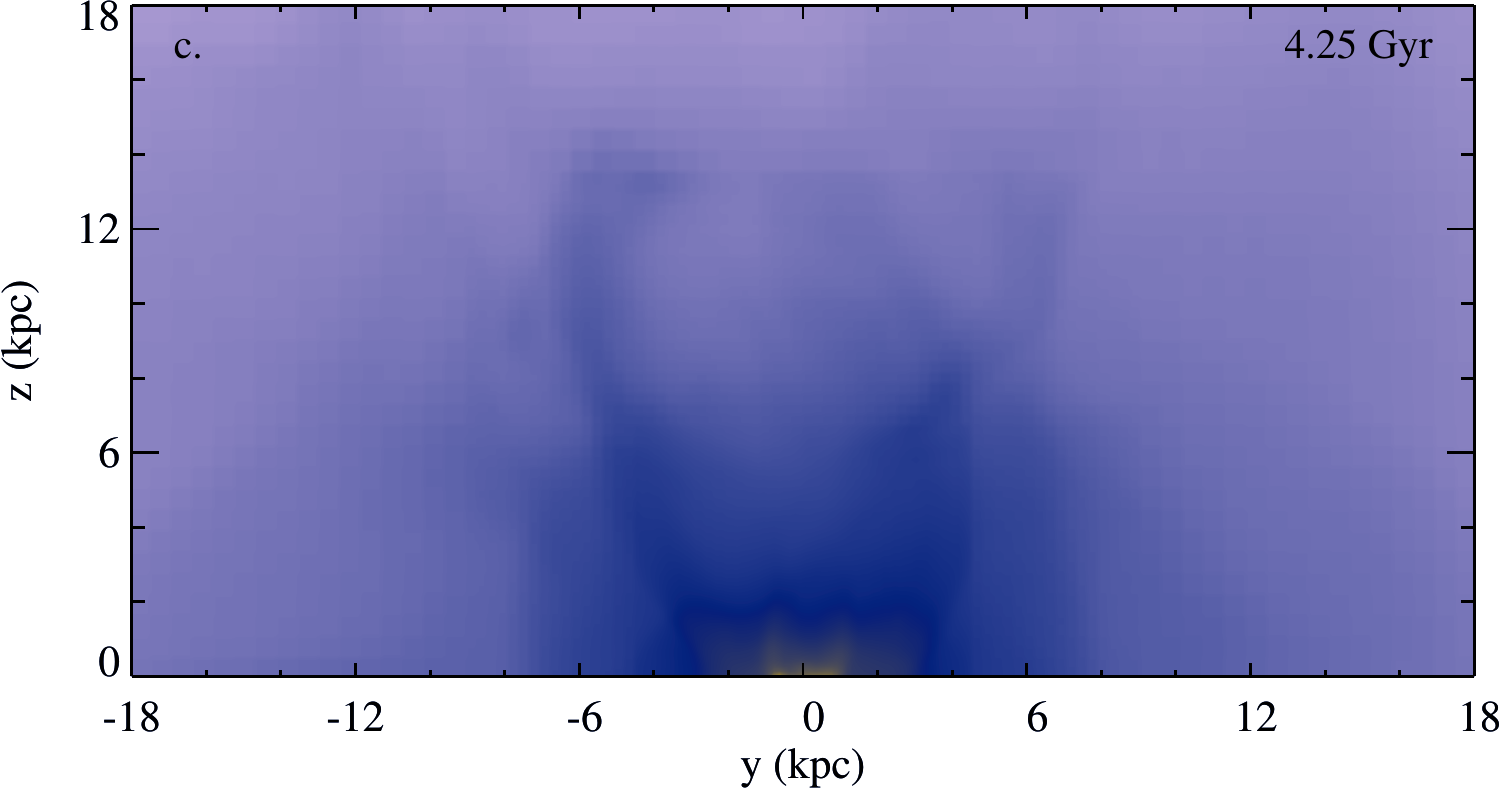}}
    \subfigure{\includegraphics[scale=0.51]{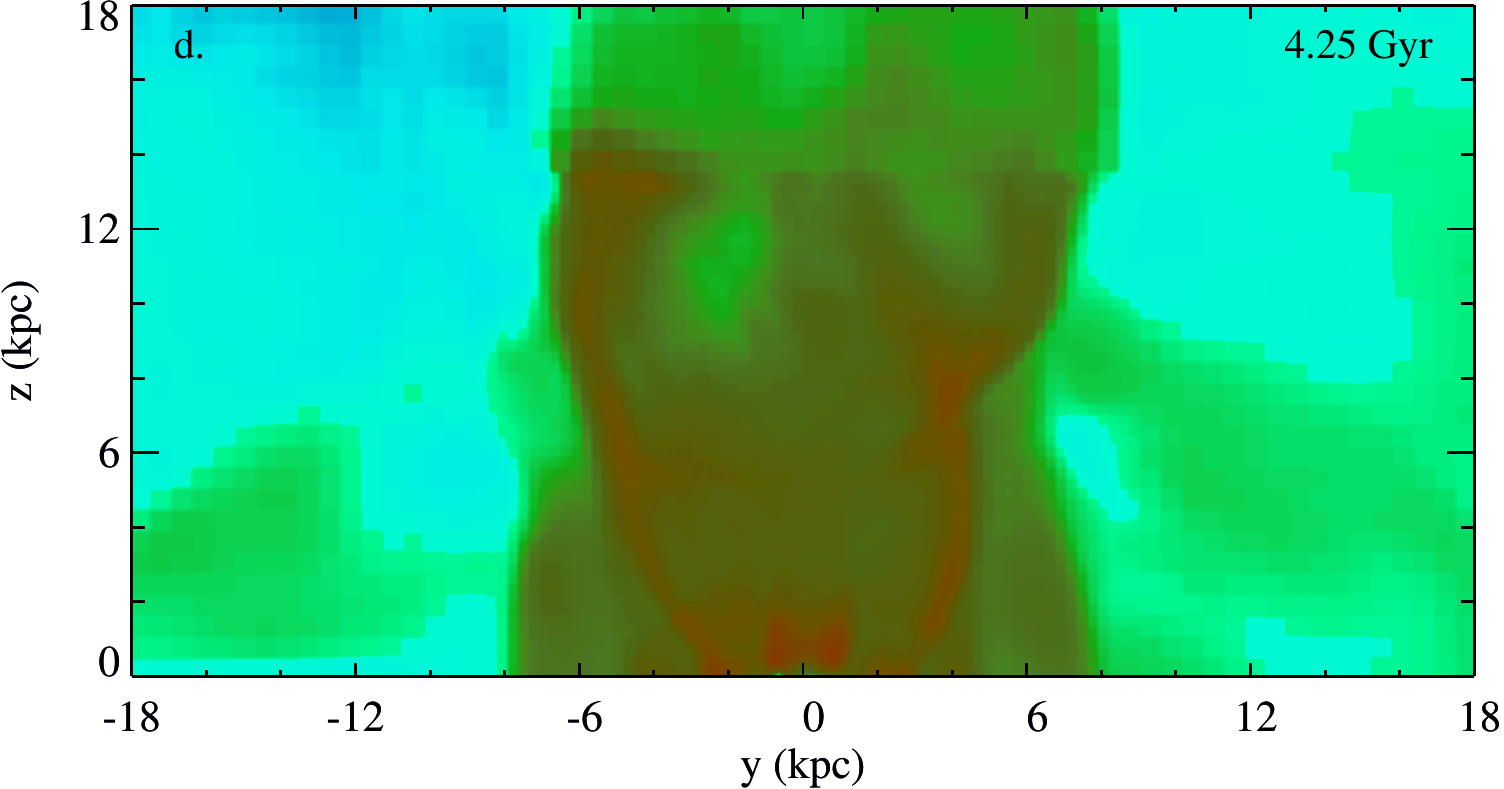}}
    \subfigure{\includegraphics[scale=0.51]{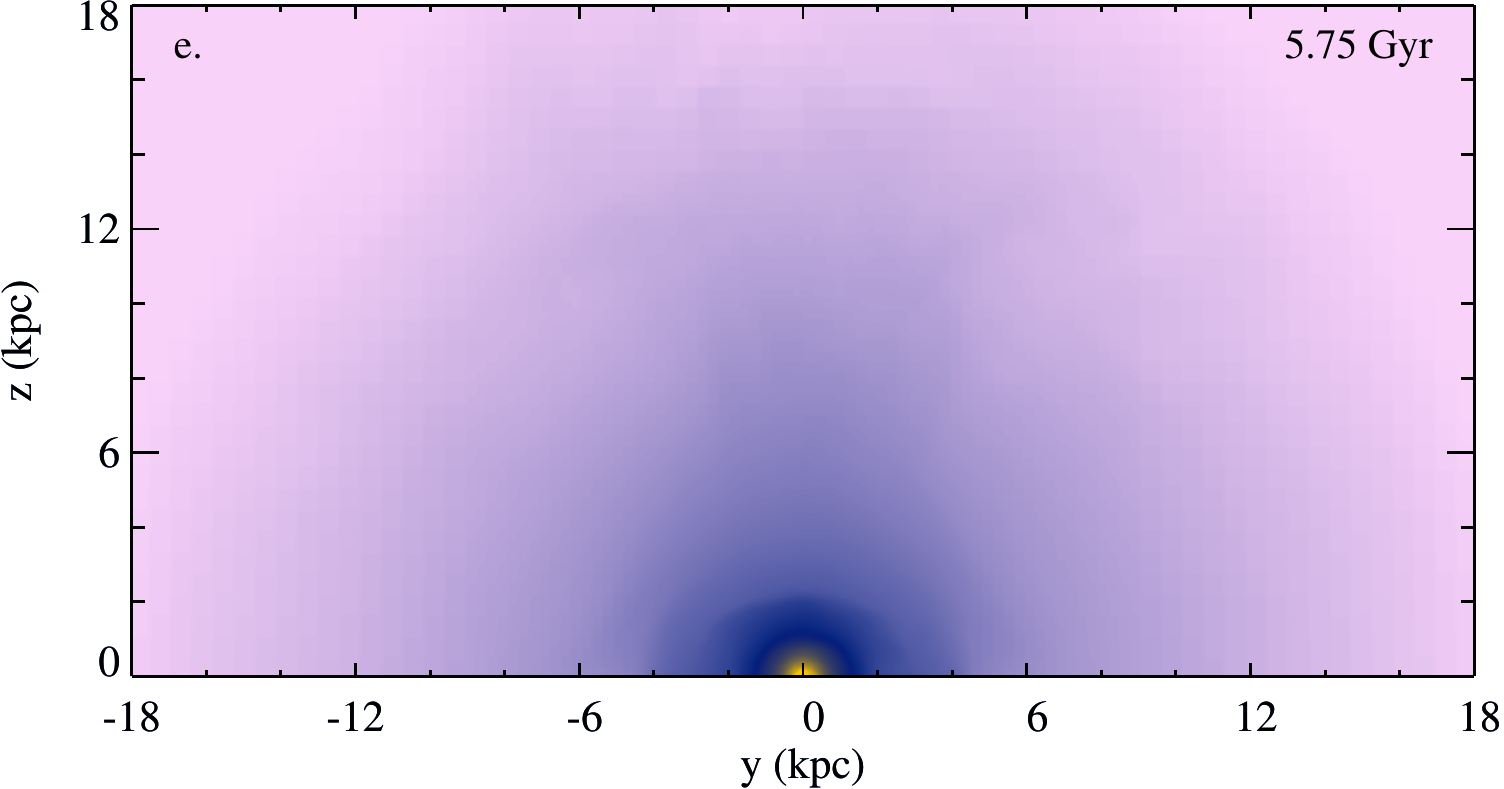}}
    \subfigure{\includegraphics[scale=0.51]{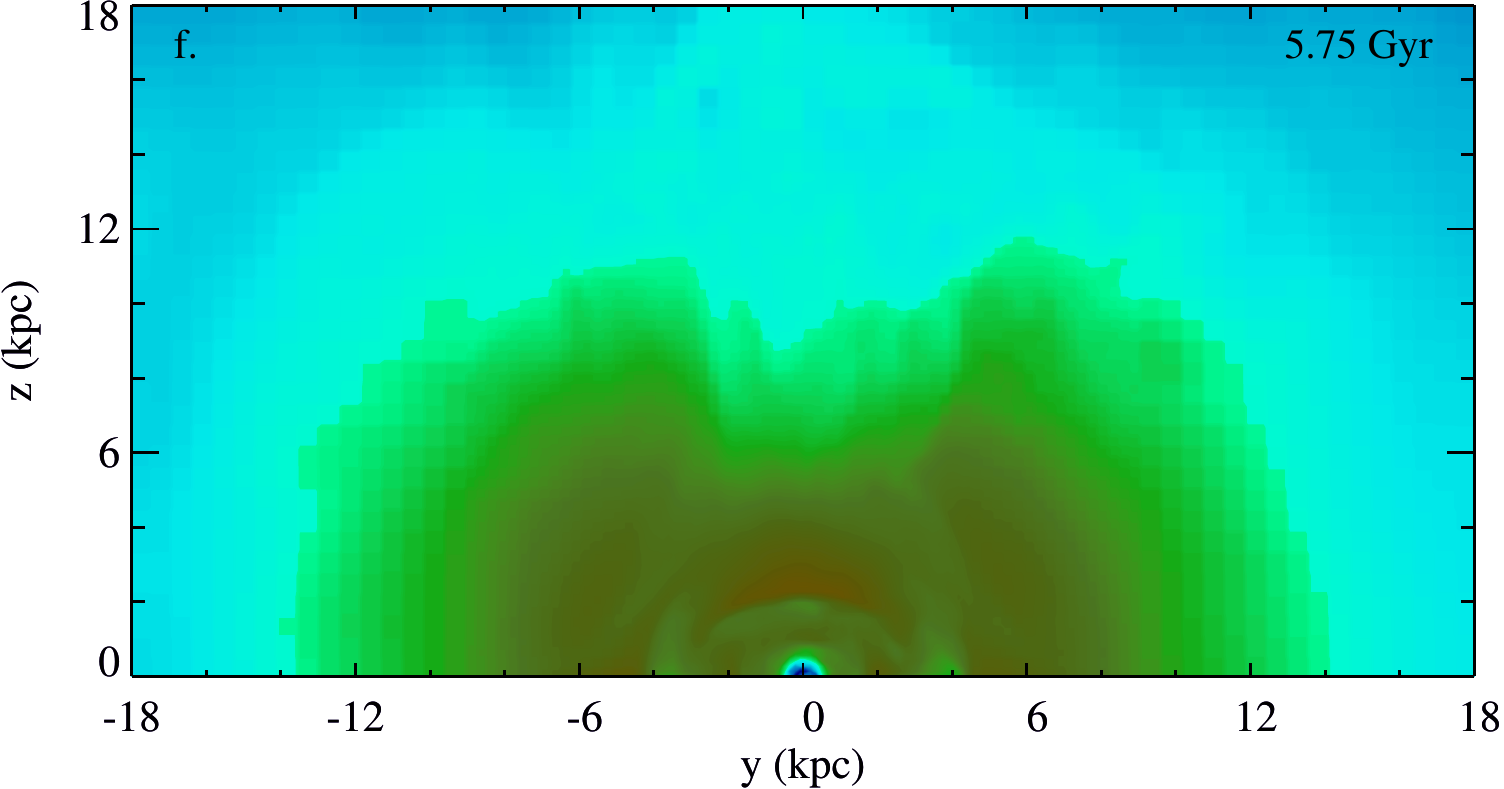}}
    \subfigure{\includegraphics[scale=0.51]{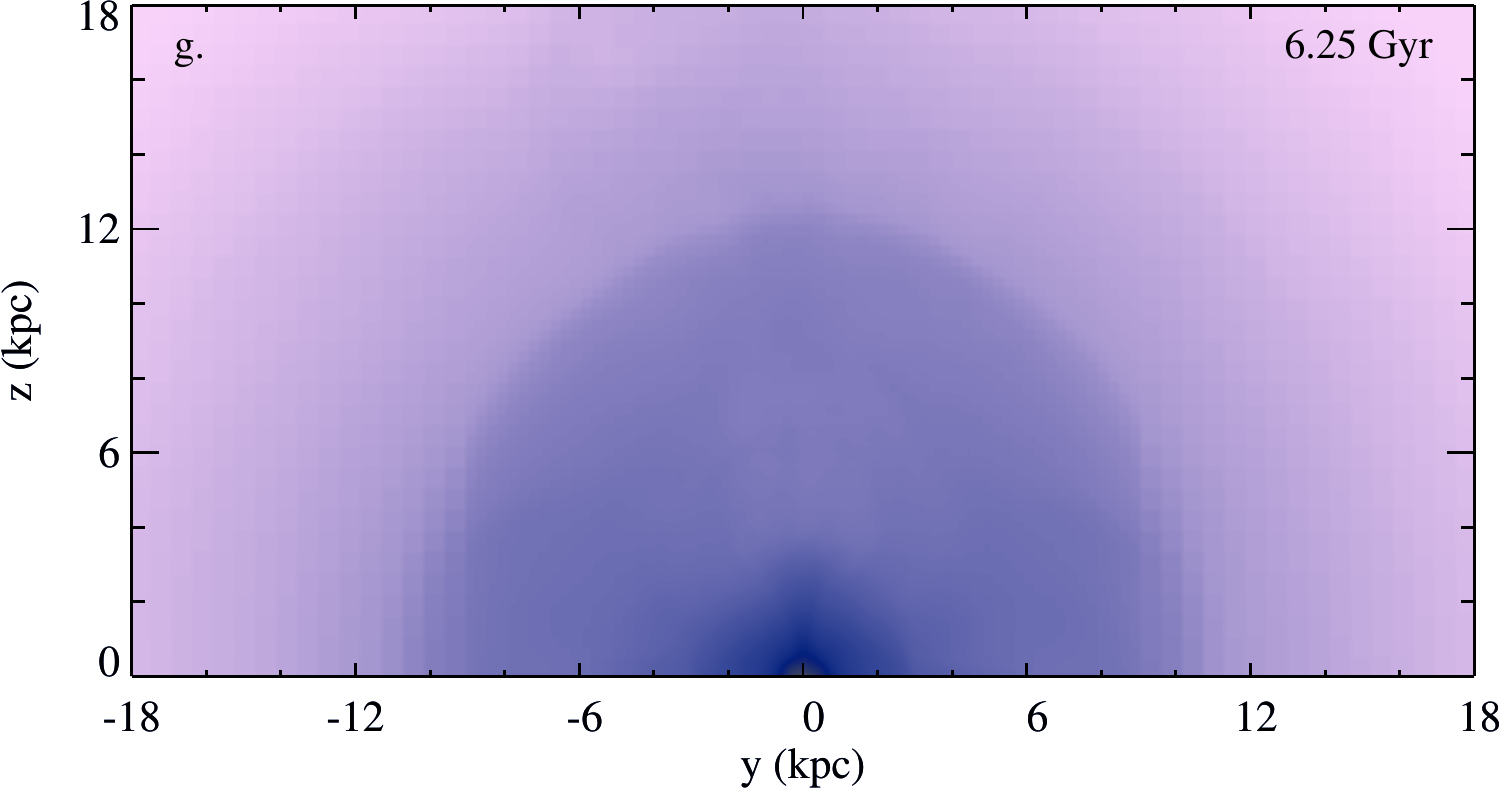}}
    \subfigure{\includegraphics[scale=0.51]{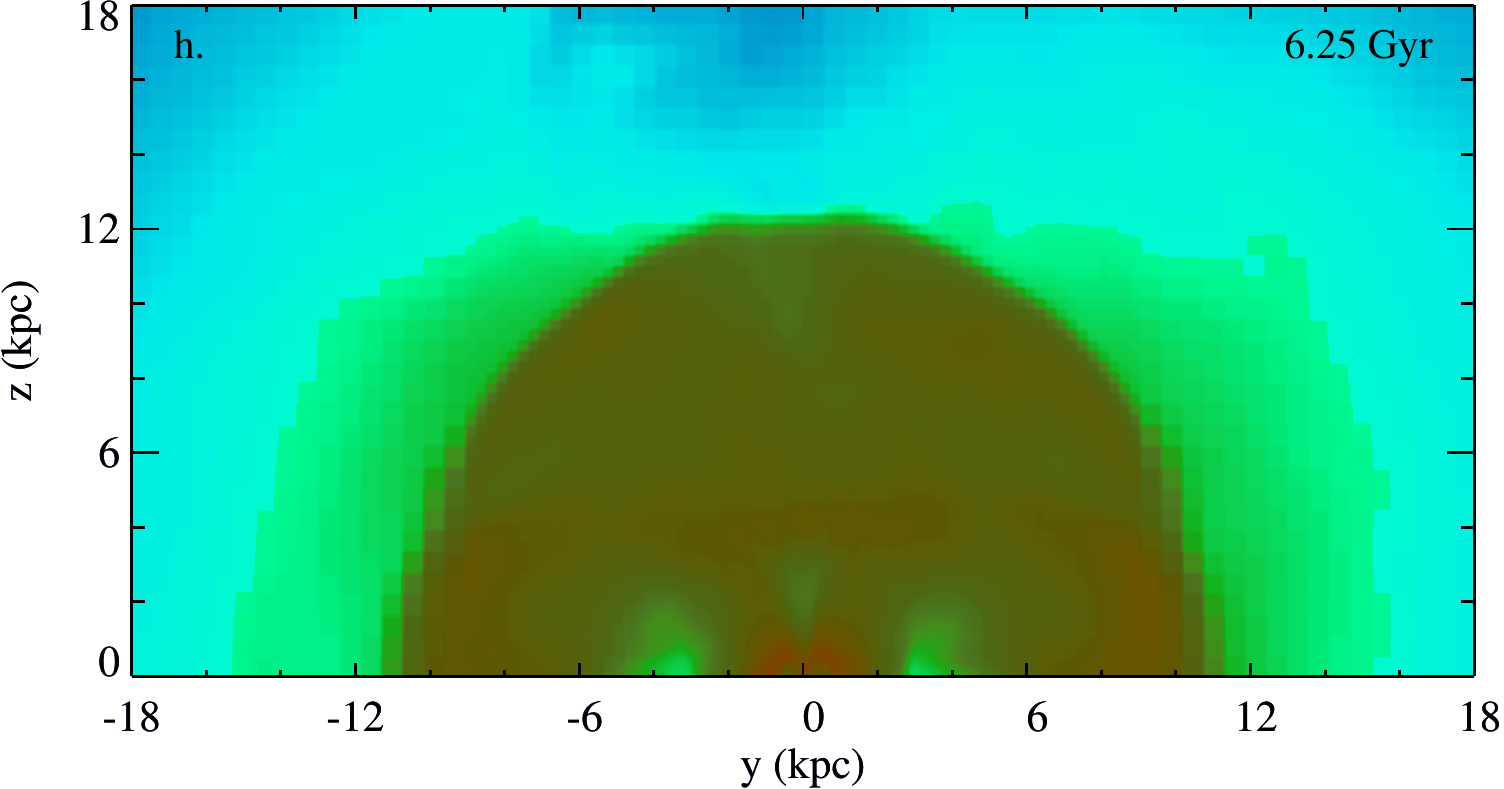}}
    \subfigure{\includegraphics[scale=0.51]{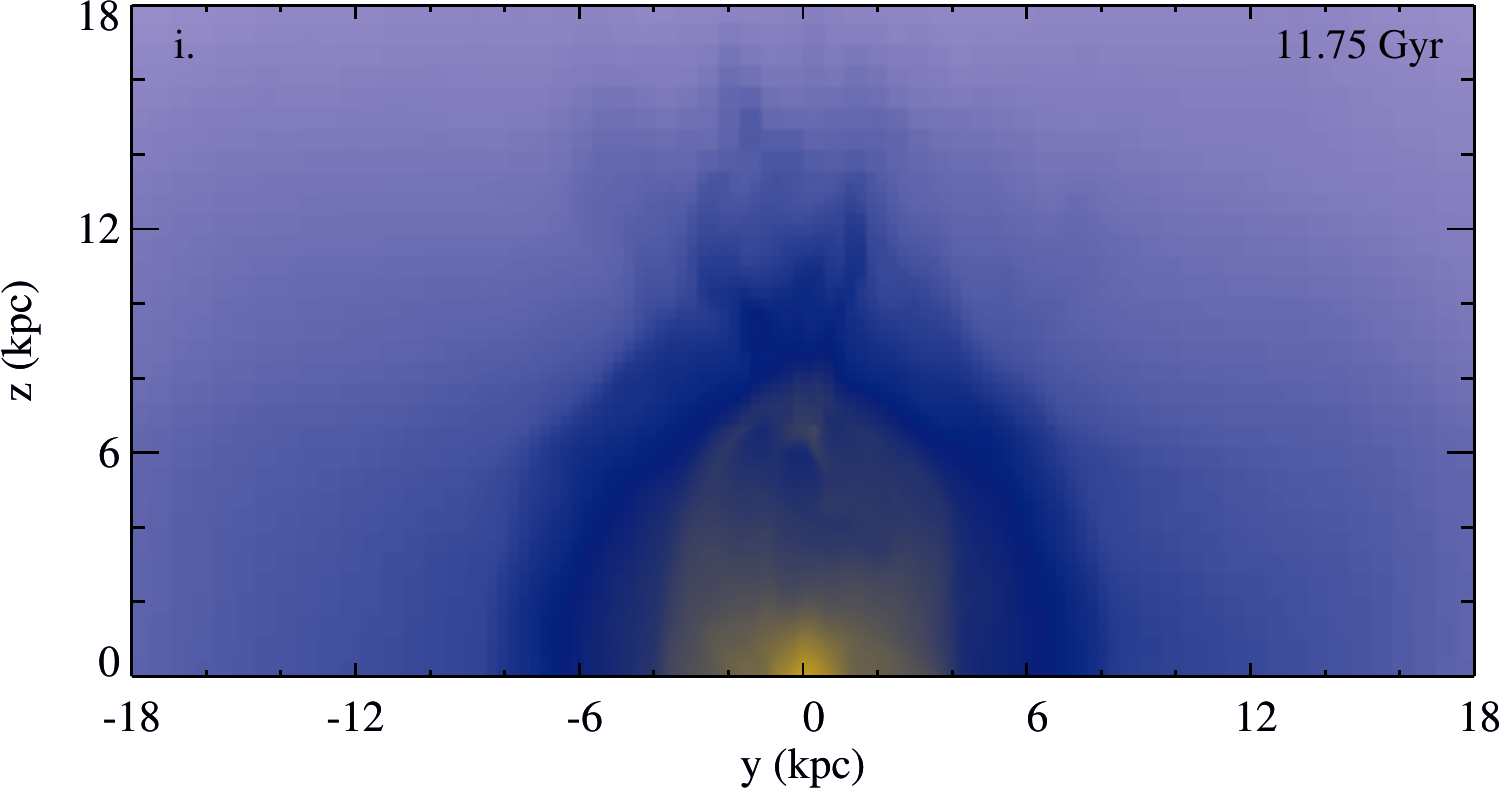}}
    \subfigure{\includegraphics[scale=0.51]{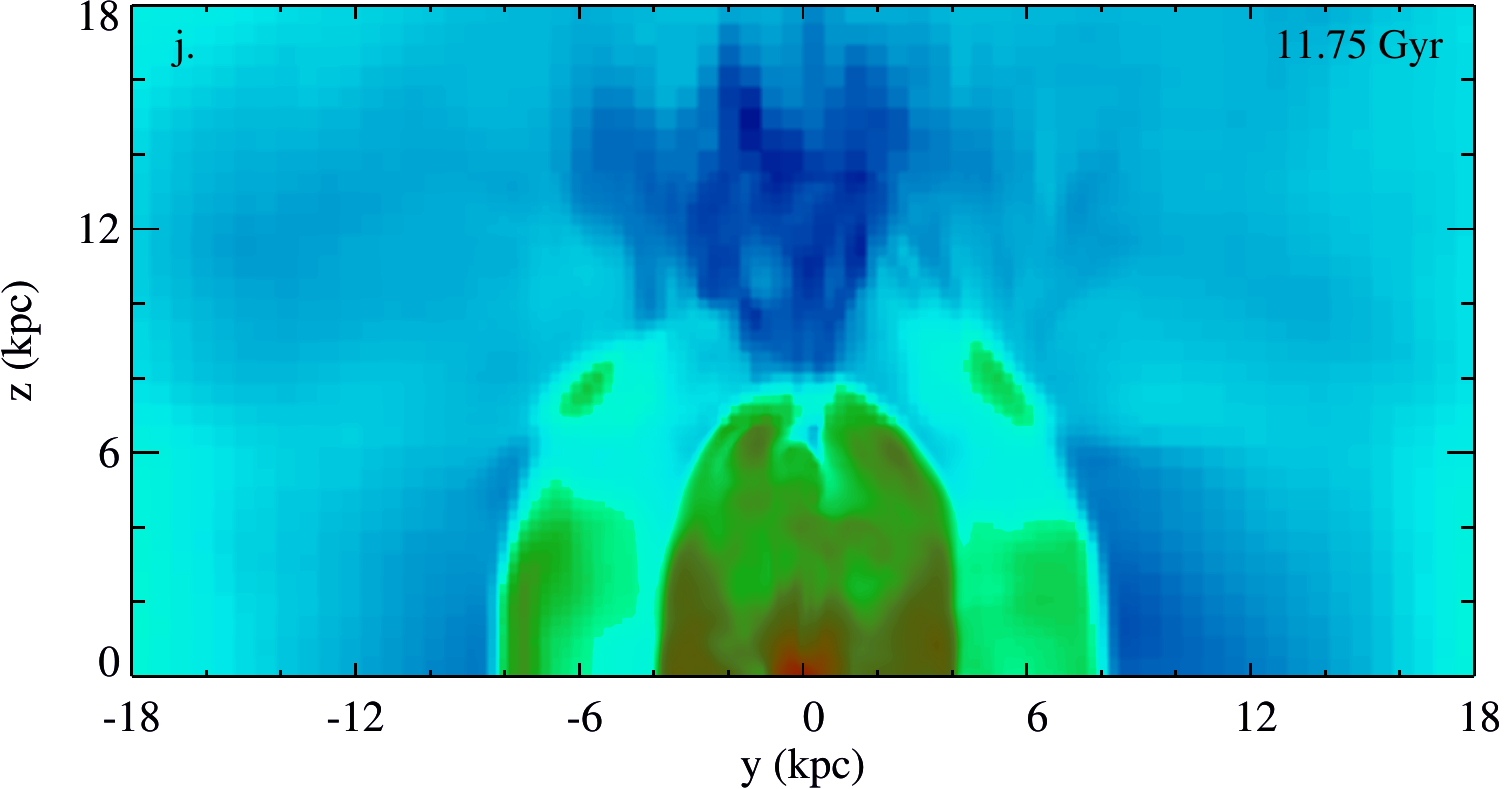}}
    \caption{Maps of X-ray surface brightness (left column) and projected spectroscopic-like temperature (right column) for model iso-3em4 at five different times (from top to bottom row): 3.75, 4.25, 5.75, 6.25 and 11.75 Gyr. See Section 3.4.1.}         
    \label{fig:iso-maps}       
\end{figure*}

In this Section we investigate the most significant X-ray features,
naturally generated by the AGN heating process,
like shocks and buoyant cavities.
A direct comparison with a specific real system is not possible
because our models are intended to be general and not tailored
on an individual object. Moreover, cavities, shocks and
perturbations are intermittently created, continually changing
the ISM appearance.
%the X-ray surface brightness
%and the spectroscopic-like temperature distribution.
The main aim of the analysis is to show
that the proposed feedback mechanism leads to ISM perturbations
similar to those observed in galaxies and groups. A feedback scenario
not satisfying basic requirements, such as the formation of cavities or
weak shocks, must be rejected.

We discuss below several X-ray snapshots, taken at different times,
of the best model iso-3em4. %($\epsilon = 3.3\times 10^{-3}$).
In Figure \ref{fig:iso-maps} (panel a in the top row) we show the X-ray surface brightness map of the central
region after 0.75 Gyr of feedback evolution. An outburst with energy
$2.85 \times 10^{56}$ erg occurred $8\times 10^6$ yr before
the time of the snapshot. The outflow, ejected with velocity $v_{\rm jet}\sim 5400$ km s$^{-1}$, 
generated a weak shock, now at a
distance $\sim 6$ kpc from the centre, with a Mach number $M\approx 1.2$
(calculated from $T_{\rm sp}$, which jumps
from $\sim 0.5$ keV in the pre-shock region to $\sim 0.6$
keV in the post-shock ISM). The outflow generated a small X-ray cavity
centred at $z=4.5$ kpc and ellipsoidal (prolate) shape with 
semiaxes 1.5 and 1 kpc.
The X-ray surface brightness depression within the cavity is about 20\%, 
a typical value for observed cavities. Relatively colder rims surround the cavity,
although the $T_{\rm sp}$ difference between the rims and the nearby gas is only
$\sim 10$\% (Fig. \ref{fig:iso-maps}, panel b). We note that the ISM adjacent to the rims has been
slightly heated by the weak shock. %(see Figure 4 E.W.TEMP MAP-014).
In the spectroscopic-like temperature map are also visible (but
probably not detectable with current X-ray telescopes) very weak
waves, separated by $\Delta r_{\rm waves} \sim 6$ kpc, corresponding to a period of $t_{\rm
waves} = \Delta r_{\rm waves}/c_s \sim 1.6\times 10^7$ yr, roughly
the typical time between significant AGN outbursts at this epoch
($t=3.75$ Gyr). We stress that these X-ray features are faint and the %t=3.754
ISM looks quite relaxed at this time.

The situation after 0.5 Gyr ($t=4.25$ Gyr; panels c and d) is different. At this epoch
the AGN is more active and perturbs the ISM in a remarkable way.
The X-ray arms and cavity have been generated by four series of
AGN outbursts of moderate power (few $10^{42}$ erg s$^{-1}$), 
occurred in the previous 10 Myr.
The energy injected during this time interval is $\sim 1.4 \times 10^{57}$ erg.
Panel c shows that the X-ray surface brightness 
map is rich of features, with two symmetric bright arms, $\sim 15$
kpc long. This structure is similar to that observed in NGC 4636,
a massive elliptical galaxy in the outskirt of the Virgo cluster
(\citealt{Jones:2002,Baldi:2009}). Thus, it might be possible that
the ISM of NGC 4636 has been shaped by multiple, moderate AGN
outflows. It is in fact difficult to reproduce all the observed features 
of NGC 4636 with a single AGN outburst (Ballone \& Brighenti, in prep.).

In the spectroscopic-like temperature map showed in panel d,
the X-ray bright arms are slightly hotter than the local ISM. 
The temperature jump is about 30\% (from 0.67 keV to 0.89 KeV)\, 
corresponding to a shock Mach number of $\sim 1.3$.

The galaxy alternates periods of quiescence, typically lasting $50-100$ Myr,
with moments during which the AGN is more
active and, consequently, the ISM more disturbed.
Other interesting features appear during the evolution. 
At $t=5.75$ Gyr a cold front is visible
in both the brightness and temperature maps, at $\sim 2$ kpc 
from the centre (panels e and f).
The front has not been caused by galaxy sloshing in our simulations,
but by the encounter between gas which is falling back toward the
centre, after an outflow, and the gas currently located in the central region.

At $t=6.25$ Gyr (panel g) the brightness map shows a
weak cavity about 3 kpc in radius, surrounded by a $\sim 10$ kpc shock
with Mach number $\sim 1.3$ (panel h). This situation is similar to that at
t=3.75 Gyr, described in panel a.
This demonstrates that the ISM cyclically evolves, with recurrent
cavities generated by the outflows and weak shocks, just as indicated
by observations.

Although not a common circumstance, a series of shocks are visible at
the same time. This occurrence is shown in panel i ($t=11.75$ Gyr),
where two ellipsoidal shocks are depicted (Mach $\sim 1.4-1.5$).
The shocks intersect the $y-$axis at $\sim 4$ and $\sim 8$ kpc
and are clearly visible both in the surface brightness and temperature maps (panel j).

\begin{figure*}  
\centering
\includegraphics[scale=0.6]{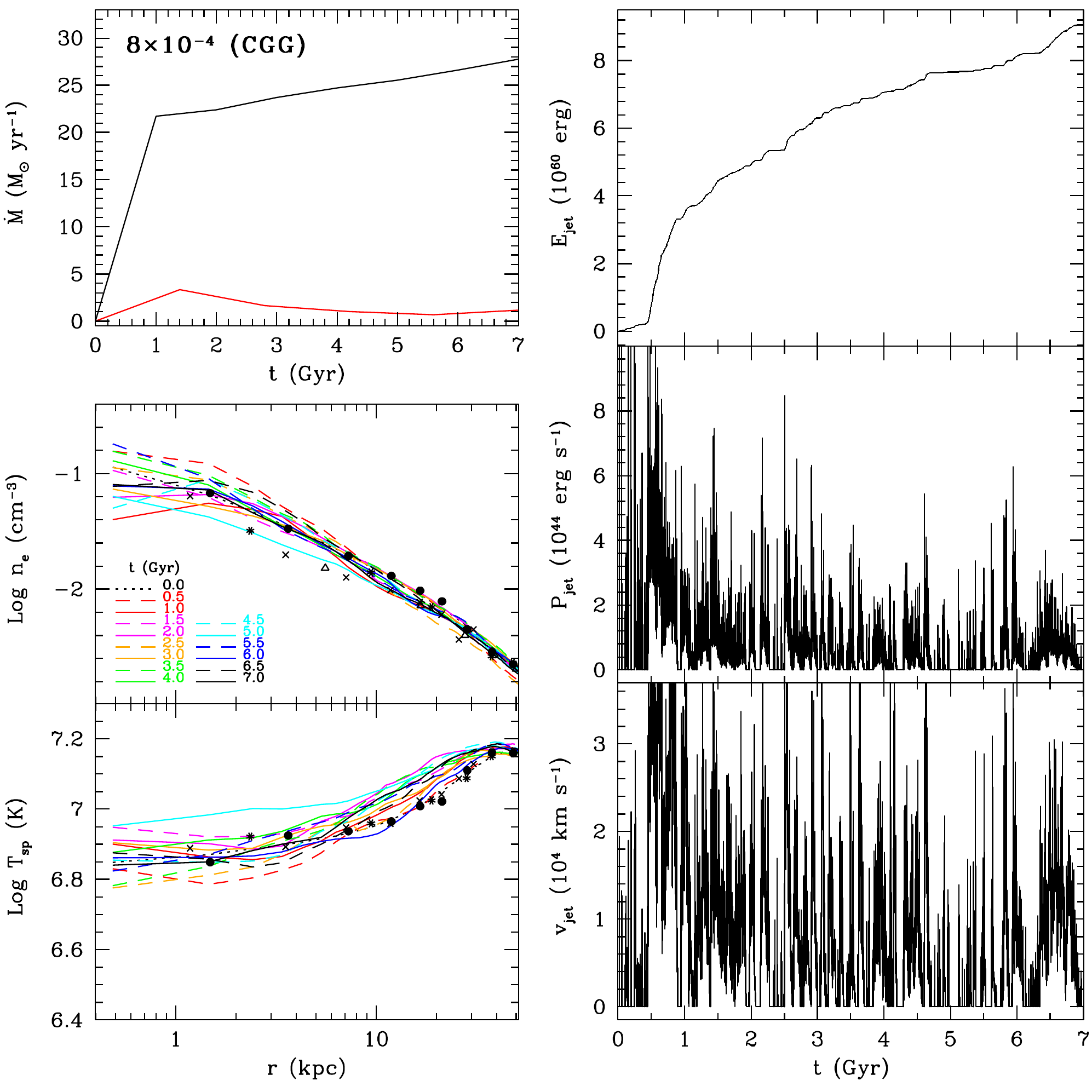}
\caption{Evolution of the elliptical galaxy with circumgalactic gas and
mechanical feedback efficiency $\epsilon=8\times 10^{-4}$.
Left column, from top to bottom panel: cooling rate as a function of time
(red: AGN feedback; black: pure cooling flow);
%(red line; cf. CF model -- black line);
electron number density and spectroscopic-like temperature at several times.
The points in the last two panels represent observational data of the bright elliptical NGC 5044 
(\citealt{David:1994,Buote:2003}).
%[[MAX: ALTRI DATI??]] Max: added
Right column: injected mechanical energy, (single) outflow power and velocity as
a function of time.}
\label{fig:cgg}
\end{figure*}

This brief description of the ISM appearance in X-ray
is intended to illustrate the richness of structures generated by the
AGN outflows. Although in the present study we do not attempt a detailed comparison with
real objects, it is important to realise the
strong similarity of simulated features, like X-ray cavities, filaments
and shocks, with the structures seen in deep X-ray observations of
ellipticals and groups
(e.g. \citealt{Biller:2004,Machacek:2006,Machacek:2011,Baldi:2009,Gastaldello:2009,
OSullivan:2011_4261,Randall:2011}).
This fact, together with the analysis in the previous Section, indicates that AGN outflows
are indeed a very robust feedback mechanism, capable to
highly inhibit gas cooling, keeping reasonable density and temperature
profiles, and giving rise to the wealth of asymmetric structures in
the ISM, seen in high-resolution deep X-ray observations.

\section{Results: circumgalactic gas}

As previously mentioned, most massive, X-ray bright ellipticals host a
very extended hot ISM that can not be explained via internal processes alone
(e.g. \citealt{Brighenti:1998,Brighenti:1999a}). These galaxies are at the
centre of large dark matter halos (\citealt{Mathews:2006}) and the ISM is
likely composed of gas shed by the stellar population, plus
circumgalactic gas linked to the cosmological evolution. It is thus important to
test if mechanical AGN outflows are successful for this class of objects, which
include many famous X-ray targets, such as NGC 4636, NGC 4472, NGC 5044,
and NGC 4649.

Interestingly, in these
cool-core objects the evidence for AGN heating seems stronger, as revealed
by the presence of X-ray cavities and other ISM perturbations
(e.g. \citealt{Biller:2004,Diehl:2008a,Baldi:2009, 
Gastaldello:2009,Dong:2010,Dunn:2010,Randall:2011,OSullivan:2011_4261}).
It might be a selection effect, but it is ironic (and instructive) to note
that just where we know the AGN is injecting energy,
the ISM close to the feedback engine is cooler
than everywhere else.
%; another signal that a simple 
%thermal/radiative blast is not the correct solution.

\subsection{AGN Feedback: $\epsilon=8\times10^{-4}$}
In this Section we present only the best feedback model for the CGG elliptical
(efficiency $\epsilon = 8 \times 10^{-4}$), able to properly solve the cooling flow problem. 
It is interesting that the
same feedback method requires now a larger efficiency to be acceptable,
in line with the results of G11b, which indicate
that in more massive systems the feedback should act more efficiently.

As in the case of the isolated galaxy, we have run a classic cooling 
flow model (no AGN heating) which serves as a reference calculation
to gauge the effects of the feedback.
We do not show the evolution of the pure CF profiles here (cgg-CF). The results
are almost identical to those presented in G11b,
although the numerical resolution there was about two times coarser ($\Delta x \approx 500$ pc),
compared to the present model.
The gas cooling rate, also indistinguishable from
that in G11b, is shown in Figure \ref{fig:cgg} (black line in the top left panel). As suggested by the
resulting bolometric $L_{\rm X} \sim 3 \times 10^{43}$ erg s$^{-1}$, the
cooling rate stays around $\sim 25$ $M_\odot$ yr$^{-1}$, revealing
again the cooling flow problem at the galactic/group scale.

The AGN feedback model, with $\epsilon = 8 \times 10^{-4}$ (cgg-8em4),
is fully presented in Figure \ref{fig:cgg}. The cooling rate (red line, first panel) has been
quenched to $\sim 1$ $M_\odot$ yr$^{-1}$ at the current epoch, implying a 
satisfactory 20-fold cooling suppression, broadly consistent 
%(albeit formally slightly larger) 
with XMM-RGS observations (\citealt{Tamura:2003}).

The spectroscopic-like temperature and density radial profiles are in
excellent agreement with those observed for NGC 5044 (left
column: middle and bottom panel). Remarkably, the cooling reduction did not come at the
expense of the cool core: the temperature still decreases toward the
galactic centre -- a fundamental quality of anisotropic mechanical feedback.

The total mechanical energy released by the AGN is $E_{\rm jet}\sim
9\times 10^{60}$ erg (first panel in second column), which in principle would correspond to a 
total mass accreted on the black hole of $\sim 6 \times 10^9$ $M_\odot$, 
given the assumed efficiency. This accreted
mass may appear substantial, but we stress again that the simulated
accretion rate is quite uncertain (SMBH can nevertheless reach masses of few $10^{10}$ $\msun$).
We note that the BH growth could in principle be reduced assuming
that a fraction of the cooling gas does not accrete onto the BH
and is ejected with the AGN outflow; increasing the efficiency would produce the same feedback power
(but adding another parameter to the scheme would not provide any further physical insight).

The most relevant astrophysical result, here, is
that self-regulated massive outflows are indeed
capable to solve the cooling flow problem on kpc or larger scales, although the
fully consistent picture -- explaining the details of the BH accretion
process and the formation of the outflows -- needs to be clarified in future via high-resolution investigations.

The time record of the outflow power and velocity is shown in the
right column in Figure \ref{fig:cgg}. Powers on the order of few $10^{44}$
erg s$^{-1}$ and velocity of several $10^4$ km s$^{-1}$ are again typical
for the stronger events. 

We conclude that massive outflows, activated when gas cooling occurs,
are able to regulate the thermal evolution of the ISM, suppressing the
cooling rate by a factor 20 or more and keeping the cool-core aspect
of massive, X-ray bright ellipticals. We now test the model via other key observational constraints.

\subsubsection{X-ray features}

\begin{figure*} 
    \subfigure{\includegraphics[scale=0.51]{art_CGG_colorbar_SBx.pdf}}
    \subfigure{\includegraphics[scale=0.51]{art_CGG_colorbar_Tspecl.pdf}}
    \subfigure{\includegraphics[scale=0.51]{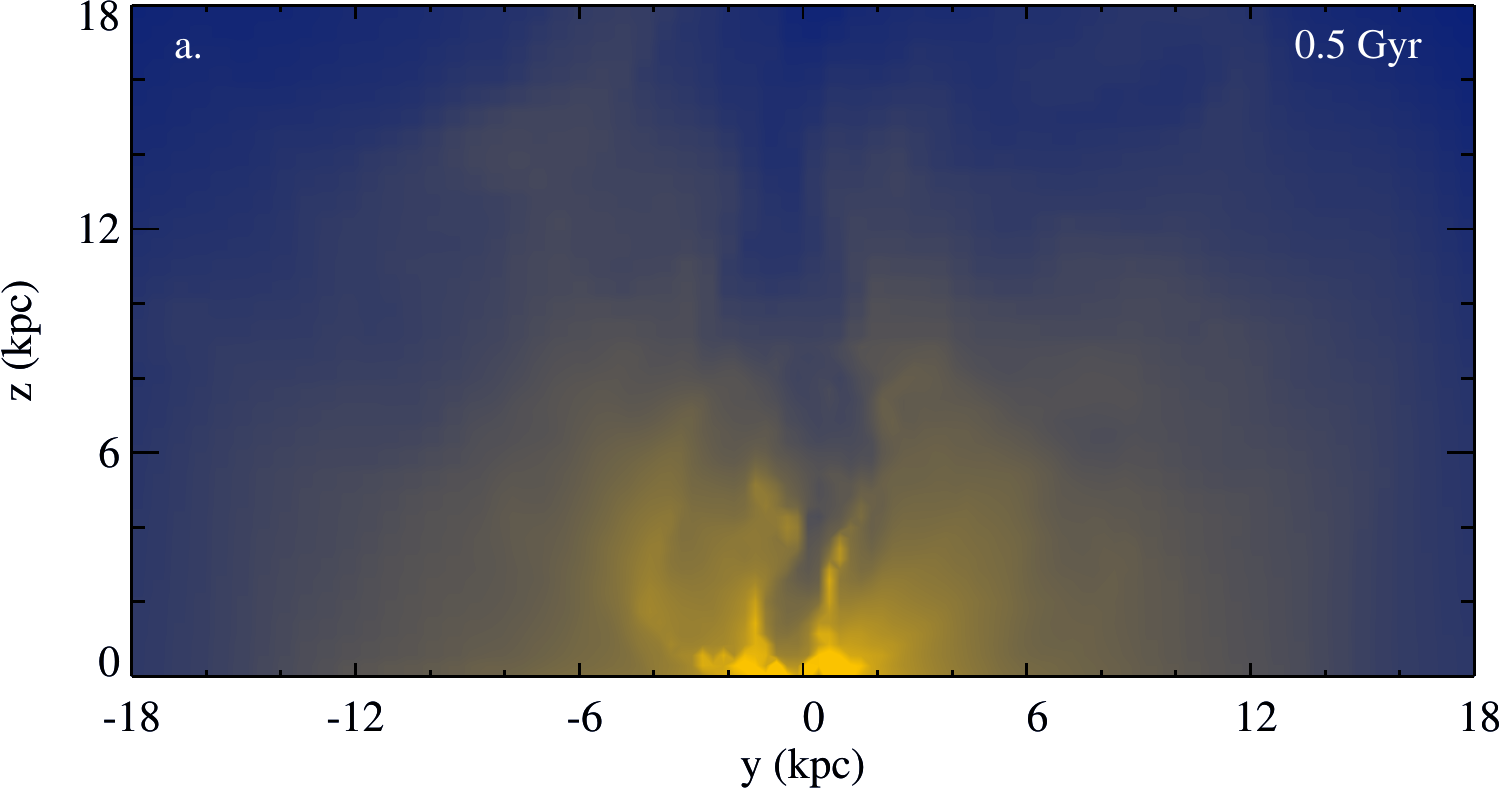}}
    \subfigure{\includegraphics[scale=0.51]{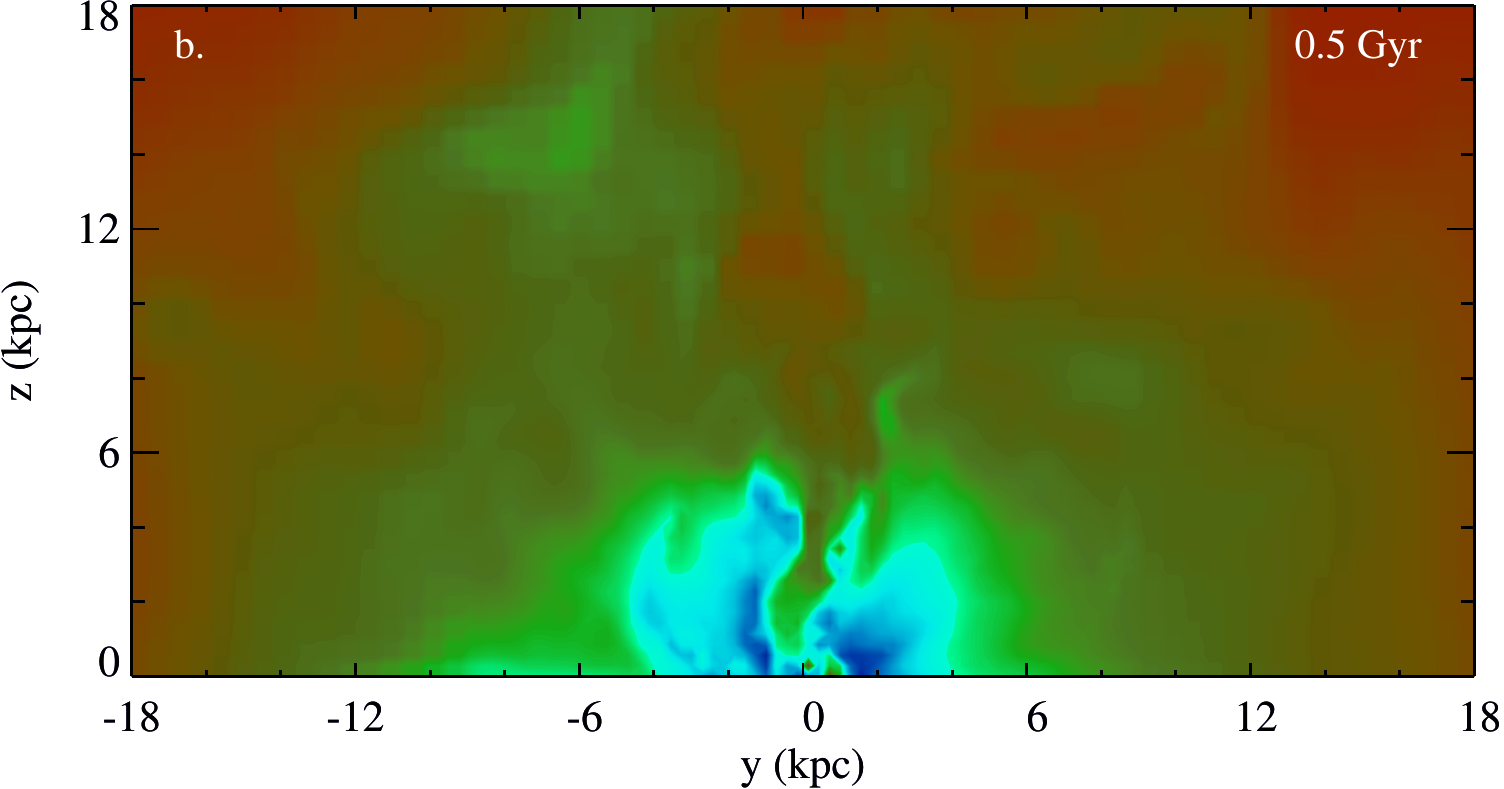}}
    \subfigure{\includegraphics[scale=0.51]{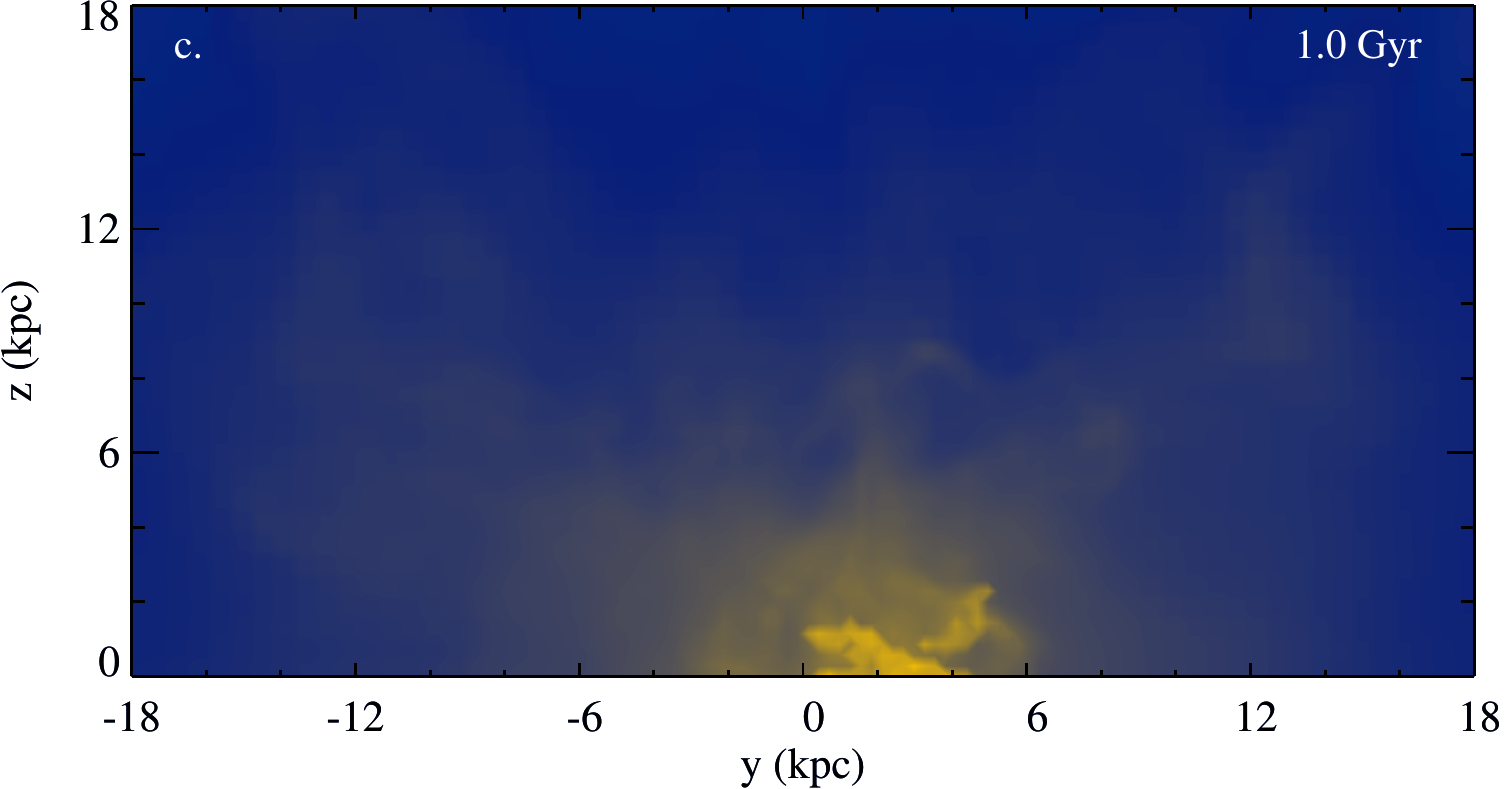}}
    \subfigure{\includegraphics[scale=0.51]{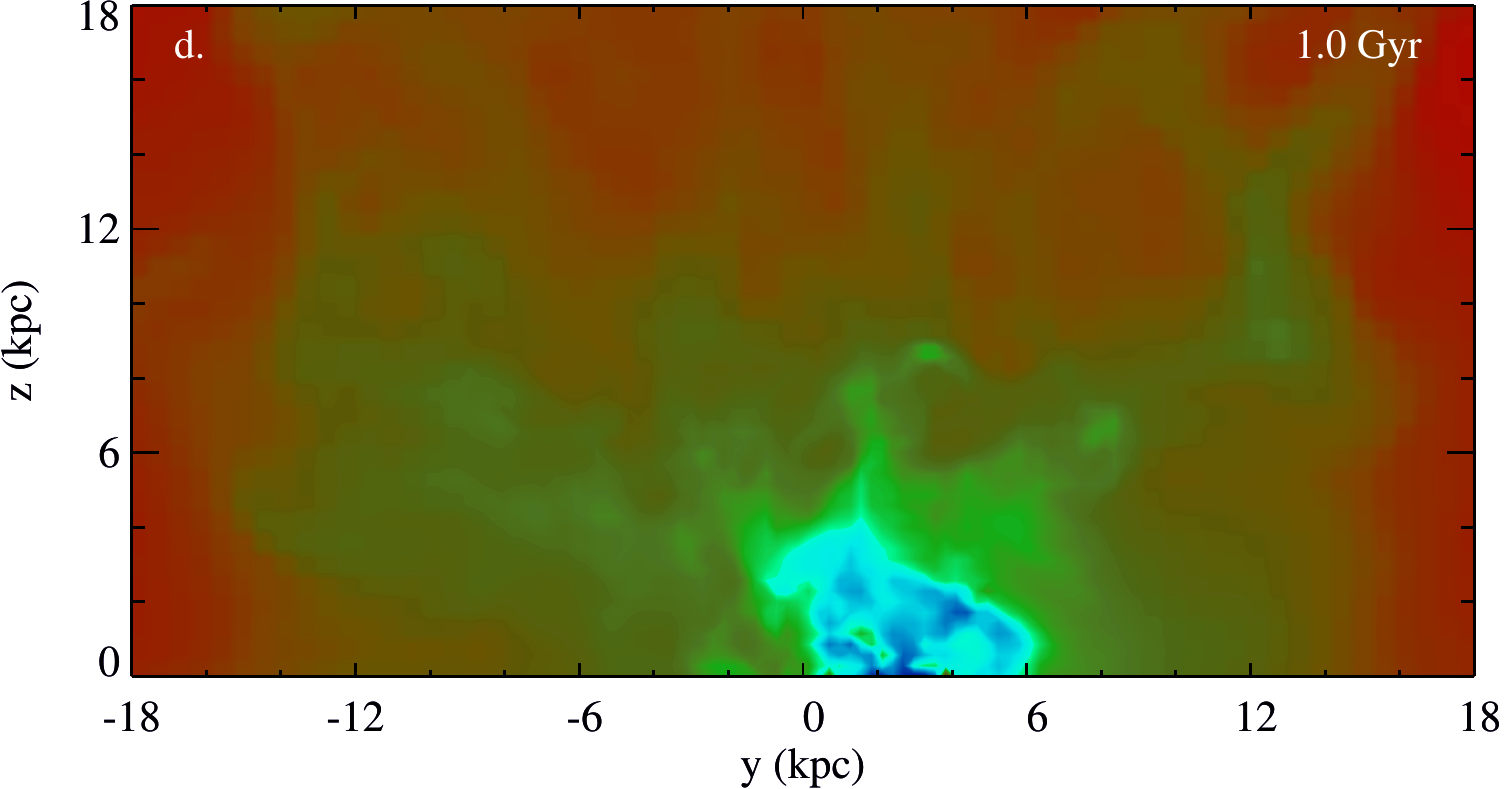}}
    \subfigure{\includegraphics[scale=0.51]{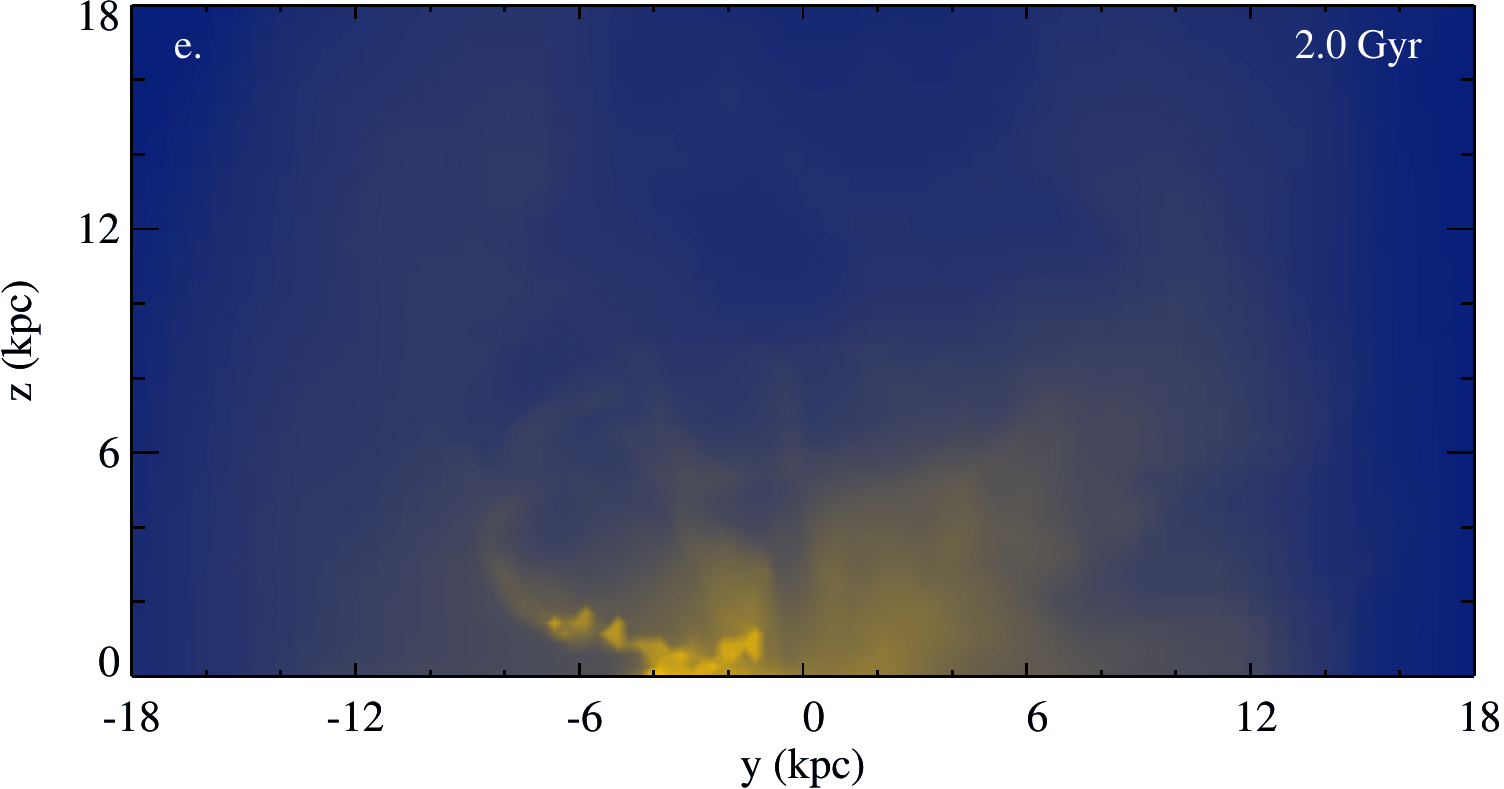}}
    \subfigure{\includegraphics[scale=0.51]{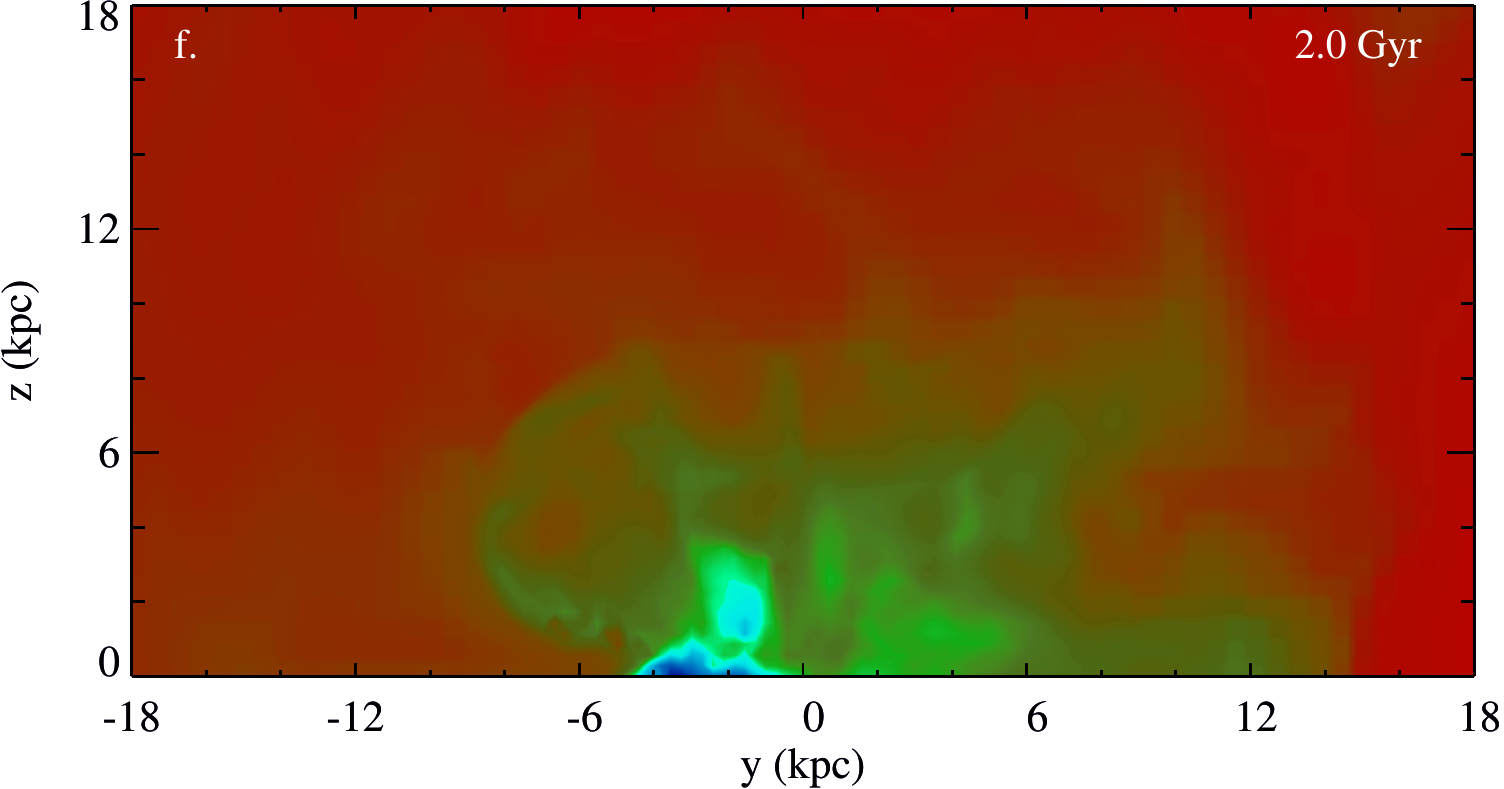}}
    \subfigure{\includegraphics[scale=0.51]{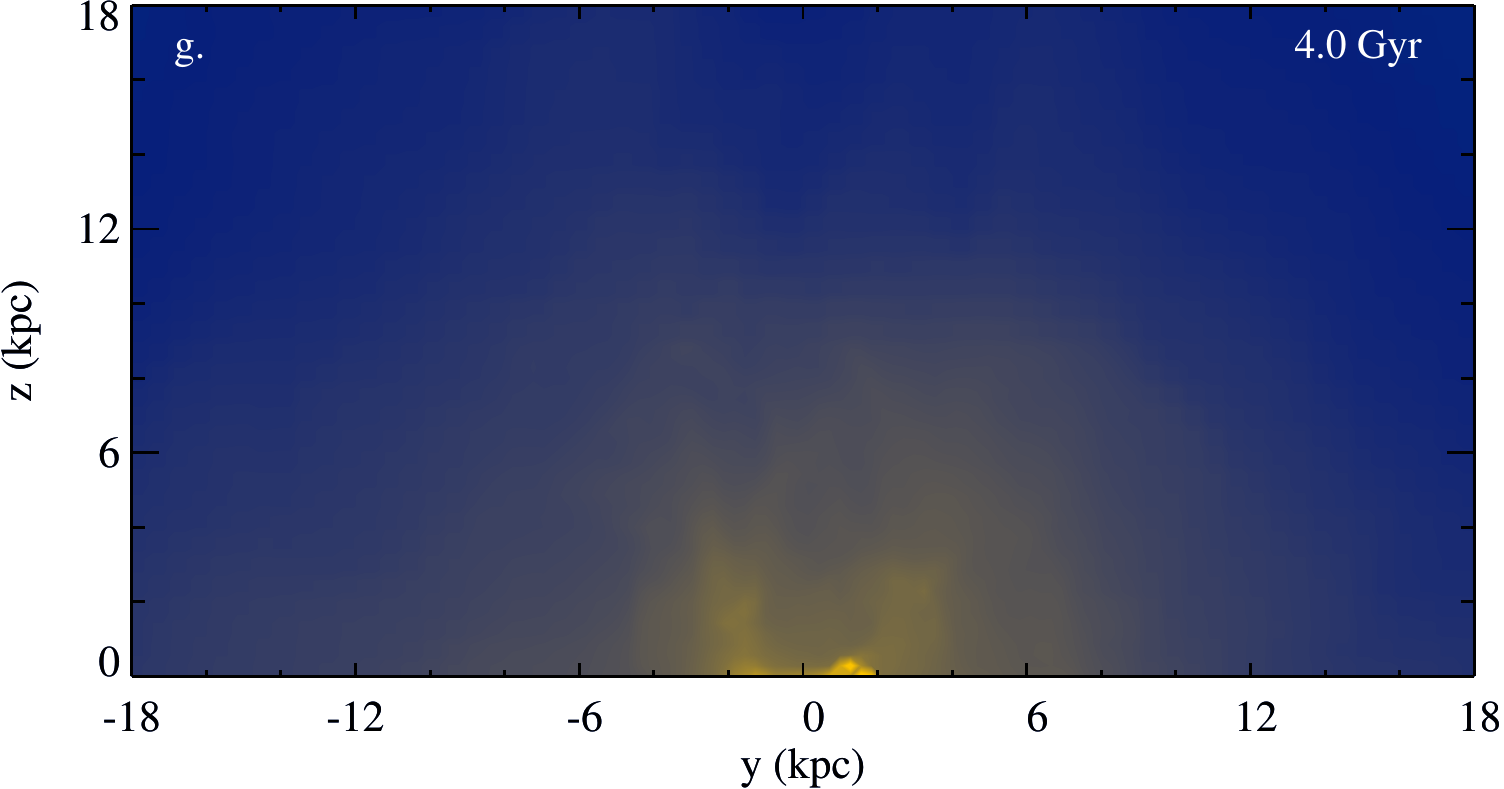}}
    \subfigure{\includegraphics[scale=0.51]{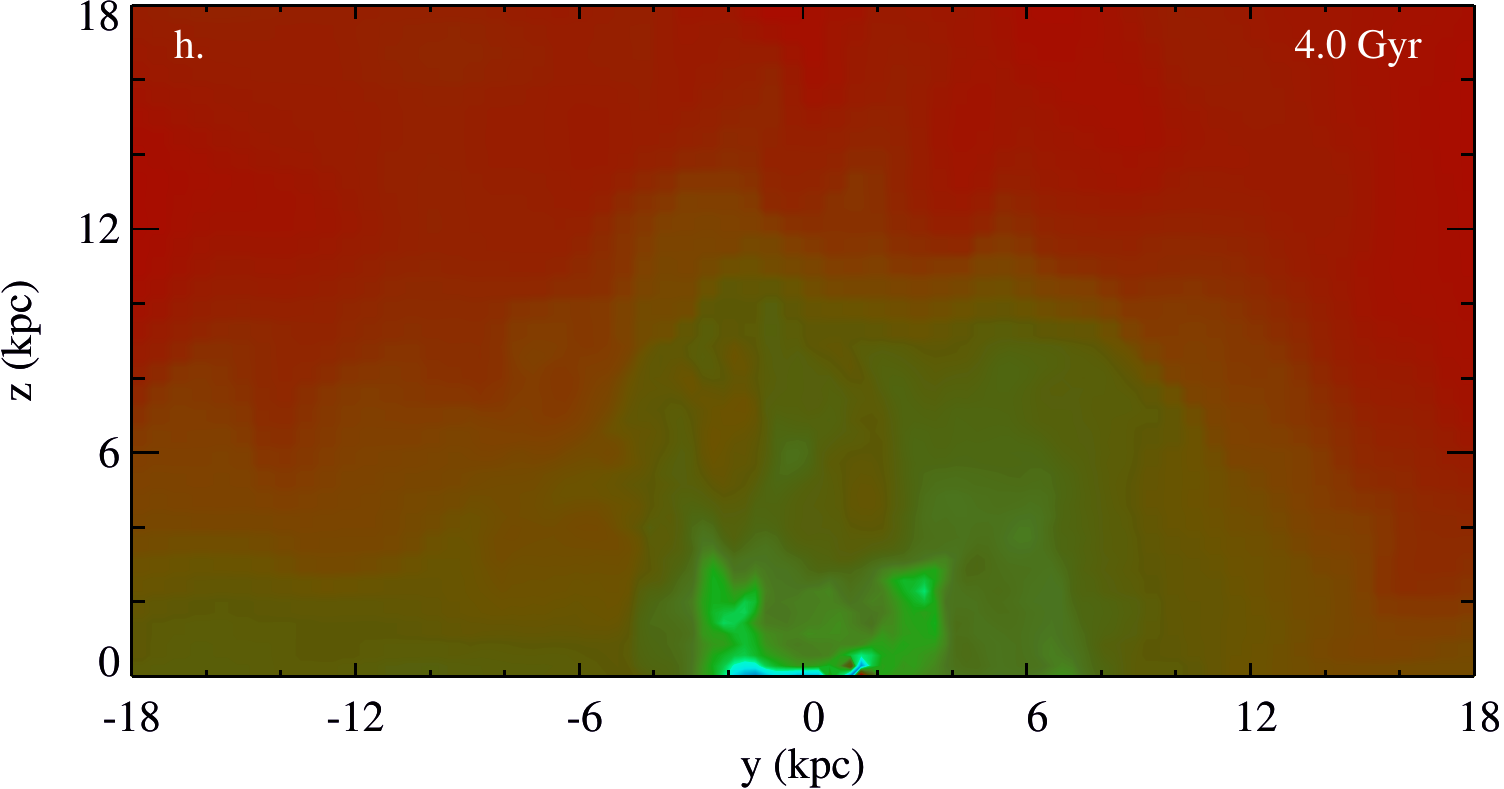}}
    \subfigure{\includegraphics[scale=0.51]{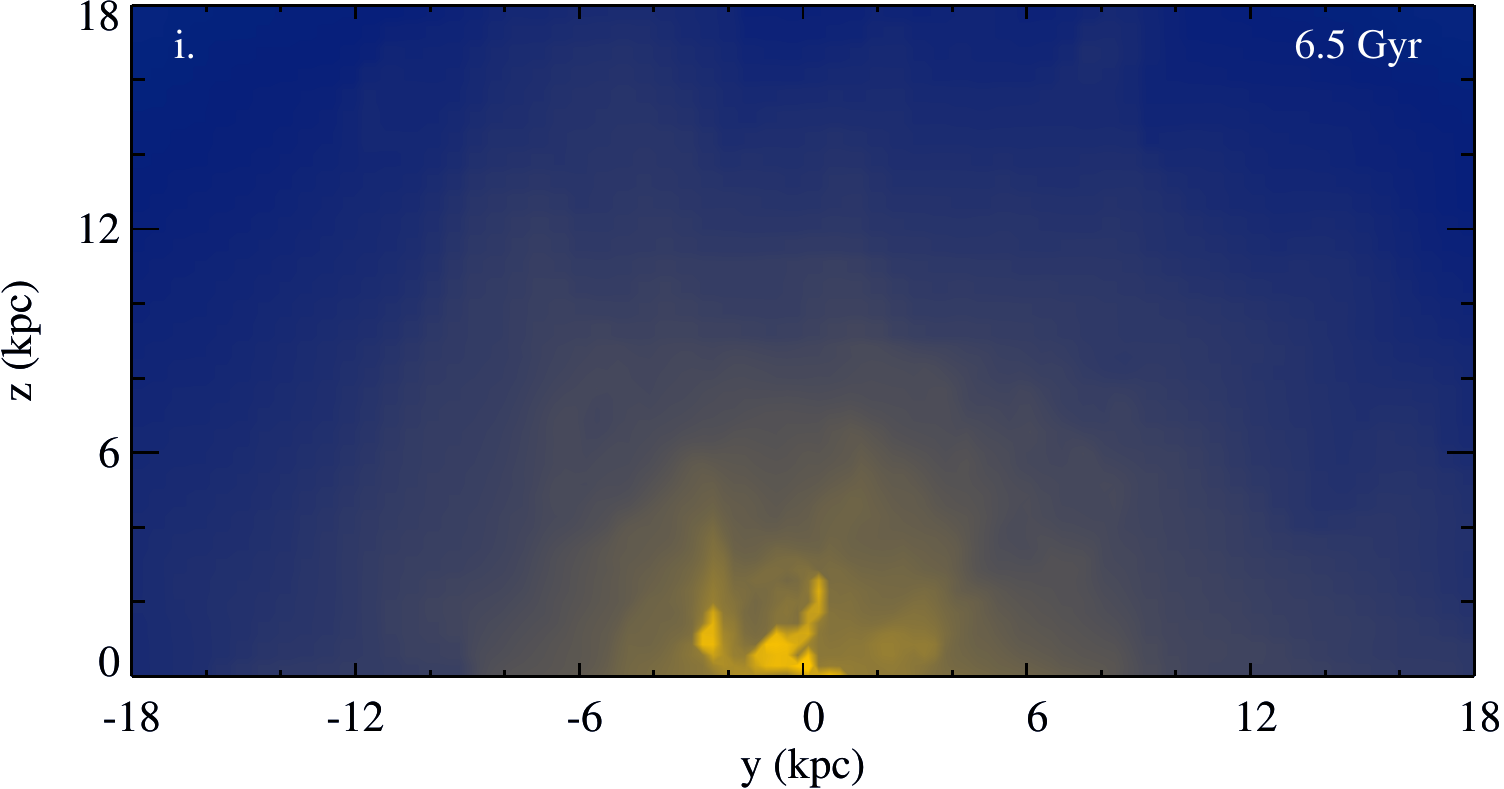}}
    \subfigure{\includegraphics[scale=0.51]{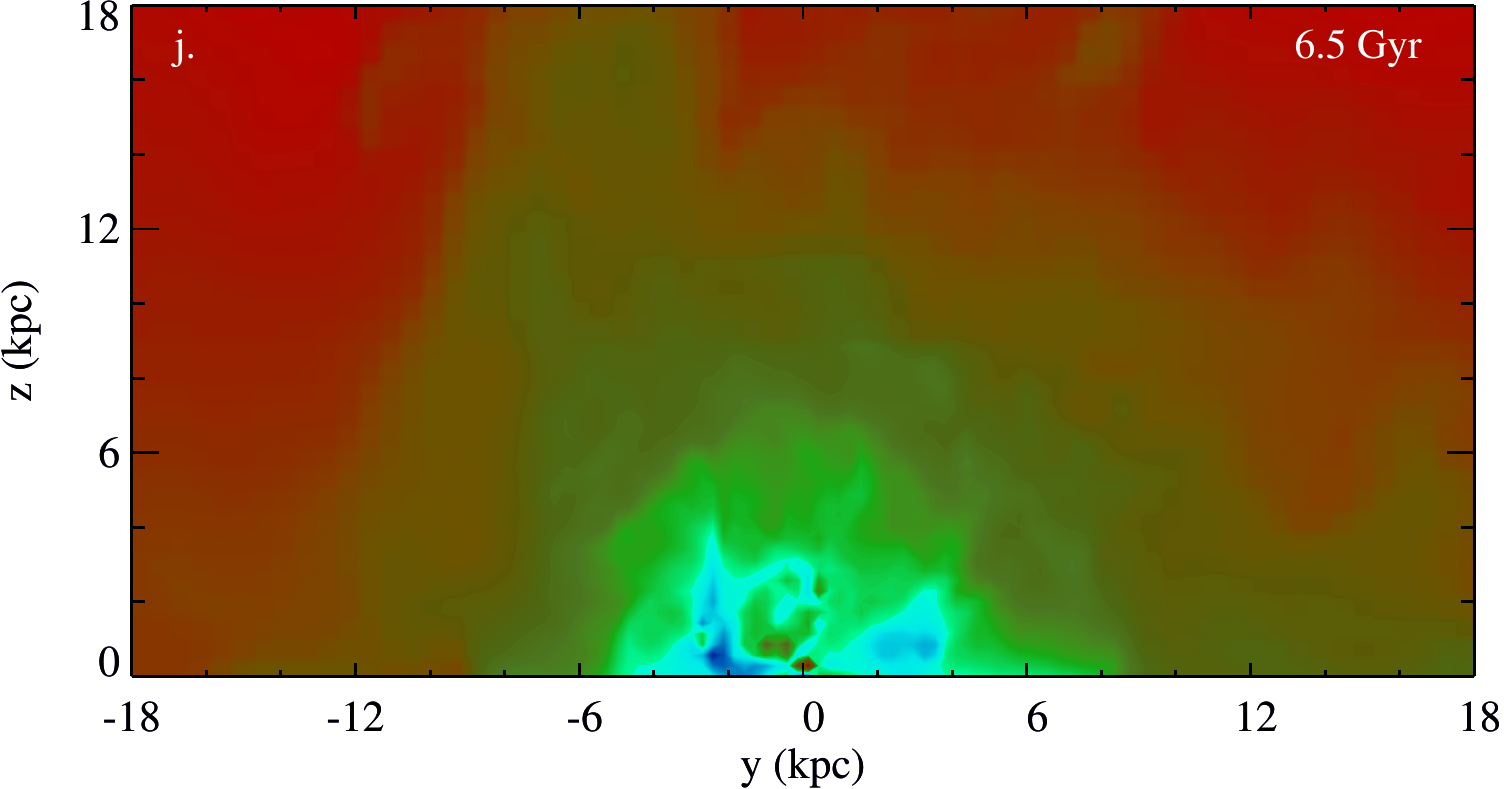}}
    \caption{Maps of X-ray surface brightness (left) and projected spectroscopic-like temperature (right) for the best model with circumgalactic gas (cgg-8em4), at five different times (from top to bottom row): 0.5, 1.0, 2.0, 4.0, and 6.5 Gyr. See Section 4.1.1.}     
    \label{fig:cgg-maps}          
\end{figure*}

\begin{figure*} 
 \addtocounter{figure}{-1}
    \subfigure{\includegraphics[scale=0.463]{art_CGG_colorbar_SBx.pdf}}
    \subfigure{\includegraphics[scale=0.463]{art_CGG_colorbar_Tspecl.pdf}}
    \subfigure{\includegraphics[scale=0.463]{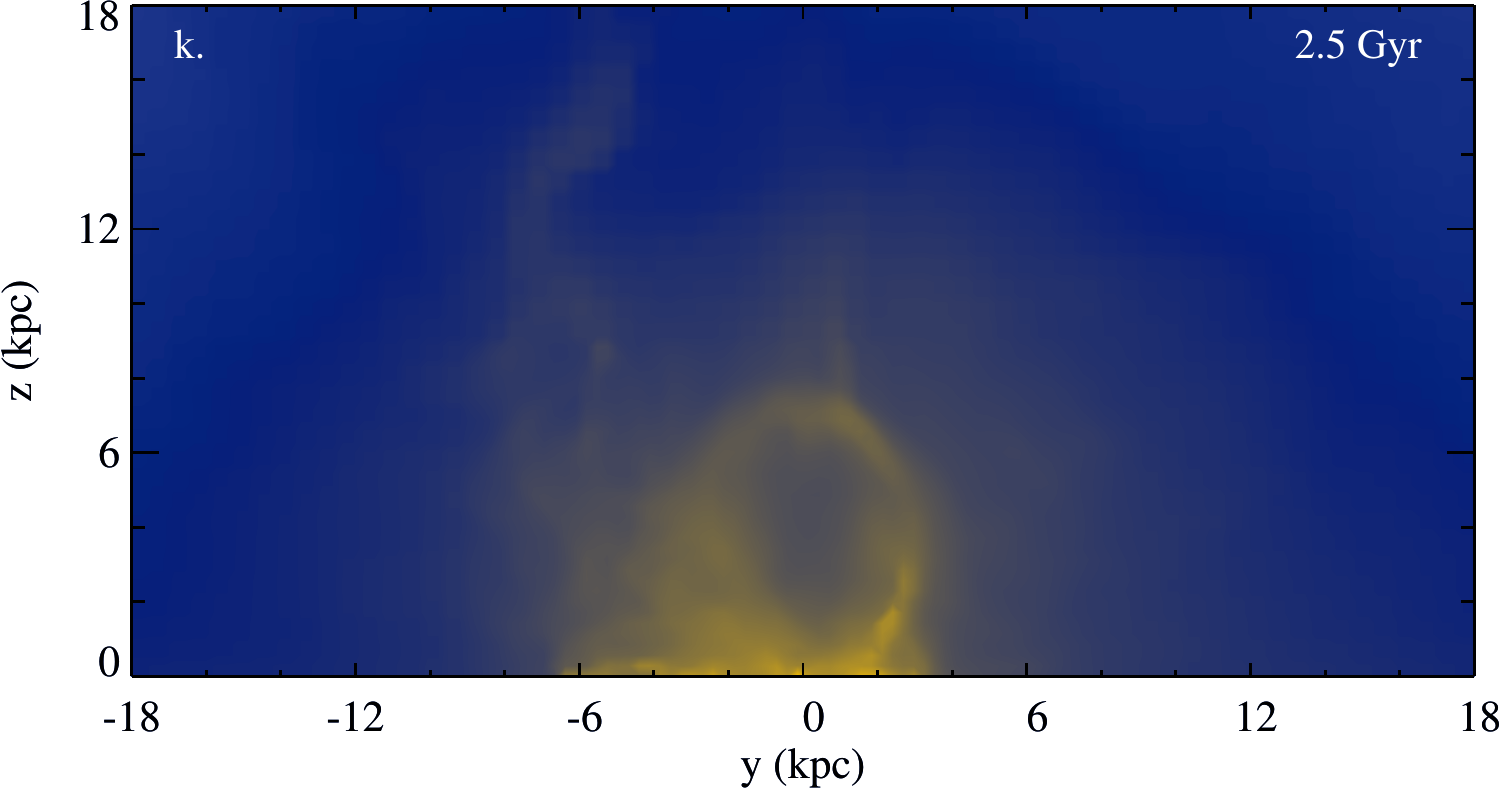}}
    \subfigure{\includegraphics[scale=0.463]{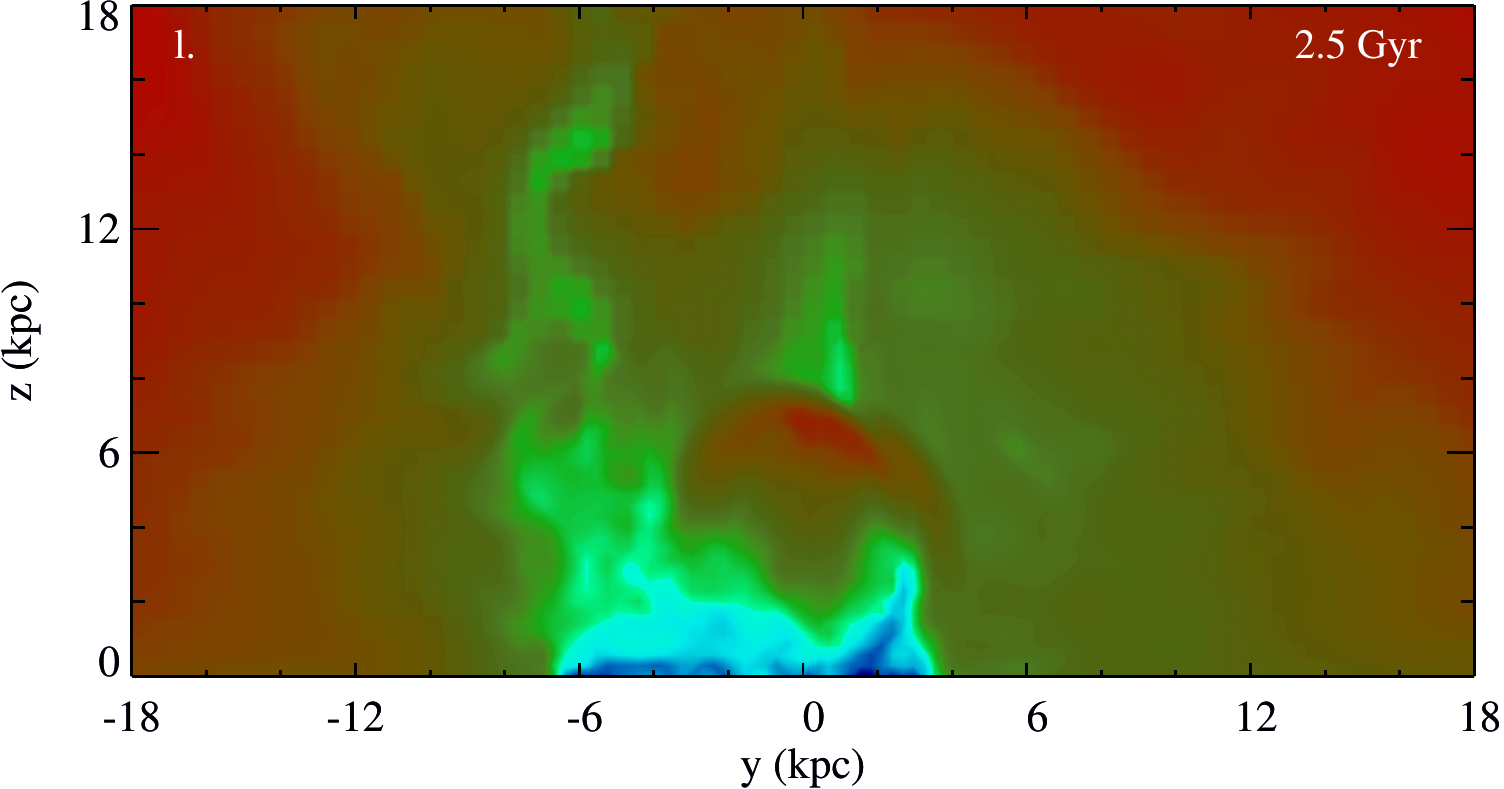}}
    \subfigure{\includegraphics[scale=0.463]{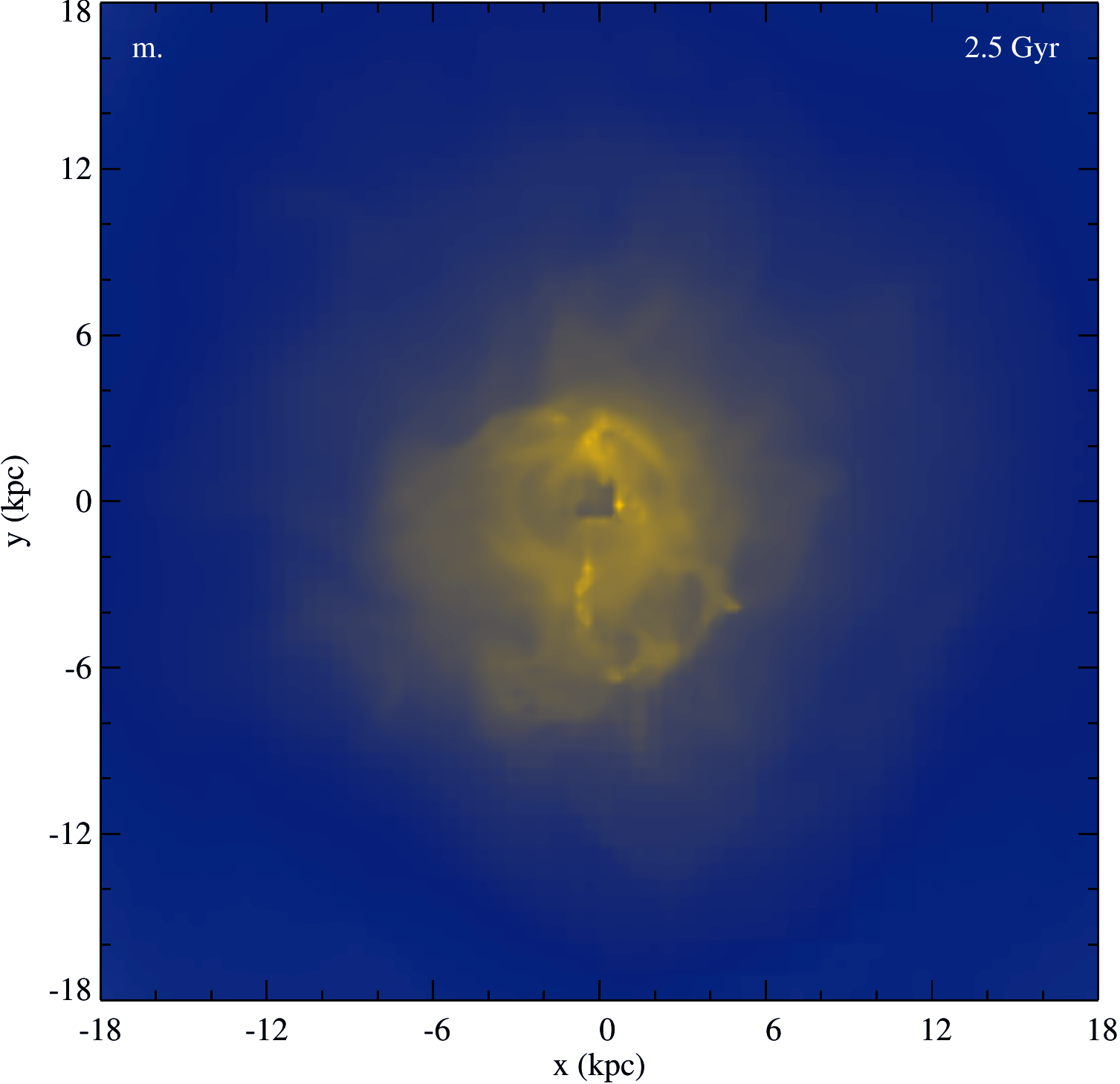}}
    \subfigure{\includegraphics[scale=0.463]{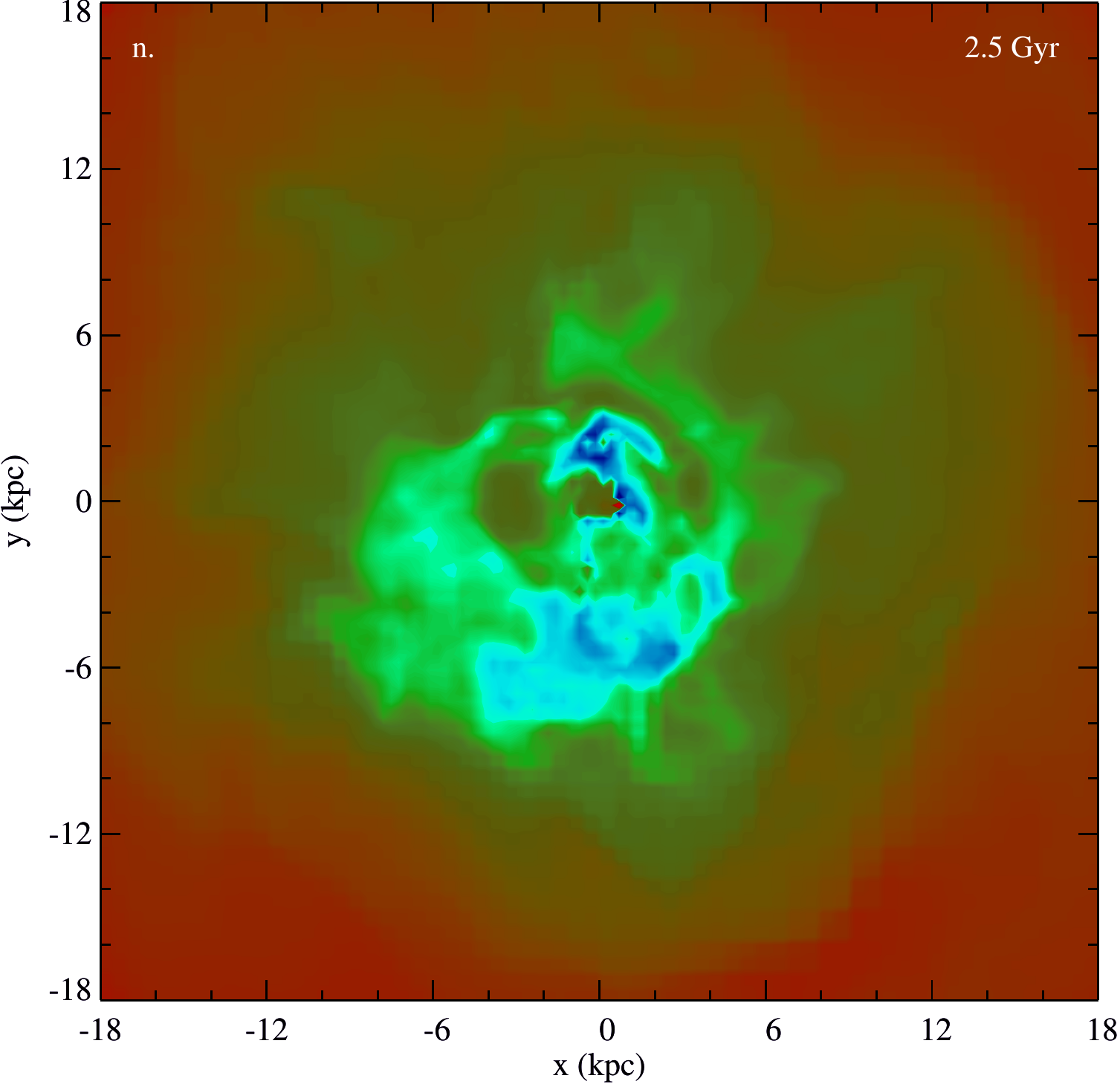}}
    \caption{{\it Continued.} For time 2.5 Gyr we present both the $x$-axis (top) and $z$-axis (bottom) projection of the SB$_{\rm X}$ and $T_{\rm sp}$ map.}                 
%    \caption{Model cgg-8em4 at $t=2.5$ Gyr: maps of X-ray surface brightness (left column) and spectroscopic-like temperature (right column), projected along the $x-$axis (first row) and $z-$axis (second row). }                
\end{figure*}
\begin{figure*} 
    \subfigure{\includegraphics[scale=0.463]{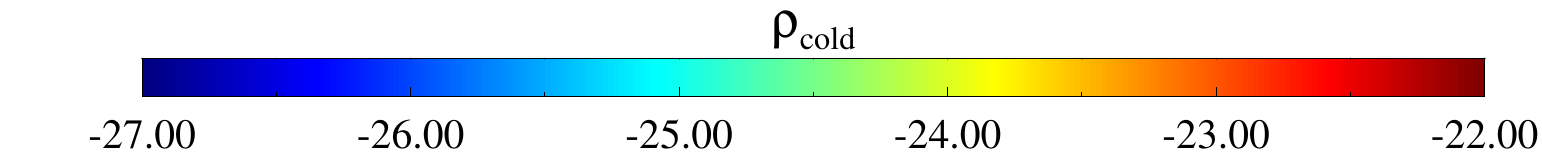}}
    \subfigure{\includegraphics[scale=0.463]{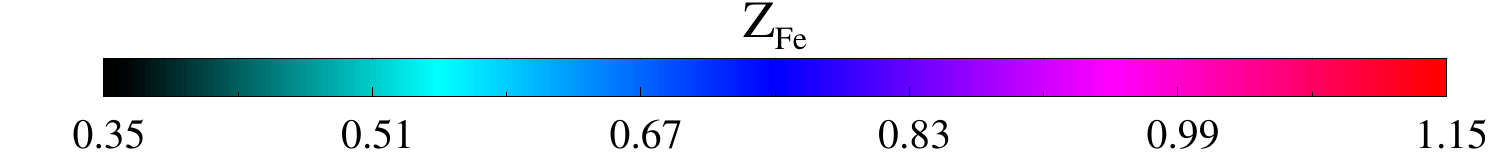}}
    \subfigure{\includegraphics[scale=0.463]{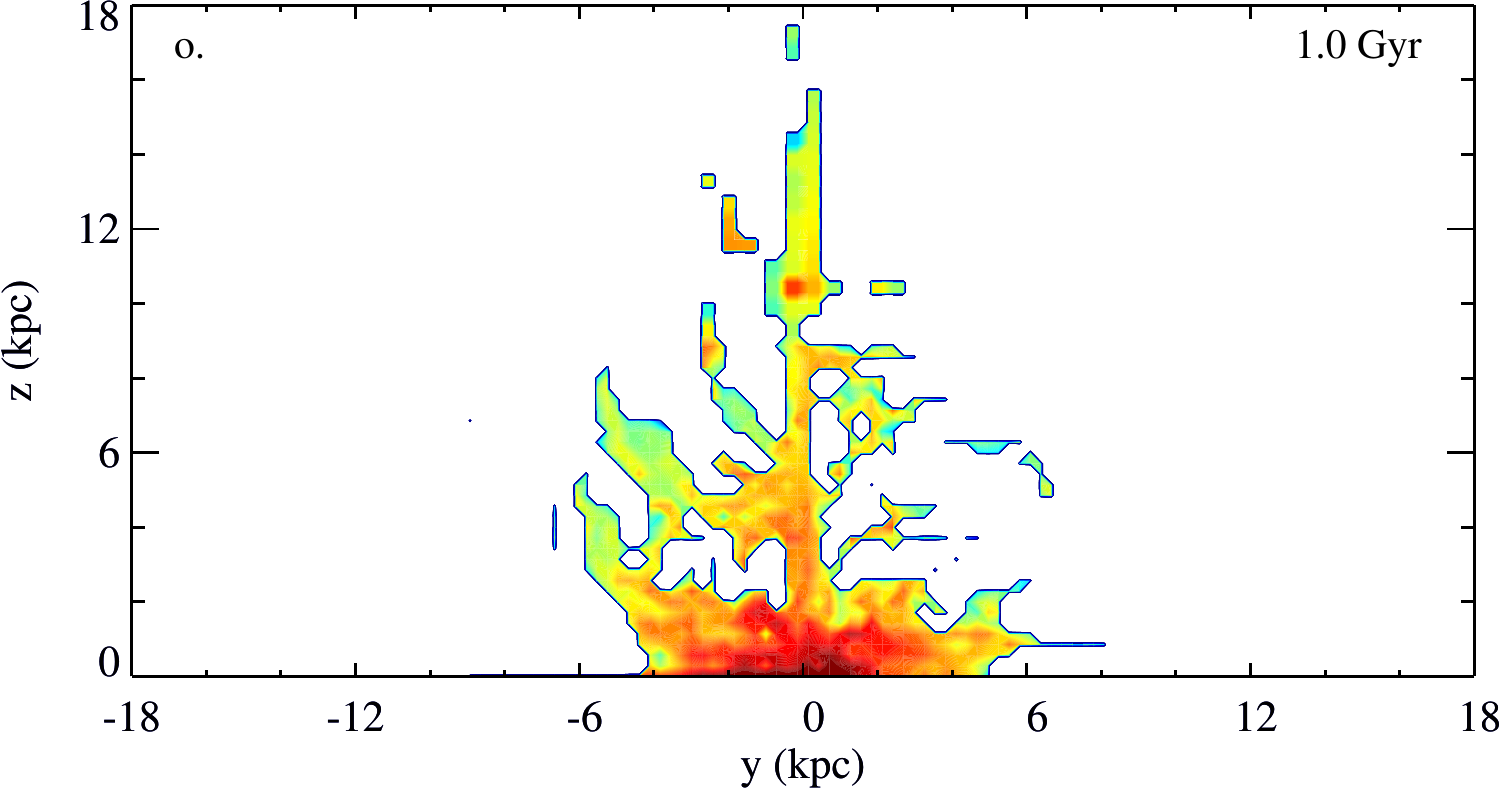}}
    \subfigure{\includegraphics[scale=0.463]{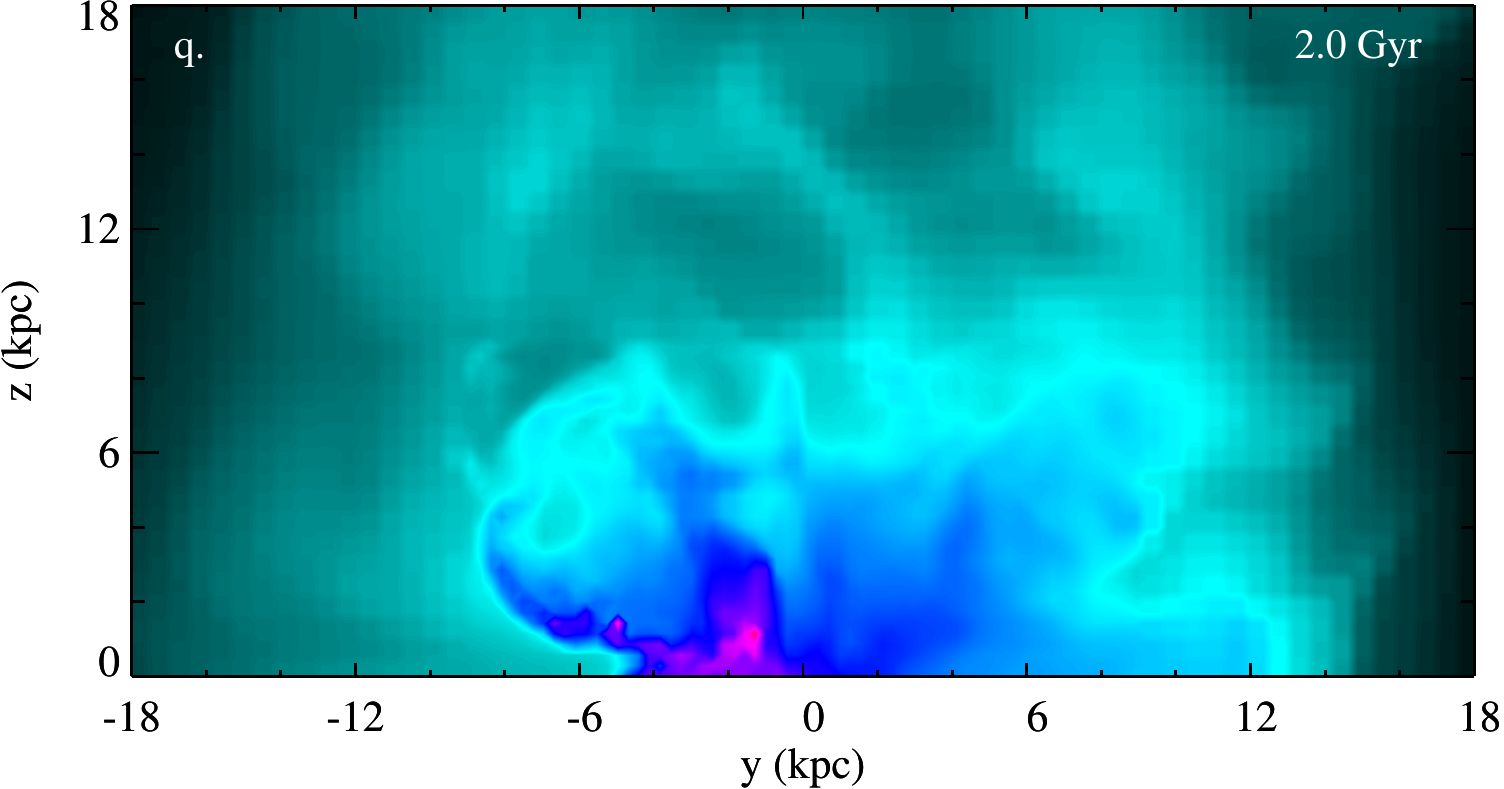}}
    \subfigure{\includegraphics[scale=0.463]{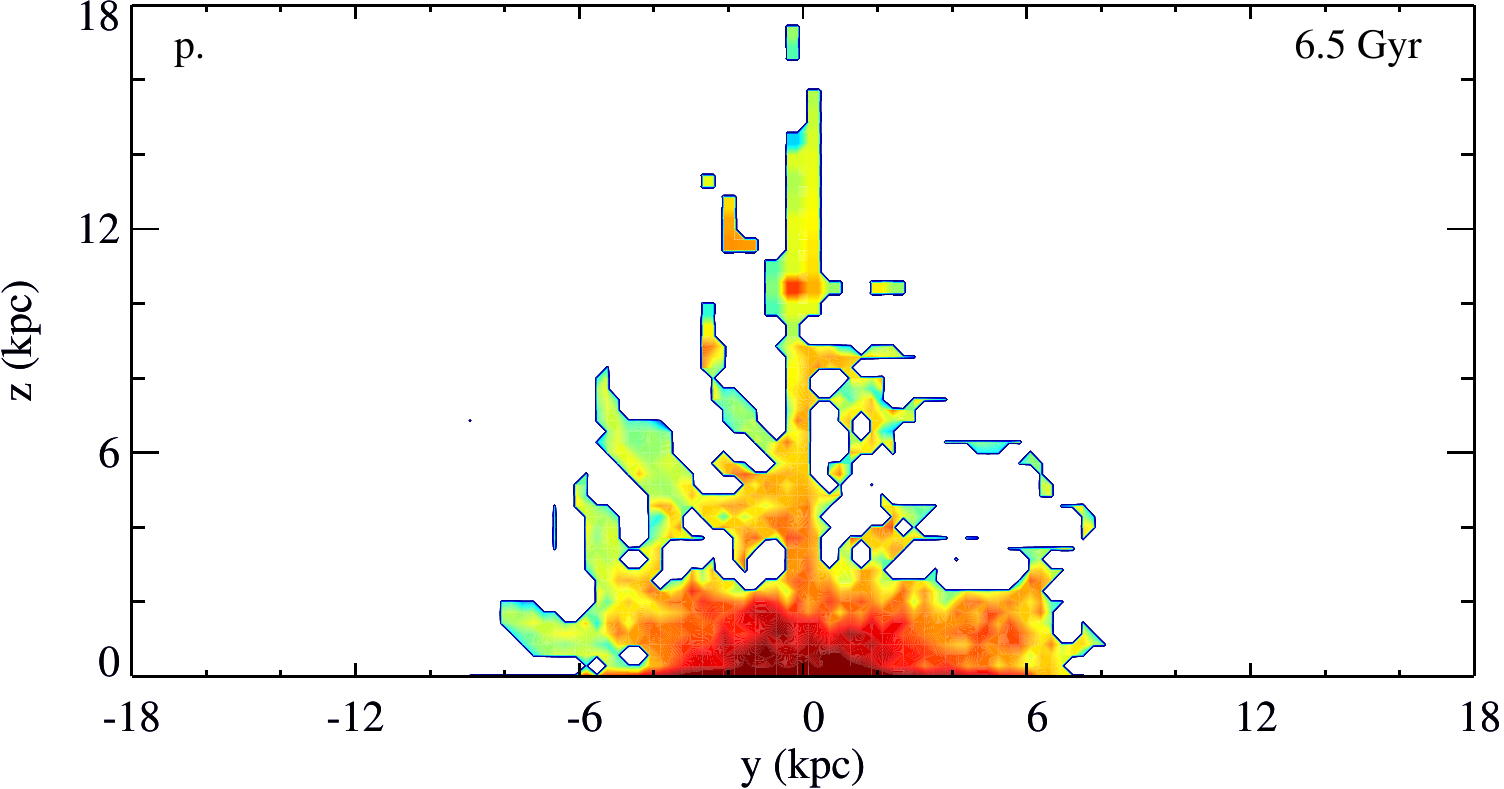}}
    \subfigure{\includegraphics[scale=0.463]{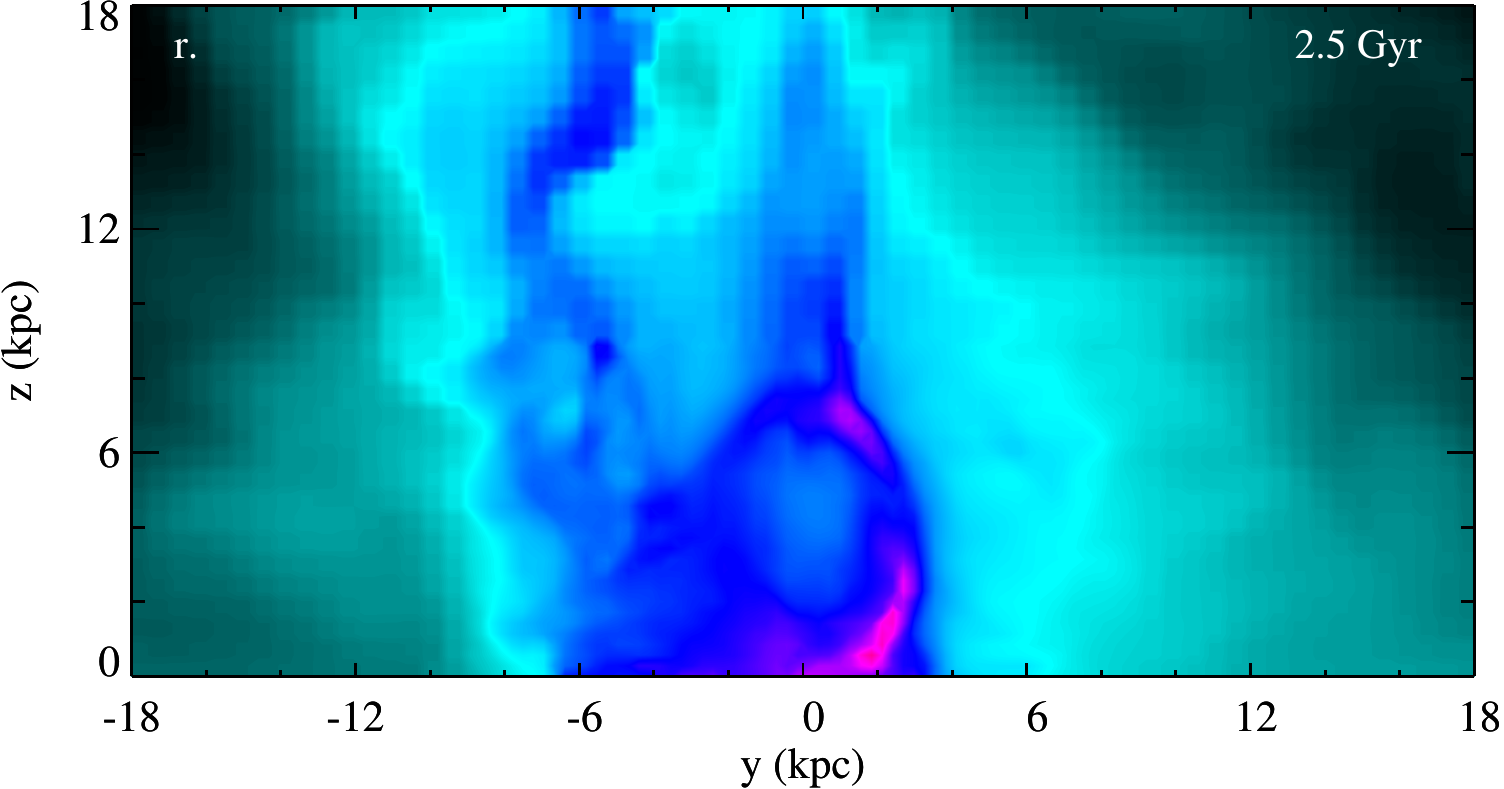}}
    \caption{Maps of total dropped cold gas ($T \leq 2 \times 10^4$ K; left) and emission-weighted iron abundance (right), both projected along the $x$-axis, for model cgg-8em4 at different times; see Section 4.1.1.}                
    \label{fig:cgg-maps2}
\end{figure*}

The variety of X-ray features induced by the AGN outflows
in the isolated galaxy (Sec. 3.4.1),
such as buoyant bubbles and shocks, is widely present also in the CGG simulation.
For the cgg-8em4 we find that the AGN activity is more frequent
and powerful compared to the isolated galaxy (cf. Figures \ref{fig:iso-AGN} and \ref{fig:cgg}). 
Nevertheless, the X-ray appearance of the two models is remarkably
similar. 

In the first row of Figure \ref{fig:cgg-maps} we show the X-ray surface
brightness and spectroscopic-like projected temperature at $t=0.5$ Gyr (panels a and b).
%[[FIGURA0002]]. 
We focus our attention on the inner $20 \times 10$ kpc of the maps, where the observations could
return better details. The outflow carved a long cavity (or channel)
$\sim 10$ kpc long and $\sim 2$ kpc wide. The brightness depression
is about a factor of 2, which should make this feature detectable in deep, high-resolution images.
For $z\lta 4$ kpc the cavity
is surrounded by sharp bright rims (or filaments), which remind the southern open
cavity in NGC 5044 (\citealt{Buote:2003,David:2009,Gastaldello:2009}). The cavity is slightly hotter 
(by $\sim 10$\%, see $T_{\rm sp}$) than the nearby gas. At $z\gta 6$ kpc broad regions ($\Delta y
\sim 3$ kpc) of enhanced surface brightness, confining the X-ray
feature, are present. This channel is typically fragmented by AGN
turbulence or strong recurrent outbursts, and would also be
  destroyed by the bulk motion (``weather'') induced by the
  cosmological accretion of matter on the system.
On the $z=0$ plane, close to the centre, the inflowing gas forms a
dense torus which is punched, but not cleared away, by the outflows.
This feature could, however, be numerically amplified by the forced planar symmetry
($z = 0$ plane), where reflection boundary
conditions are implemented.

At $t=1$ Gyr (panel c, Fig. \ref{fig:cgg-maps}) the SB$_{\rm X}$ map shows a region of %0004
enhanced emission, several kpc in size, close to the centre of the
galaxy ($0\lta y \lta 5$ kpc). 
They are caused by relatively dense and cold gas, some of
which is actually cooling to low temperatures ($T \lta 10^5$ K) and dropped out from the flow (Figure \ref{fig:cgg-maps2}, panel a).
This is a very interesting phenomenon, because the off-centre cooling
might explain the widespread presence of cold ($T\sim 10^4$ K) gas in
large ellipticals with conspicuous hot gas (see references quoted in
the Introduction). %Max: references? ...
Despite being linearly stable against thermal instability
(e.g. \citealt{Balbus:1988,Balbus:1989,Loewenstein:1989,Malagoli:1990}),
the non-conducting hot gas may well cool 
% Max: why umagnetized? Se magnetizzato puo' generare TI, vedi McCourt
% (tolto)
if the amplitude of the perturbations is sufficiently large
(\citealt{Reale:1991,Yoshida:1991,McCourt:2012,Sharma:2012}).  High
resolution X-ray maps show indeed a wealth of irregularities in the
hot gas, which certainly imply large perturbations in the gas density.
Localised gas density enhancements are naturally generated in regions
of converging flows, resulting in local cooling times much lower than
the free-fall times. This allows the gas to reach very low
temperatures, $T\approx 5\times 10^4$ K, when it is removed from the
grid (see also \citealt{Brighenti:2002}).  The process of multiphase
gas formation (in cluster cores heated by AGN jets) has been
investigated in depth by our recent work, \citet{Gaspari:2012},
showing that the non-linear perturbations, associated with the AGN jet
feedback,  commonly induce the condensation of hot gas into a
filamentary extended cold phase (as also indicated by Figure \ref{fig:cgg-maps2},
left panels).

Cooling gas at large distance from the galactic centre is seen also at $t=2$ Gyr (panel e),
%[[FIGURE0008]], 
in an enhanced SB$_{\rm X}$ region defined by $z\lta 1$ kpc and $R
\lta 7$ kpc. Several cavities surrounded by relatively cold, bright
rims buoyantly rise on various scales (panel f), while very weak
shocks rapidly vanish as sound waves at large radii.  The cold rims
are metal-rich, as expected, being formed by gas originally at the
centre of the galaxy. In the present model, in fact, we followed the
chemical enrichment of the hot gas halo, with emphasis on the
evolution of the iron expelled by SNIa.  The map of the
(emission-weighted) iron abundance is shown in Figure
\ref{fig:cgg-maps2} (panel b).  The AGN outflows generate asymmetries
in the Fe distribution: the relatively metal-rich gas originally in
the centre is uplifted along the direction of the outflow and
cavities. This results in regions of enhanced iron abundance (by
$20-30$\%) which trace the dynamics of the recent outflows.
Beside this global behaviour, kpc-sized inhomogeneities are
also evident in the abundance map. The ability of mechanical
  feedback to dredge up cold, metal-rich gas from the centre of the
  system could in principle be used to discriminate between outflows
  or thermal feedback mode.
% where the iron enrichment due to SNIa is more important. The
% abundance typically increases by $\sim 20-30$\% in the rims.

At $t=4$ Gyr (panels g and h)
%[[FIGURE0016]] 
a complex structure is visible in the X-ray surface brightness image,
mainly due to the generation of multiple cavities and bright
filaments. As usual, brightness contrasts are accompanied by
temperature and metallicity variations, here on the order of
10\%. These images show that the recurrent AGN outflows naturally
generate and drive turbulence in the core, with typical velocities of
few hundreds km s$^{-1}$ (see Section 5.1.1). The chaotic turbulent
environment
%The action of the outflows in a turbulent environment %`live' environment
naturally produces the complex irregular features, at the same time
increasing the level of thermalisation in the central region via jet
fragmentation.

At $t=6.5$ Gyr (panels i and j), almost at the end of the simulation,
%[[FIGURE0026]]. 
we observe irregular regions of high surface brightness in the central
few kpc, where the gas cools to very low temperatures producing
diffuse H$\alpha$ emission.  Figure \ref{fig:cgg-maps2} (panel c)
provides the total dropped cold mass (integrated in time), indicating
again that the cold phase condenses out of the hot flow not only at
the very centre, but in an extended region of radius $7-15$ kpc.
The dropout term in the equations used prevents a detailed
  analysis of the thermal instability; we refer the reader to
  \citealt{Gaspari:2012}, where fully multiphase simulations are presented).
Quantifying the exact amount of gas cooling far from the
centre is a difficult problem (due to numerical diffusion and
resolution). Nevertheless, the result is qualitatively robust:
off-centre cooling exists, as indicated by observations (e.g.~\citealt{McDonald:2011}), and it can be
triggered by the AGN feedback process.

We conclude the snapshot analysis for the CGG model showing
%[[FIGURE0010]], 
a textbook example of X-ray cavity confined by bright rims (panels k
and l).  The largest brightness contrast between the X-ray hole and
the rims is about a factor of two.  From the spectroscopic-like
temperature map we see that the lower part of the rims ($z\lta 3$ kpc)
are colder, in projection, than the nearby ISM, while the upper part
is slightly hotter, because of shock heating.  Again, the rims, being
originated from central gas, are slightly more metal-rich than the ISM
at the same radius, by 15-25\% in the iron abundance
(Fig. \ref{fig:cgg-maps2}, panel d).  For this time, we also show
the %brightness and temperature maps
projection along the $z$-direction (Fig. \ref{fig:cgg-maps}, panels m
and n).
%[[QUESTA E' COMPOSTA DELLE 2 MAPPE QUADRATE DI SB e T]]. 
Several structures and cavities are
visible, together with X-ray bright filaments. In this projection the
$T_{\rm sp}$ is characterised by an arc-shaped %Max: banana sembrava strano...
region about $\sim 20$\% colder than the neighbouring gas.
This comparison between two different projections warns how
complicated is the interpretation of real data, especially when the
inclination of the outflow, with respect to the plane of the sky, is uncertain.

The generation of buoyant bubbles, together with all the previous
consistent observational features (extended multiphase gas, metals
asymmetries, turbulence, weak shocks), strengthens the key role of
mechanical AGN outflows in regulating the thermal and dynamical
evolution of elliptical galaxies, with or without circumgalactic gas.

\section{Discussion} %\& Conclusions}
%\subsection{Summary of results for Elliptical galaxies}
\subsection{AGN feedback in elliptical galaxies}
In the previous Sections we have investigated the effect of AGN
feedback on the evolution of the ISM in massive elliptical galaxies.
In designing the 3D hydrodynamic simulations we have been 
guided by several ideas.
\begin{enumerate}
\item
The AGN feedback in low-redshift elliptical galaxies acts mainly through a
(radiatively inefficient) mechanical and directional process, in the
form of jets or winds. 
%(especially at low and moderate redshifts).
\item These outflows interact with the ISM and heat the medium on
  scales of several kpc, as indicated by X-ray observations of
  cavities and shocks.
\item The AGN reacts to gas cooling: cold gas condenses out of the hot
  phase in the central region, and it is assumed accreted onto the
  black hole in a few dynamical times, triggering the AGN feedback.
  The uncertainties on the the real accretion rates force us to adopt a
  parametrisation as simple as possible (the only governing parameter
  being the mechanical efficiency $\epsilon$).  Given the
  uncertainties mentioned in Section 2.1.1, we are not in the position
  to investigate important issues such as the exact BH growth or the
  details of the AGN duty cycle.
\item
The feedback mechanism must work for objects of any scale, from
isolated galaxies to massive galaxy clusters.
\item
The calculated evolution of the hot gas must be checked against
many available observational constraints, for a long time span (several Gyr).
\end{enumerate}
In G11a,b we have shown that AGN outflows are able to solve the
cooling flow problem for both massive clusters and groups, for more
than 7 Gyr.
%indicating that they can meet point (iv) too.
We are left now to demonstrate that these mechanical massive outflows,
self-regulated by cold accretion, represent also the dominant feedback
mechanism on the galactic scales.

In Section 3 we have discussed several models for the isolated
elliptical, while the more astrophysically relevant case, i.e. the
galaxy with circumgalactic gas, is tackled in Section 4.  The two key
tests that the models must pass are the agreement with the low cooling
rates allowed by observations ({\it no overcooling}), and the moderate gas temperature in the
central region of galaxies ({\it no overheating}). The former sets a lower
limit for the feedback mechanical efficiency, the latter provides an
upper bound.

In Figures \ref{fig:iso-AGN} and \ref{fig:cgg} we showed that assuming
an efficiency $\epsilon \gta 3\times 10^{-4}$ ($\epsilon \gta 8\times
10^{-4}$) for the isolated (CGG) model, the cooling rate drops to
values close the observational limits reported in the Introduction.
The previous values for the mechanical efficiency are lower than those
working for clusters (G11a) and similar to those quoted in
\citet{Ciotti:2007}, although for a different feedback scheme.
%still in an isolated elliptical galaxy. 
%These number depends somewhat on the
%parameters of the feedback scheme and the numerical resolution
%adopted. 
Because our estimated accretion rate onto the
BH is likely an upper limit, a firm result of our
calculations is that the
mechanical efficiency must be $\epsilon \gta \, {\rm few}\, \times
10^{-4}$ in order to 
prevent significant gas cooling in massive elliptical galaxies.
It is more
difficult to place an upper bound for the efficiency.
We have shown in Section 3 that when $\epsilon \gta 10^{-3}$ the AGN
feedback is too intense and the ISM in the central region is
overheated. However, larger mechanical efficiencies would be allowed
if the accretion rate is grossly overestimated in our simulations. 

While the exact details of feedback engine (such as the duty cycle)
are uncertain and must be investigated by high-resolution, specialised
simulations, the astrophysically most relevant and solid result of
this work is that AGN outflows can {\it solve the cooling flow problem} also in
elliptical galaxies, drastically quenching the cooling rates
and properly maintaing a quasi thermal equilibrium in the core. This
is a remarkable result, given the difficulties encountered with the commonly used thermal feedback
(regardless of its origin, i.e. artificially inflated
bubbles or radiative heating).

They also agree with a number of other key and independent
observational constraints, as illustrated in Sections 3 and 4. A crucial requirement for
an AGN feedback scenario is the ability to produce (weak) shocks,
X-ray cavities and abundance inhomogeneities, common features shown in
deep X-ray observations (see references in the Introduction). These
properties point toward a feedback mechanism able to distribute energy
and momentum in a directional way, on scales of several
kpc. Observationally, these ISM perturbations are best studied in
X-ray bright objects, often giant ellipticals with CGG or
groups. About 25-50\% of galaxies/groups observed at high resolution
show evidence of cavities (\citealt{McNamara:2007,Dong:2010}), a
number likely underestimated. In galaxy clusters the detection
fraction rises to $\gta 2/3$ (\citealt{Dunn:2006}). Moreover, cavities
are more likely present in cool-core systems (\citealt{Dong:2010}),
implying that the feedback action must preserve the positive
temperature gradients.  Elliptical weak shocks are also commonly found
(\citealt{Gitti:2012}), surrounding stable buoyant cavities, a natural
by-product of jets and outflows (G11a,b).

In Sections 3.4.1 and 4.1.1 we have illustrated the wealth of
observable features created by the (anisotropic) mechanical feedback
scenario.  Deep X-ray observations of galaxies (e.g. NGC 4472,
\citealt{Biller:2004}; NGC 4374, \citealt{Finoguenov:2008}; NGC 4636,
\citealt{Baldi:2009}) and groups (e.g. NGC 5044, \citealt{David:2009};
HCG 62, \citealt{Gitti:2010}; NGC 5813, \citealt{Randall:2011}; NGC
5836, \citealt{Machacek:2011}) reveal a wealth of features certainly
linked to the AGN feedback
outbursts. %by subrelativistic collimated outflows.
In particular, the two typical fingerprints naturally left by
collimated AGN outflows, i.e. X-ray cavities surrounded by a weak
shock, are seen in many snapshots of Figures \ref{fig:iso-maps} and
\ref{fig:cgg-maps}, often resembling observations also from a
quantitative point of view (e.g. elliptical cocoons, Mach number,
bubble size and temperature). It is impossible to make a 1:1
comparison with real objects, catching the exact evolutionary moment,
with the same instantaneous feedback values.
%because of the complex and variegated evolutionary sequence. 
Unfortunately, we are not in the position to calculate the time
fraction in which the simulated cavities or shocks are visible; this
would in fact require an analysis of the flow very finely spaced in
time -- subject of a future investigation.
% Moreover, the uncertainties in simulating the accretion/outflow
% process and so the duty cycle of the AGN, would make this result
% rather uncertain anyway.

Typically, the cavities inflated by the outflows have size of few kpc
and are surrounded by bright rims (e.g. Figure \ref{fig:iso-maps},
$t=3.75$, $4.25$, and $11.75$ Gyr; Figure \ref{fig:cgg-maps}, $t=0.5$,
2.0, 2.5, and 4.0 Gyr). The rims are often slightly cooler and
metal-richer compared to the nearby gas, although there are few
exceptions. %(Figure 3, $t=4.25$ Gyr).
It is worth mentioning that dedicated simulations tailored for NGC
4636, indicate that it is possible to explain most of the properties
of the cavities detected in NGC 4636, assuming the feedback proposed
in this work (Ballone \& Brighenti, in preparation).

In the case of the galaxy with CGG (similar to a group environment),
the frequent activity of the AGN causes the common presence of a
low-density channel along the $z-$direction carved by the outflows,
$\lta 1$ kpc wide and several kpc long. This feature is very difficult
to detect in the surface brightness map.  In Figure \ref{fig:cgg-maps}
we see that only at the beginning of the calculation, at $t=0.5$ kpc,
the tunnel is visible, with a brightness contrast of $\sim 2-3$
(rapidly decreasing beyond $z=5$ kpc). Most of the time, the thin
channel is wiped out in the brightness image by projection effects (see also
G11b). Occasionally, the tunnel may resemble a X-ray cavity
(Fig. \ref{fig:cgg-maps}, $t=2$ Gyr). A dedicated analysis through a
very deep X-ray observation in the core of the galaxy may detect this
feature or, at least, set constraints on its presence (see, for instance, the SB$_{\rm x}$ contours
in the inner 3 kpc of NGC 5813, \citealt{Randall:2011}). The channel is
nevertheless a transient feature, easily fragmented and destroyed by
the AGN turbulence (and by the ICM ``weather'' generated by 
cosmological accretion, not considered here), 
a key element for a proper deposition and
thermalisation of mechanical energy in the core (Sec. 5.1.1).

The proposed AGN feedback mechanism has another impact on the gas
cooling process, beside significantly lowering the total rate. It
promotes spatially distributed cooling (via thermal instabilities), over
an extended region of size $\sim 15$ kpc centred on the BH. Figure
\ref{fig:cgg-maps2}, left column, show the density map of total cold
gas which has cooled and dropped out of the flow, highlighting both
the extended and concentrated toroidal distribution of cooler gas (consistent
with \citealt{Gaspari:2012}).
%It demonstrates that most gas cools in a toroidal region, 
%with size $R\lta 5$, $z\lta 2$ kpc. 
Our code does not follow the
dynamics of the cooling gas, but it is plausible that some of the
cooled gas accretes on the BH, and some is instead used to form new stars and
the emission line nebulae commonly observed in elliptical galaxies
(see Introduction). At the moment of cooling, the
gas has a chaotic motion, with typical velocities of $100-200$ km
s$^{-1}$, concordant with H$\alpha$ spectroscopic studies (\citealt{Caon:2000}).

The mechanical feedback is also able to reproduce the anisotropic
distribution of metals, commonly observed in AGN-heated cores
(e.g. \citealt{Rasmussen:2009,David:2011,Kirkpatrick:2011,OSullivan:2011enrich}). The
powerful mechanical jets can indeed uplift up to $\sim10$ kpc the
metals processed by SNIa and stellar winds in the nuclear region (with
a peak metallicity of $Z_{\rm Fe}\lta1\ Z_{\odot}$;
\citealt{Mathews:2003}).  In the more quiescent phases, AGN turbulence
tends instead to stir and diffuse the metals, restoring the
homogeneous distribution (Figure \ref{fig:cgg-maps2}, right panels).

\subsubsection{AGN turbulence}

Another test the feedback model must pass, although not stringent as
those previously discussed, is the generation of turbulence (or bulk
motion) in the ISM, especially in the central region.
Deep X-ray observations start to provide
reliable estimates on the turbulent pressure
(\citealt{Werner:2009,Sanders:2011,dePlaa:2012}). The latter authors, in
particular, give estimates of the ISM characteristic turbulent velocities,
defined as $v_{\rm turb}\equiv\sqrt 2\,\sigma_{\rm turb}$ 
%(with $\sigma_turb$ the Gaussian velocity dispersion along line of sight, 
(linked to line broadening),
in the central $\sim 10$ kpc for two exemplary groups:
$320<v_{\rm turb} < 720$ km s$^{-1}$ (NGC 5044) and
$140<v_{\rm turb} < 540$ km s$^{-1}$ (NGC 5813).
%This allow a direct comparison with our CGG model,
%which was designed to agree with NGC 5044. \citet{dePlaa:2012}
%find a hot gas turbulent velocity in the range .
In order to compare the prediction of our CGG model with this
observation, we calculate the 1D gas velocity dispersion along the line of sight $l$
%($x$- or $z$-axis),
weighted by the emission-measure, as
\begin{equation}
\sigma_{\rm turb}^2\,=\,\frac{\int\rho^2 (v_l - \bar v_l)^2\,dl}{\int{\rho^2}\,dl }\, ,
\end{equation}
where the mean (emission-weighted) velocity $\bar v_l$ is typically close to zero, 
lacking large bulk motions.

In the (projected) circular region $0 < R < 5$ kpc, the time average
velocity dispersion is $\sim 200$ km s$^{-1}$, when the system is
viewed along the outflow $z$-axis. In the phases where the AGN is more
active $\sigma_{\rm turb}$ can reach values as high as $400$ km
s$^{-1}$. Note that the outflow, with velocity often
exceeding $10^4$ km s$^{-1}$, poorly contributes to $\sigma_{\rm turb}$, 
because of the very low density.  The gas turbulence becomes
weaker at larger distances from the centre. In the ring $5 < R < 20$
kpc, $\sigma_{\rm turb}$ decreases to $\sim 60-100$ km s$^{-1}$, while
in the region $20 < R < 40$ kpc, $\sigma_{\rm turb} \sim 40-70$ km
s$^{-1}$.  One missing element of our models is the effect of
cosmological merging and inflow, which may trigger additional
turbulence, although more efficiently at large radii ($>0.1$ the
virial radius; e.g. \citealt{Vazza:2011}).

When the galaxy is observed along a line of sight perpendicular to the
outflow direction (e.g. the $x-$axis) the mean central $\sigma_{\rm
  turb}$ has lower values, $\sim 100$ km s$^{-1}$, with peaks about a
factor of two larger.  For the standard cooling flow model presented
in Section 3.1 we find instead $\sigma_{\rm turb}\lta 45$ km
s$^{-1}$. This quantity is certainly not associated with turbulent
motions, while it represents the steady radial inflow velocity due to
the massive cooling flow.

We conclude that subrelativistic outflows are able to generate, in the core, the
observed turbulent velocities,
in the range $v_{\rm turb} \equiv \sqrt2\sigma_{\rm turb} \approx 100-400$ km s$^{-1}$.
The non-thermal pressure is generally a fraction of the thermal energy, 
around one third ($E_{\rm turb}/E_{\rm th}\approx0.2\,(v_{\rm turb,
200}^2/T_7)$, where the turbulent velocity is expressed in units of
200 km s$^{-1}$, and the gas temperature in units of $10^7$ K).

\subsection{Comparison with other AGN feedback scenarios}

It is interesting to compare our purely mechanical feedback with other
types of AGN heating processes.  To our knowledge, only very few
investigations focused on AGN feedback in ellipticals\footnote{We refer the reader to G11a and G11b for the comparisons with different AGN feedback models
in other types of environment, as galaxy clusters (e.g.~\citealt{Vernaleo:2006,Cattaneo:2007}).}. 
For example,
Ciotti, Ostriker and collaborators (1997, 2010) have investigated a
type of feedback triggered by AGN radiation (mainly Compton heating,
radiation pressure, plus an approximation of a broad-line-region
wind), through 1D hydrodynamic simulations.  The best models have
been analysed in detail by \citet{Pellegrini:2012} and compared to
observations.  The key feature of these models is the formation of a
cold shell at $r\sim 1$ kpc, whose evolution governs the whole
quasar-like feedback heating (in order to be efficient, radiative
feedback requires a very dense - optically thick - absorbing medium).
We note that in 1D simulations a cold shell can numerically form under
several circumstances, with or without AGN heating (see for instance
the `galactic drip' phenomenon described in \citealt{Mathews:1997} and
\citealt{Brighenti:2002}).
%where the drip was instead induced by a central heating.  
It is not clear how 1D calculations are able to realistically follow
the formation and evolution of the shell.  Subsonic, non-radial gas
velocity appearing in multidimensional models tend to suppress this
feature (\citealt{Brighenti:1997}). The cold shell is also
Rayleigh-Taylor unstable and should fragment in about a dynamical
time, $\sim \,{\rm few} \times 10^6$ yr. Another complication is the
difficulty for hydrodynamic simulations to reproduce the real
complexity of the observed multiphase ISM in the inner $<$ kpc (see
Introduction).

Given these uncertainties, it seems more appropriate to compare our
results with those presented in \citet{Novak:2011}, where the 2D
version of the previous feedback scenario for an isolated elliptical
galaxy is investigated.
In these simulations conical
outflows, representing broad absorption line (BAL) winds, 
are generated near the centre of the
grid, making the computed scenario qualitatively similar to our
models. %that proposed in \citet{Brighenti:2006}, G11a,b.
\citet{Novak:2011} found that the accretion process significantly
changes in 2D, reducing the effectiveness of the feedback mechanism
with respect to isotropic 1D models.  They explored the effect of
varying the mechanical efficiency, finding that an acceptable BH
growth is attained when $\epsilon \gta 10^{-4}$, consistent with the
preferred values found in our work.  About $10^{10}$ $M_\odot$ of
stars form in Novak's ellipticals, almost independently on the wind
efficiency, implying 
%rather high cooling and star formation rates
a mean star formation rate of $\sim0.8$ $M_\odot$ yr$^{-1}$.
It would be interesting to estimate the central Balmer indexes
(dependent on the star formation history) and compare them with the
available observations (e.g. \citealt{Kuntschner:2010}).

Unfortunately, \citet{Novak:2011} do not show azimuthally averaged
temperature and density profiles, thus it is not possible to assess
the effect of the AGN feedback on these key observable constraints.
Nevertheless, together with our calculations, the results presented in
\citet{Novak:2011} corroborate the potential importance of bipolar
outflows as a primary source of AGN feedback.
%, although more detailed investigations are clearly needed. 
%At the moment it is not clear if radiative heating is also essential
%in Novak's models.

\citet{Debuhr:2012} carried out 3D SPH simulations of AGN outflows
driven by quasar radiation during a major merger. They noted that
radiation pressure alone does not produce any substantial feedback,
requiring an additional momentum kick with velocity $\sim10^4$ km
s$^{-1}$ (like BAL winds), similar to our best models.  Large values
of the optical depth are usually required to boost such fast outflows
($\tau\sim5-10$).  The strong isotropic outburst produces a clear
galactic outflow, drastically reducing the mass inside 3 kpc (two
orders of magnitude). This is a typically feature associated with
strong isotropic heating (either mechanical or thermal), which unbinds
and shocks the central gas to unobserved levels in normal ellipticals.
Bipolar mechanical
outflows are instead gentler, capable to reproduce the observed
anisotropic X-ray features as cavities (Figure \ref{fig:iso-maps}).

As a concluding remark, we note that the radiative feedback displays a
strong impact on the ISM only in the presence of quasar-like objects,
which are certainly rare in the local universe, with a rapid decline
below redshift $\sim1.8$ (\citealt{Schneider:2005}).

\subsection{Comparison with galaxy clusters and groups}

After a systematic study of AGN outflow feedback, from the cluster
scales down to the galactic systems (\citealt{Gaspari:2009,Gaspari:2011a,Gaspari:2011b,
Gaspari:2012}), it is worth to
wrap-up and emphasise the main similitudes and differences.

Despite the very different environments of clusters, groups and
ellipticals, purely mechanical AGN feedback, in the form of massive anisotropic
outflows, is anyway able to suppress the cooling flow down to a factor
$\la5-10\%$. The self-regulation process is fundamental for preserving
the cool-core structure, in a state of quasi thermal equilibrium.

In order to have a proper self-regulation, the AGN feedback seems to
require mechanical efficiencies decreasing with the halo mass:
$\epsilon \sim5\times10^{-3}$ (cluster), $\sim8\times10^{-4}$ (group) and
$\sim3\times10^{-4}$ (isolated elliptical). We stress that this
$\epsilon$ does not probably represent the actual efficiency.
The accreted mass on the black hole may be lower,
while the feedback power could be maintained on the same level by
increasing the efficiency. Nevertheless, the
increasing efficiency with system mass trend could be real.
% \footnote{In lighter systems we progressively adopted higher
% resolution (from $\sim2.7$ kpc to $150$ pc), thus getting closer to
% the actual accretion rate.}.
In fact, the same mechanical feedback successful in massive clusters 
($M_{\rm vir}\sim10^{15}$ $M_\odot$) needs to work on a relatively 
less powerful regime in order to prevent the core disruption in groups
or galaxies ($\la4\times10^{13}$ $M_\odot$), which are less bound
objects. 
Even with a lower $\epsilon$, the feedback in lighter halos has still
more dramatic consequences on the temperature and density profiles, 
with the stronger outbursts easily perturbing the central region. 
The isolated galaxy is the exemplary case, in which single outflows of
power $\sim10^{44}$ erg s$^{-1}$ and $\epsilon=3.3\times10^{-4}$ can
eject the gas outside the small system, stopping cooling for several
tens Myr. Applying a feedback 20 times stronger, as required for clusters,
would eject the whole ISM and empty the galaxy for several Gyr.
The physical reason for the mechanical efficiency to be linked to the 
environment/potential well is unclear and needs to be clarified with
future investigations. The fact that the most massive black holes reside 
at the centre of clusters (\citealt{McConnell:2011}) could play a crucial role.

The cooling flow problem in our simulations is solved through AGN outflows
for more than 7-10 Gyr, despite the exact details of the feedback
mechanism (cf. G11a and \citealt{Gaspari:2012}). 
In the cluster calculations the resulting feedback was impulsive, with
strong outbursts sometimes exceeding Eddington power. 
Increasing the resolution leads to a higher outflow frequency (and
lower powers), because the central inflowing material is able to
accrete more easily along the direction perpendicular to the jet axis. 
% where a rotationally-supported torus would usually form (if the cold
% gas is not dropped out).
The AGN outburst frequency is also correlated to the efficiency, e.g.
in the group the outflow injection is almost continuous with a duty
cycle around 0.8. 
The isolated galaxy is a slightly different case, because the absence of the
circumgalactic gas helps the feedback action, leading to a mildly
impulsive behaviour.  

The galaxy simulation cgg-8em4 (Sec. 4) can be regarded as a
convergence test of the best feedback group model in G11b. The
increase in resolution is almost a factor 2 and we find indeed
consistent results.  Cooling rates and profiles are similar,
concordant with observational data. Also the jet powers
($\sim3-5\times10^{44}$ erg s$^{-1}$) are similar, showing an
analogous evolutionary pattern and frequency. It is worth noting that 
the jet active region is $R \times z \simeq 1\times1$ kpc in the CGG model,
while $1\times2$ kpc in the group run presented in G11b. Therefore, a slightly longer jet does
not alter the feedback properties (the same can be said for the width,
as tested in previous models).

In all the best models the self-regulated outflow velocities are
around $10^4$ km s$^{-1}$, with rarer events reaching several
$10^4-10^5$ km s$^{-1}$ in the strongest phases. The mean mass outflow
rate decreases with the mass of the system: several tens $M_\odot$
yr$^{-1}$ (cluster), few $M_\odot$ yr$^{-1}$ (group), and $\la 1$
$M_\odot$ yr$^{-1}$ (isolated galaxy). More massive and slower outflows
usually show a higher piercing power, producing a narrow
unidirectional channel. This is more evident in almost continuous
feedback models, like galaxy groups or galaxies with CGG. However, the
projected X-ray surface brightness maps mask this narrow feature
($\sim 1$ kpc) most of the times. Moreover, the turbulence generated
by the feedback can easily alter and fragment the jet path, producing
tiny bubbles. On the contrary, relatively lighter and faster outflows (still subrelativistic) 
thermalise the energy more efficiently, inflating big buoyant bubbles. 

In the
cluster regime the bubbles have usually a radius of few tens kpc,
with emission-weighted temperatures slightly hotter than the ambient
medium; the reduced power in groups and ellipticals produces more
`delicate' cavities (with radius several and few kpc, respectively),
showing relatively cold rims and mild internal temperatures; the jump
in SB$_{\rm X}$ is typically $20-40\%$. In all three systems,
different or fragmented jet outbursts generate also a series of weak
shocks with Mach number $1.1-1.3$, visible as faint ripples in the
temperature and brightness profiles. Another fundamental feature they all have
in common is the uplift of relatively cold and metal-rich gas, in
particular during more powerful episodes, restoring the lost potential energy and 
creating the asymmetrical distribution in the iron abundance maps ($10-20\%$
contrast with the background).

% The heating linked to supernovae Ia and stellar winds in the central
% galaxy ($M_{\ast}\approx7\times10^{11}$ $\msun$) has been modeled in
% the same way in all three objects, employing identical SNIa and
% specific SW rate, $\propto0.06\, t^{-1.1}$ SNu and $\propto
% 4.7\times10^{-20} t^{-1.3}$ s$^{-1}$, respectively. In the cluster
% and group runs the heating is a few orders of magnitude lower
% compared to the injected AGN energy (few $10^{61}-10^{62}$ erg) and
% thus energetically irrelevant. In the isolated elliptical the AGN
% energy barely reaches $10^{60}$ erg; here the stellar feedback has a
% significant -- although not resolutive -- role, as also emerges from
% the CF run (the cooling rate decreases from 6 to 1 $\msun$ yr$^{-1}$
% even without AGN jets).

The most successful models are those where the accretion is linked to cold
gas. Hot gas accretion, based on the original Bondi prescription, usually provides
accretion rates one or two orders of magnitude lower, requiring very
high mechanical efficiencies, $\epsilon\ga0.1$, in order to make the feedback effective.
Moreover, its continuous nature poses serious doubts on the cyclic
production of typical features, such as buoyant cavities. Linking the accretion to the cooling rate 
induces instead a natural self-regulation, 
with some kind of recursive cycle and peaks of activity.
The success of the feedback is essentially independent of the dropout
parameters, such as $T_{\rm cut}$ (e.g. $2\times10^4$ or
$5\times10^5$ K), the dropout function (exponential or step), %Max: corretto
or the delay associated with the free fall of cold blobs.
The simulations carried out in \citet{Gaspari:2012}, with the accretion traced only by the
inflow in the nuclear region, confirm that the bulk of AGN fuelling
is associated with the cold -- and not hot -- phase (check $\dot{M}_{\rm acc}$ in fig. 3), whose condensation is linked to jet-induced thermal instabilities.\\

\section{Cardinal conclusions}

For all the reasons argued above 
and based also on our previous theoretical numerical studies on AGN heating,
we find that the best feedback mechanism able to {\it solve the cooling flow problem} in every virialised system with a substantial hot gas halo, requiring
\begin{itemize}
\item
no overcooling ($\dot{M}_{\rm cool}<5-10\%\ \dot{M}_{\rm classic\,CF}$),\\
\item
no overheating (preserve the cool-core structure),\\
\end{itemize}
and reproducing several observational features, such as
\begin{itemize}
\item
buoyant underdense bubbles,\\
\item
elliptical cocoons,\\
\item
weak shocks/sonic ripples (Mach $\lta 1.5$),\\
\item
dredge-up of metals and cold gas,\\
\item
subsonic turbulence (hundreds km s$^{-1}$),\\
\item
extended filamentary and nuclear cold gas,\\
\end{itemize}
should possess the following key characteristics:
%must possess the following fundamental characteristics:
\begin{itemize}
\item
mechanical, \\ %(for gradual thermalisation),\\
\item
anisotropic, \\ %(for generating cavities and allowing moderate accretion),\\
\item
driven by massive and subrelativistic outflows,\\
\item
self-regulated by cold accretion,\\
\item
and less efficient in lighter halos.\\
\newline
\end{itemize}

%After this extensive theoretical study on AGN heating, we can conclude that
%anisotropic mechanical feedback is a robust model, consistent with 
%several observational constraints for a wide mass range of systems.
%Not only AGN outflows can solve the cooling flow problem, drastically
%quenching the cooling flow and at the same preserving the cool core,
%but they also provide a natural explanation for the complex
%heating cycle commonly observed in galaxies, groups and clusters (see references quoted in the Introduction): early strong shock injection, inflation of buoyant underdense cavities, 
%weak shock/sonic ripples, driving of turbulent and stirring motions, dredge-up of cold
%and enriched material.

\section*{Acknowledgments}
The software used in this work was in part developed by the DOE NNSA-ASC OASCR Flash Center at the University of Chicago. We acknowledge the NASA awards SMD-10-1609, SMD-11-2209 (Pleiades), and
the CINECA awards HP10BPTM62, HP10BOB5U6 (SP6). 
Partial support for this work was provided by NASA under grant NNH09ZDA001N,
issued through the Office of Space Sciences Astrophysics Data Analysis
Program. We thank also J. de~Plaa, E. Churazov and P. Sharma for helpful comments,
and the referee for a constructive report.
%We would like to thank P. Temi as the principal support scientist
%at the NASA/Ames base. 

\bibliographystyle{mn2e}
\bibliography{biblio}

\label{lastpage}

\end{document}